\documentclass[letterpaper]{article} 
\usepackage{aaai24}
\usepackage{times}  
\usepackage{helvet}  
\usepackage{courier}  
\usepackage[hyphens]{url}  
\usepackage{graphicx} 
\usepackage{array}
\usepackage{adjustbox}
\usepackage{natbib}  
\usepackage{caption} 
\usepackage{algorithm}
\usepackage{algorithmic}
\usepackage{multirow}
\usepackage{xcolor}
\usepackage{subfig}
\usepackage{subcaption}
\usepackage{newfloat}
\usepackage{listings}
\DeclareCaptionStyle{ruled}{labelfont=normalfont,labelsep=colon,strut=off} 
\floatstyle{ruled}
\newfloat{listing}{tb}{lst}{}
\floatname{listing}{Listing}
\newcommand{\paragraphb}[1]{\noindent{\bf #1} }

\title{Memory Triggers: Unveiling Memorization in Text-To-Image Generative Models through Word-Level Duplication}
\author {
    Ali Naseh,\textsuperscript{\rm 1}
    Jaechul Roh,\textsuperscript{\rm 1}
    Amir Houmansadr\textsuperscript{\rm 1}
}
\affiliations {
    \textsuperscript{\rm 1} University of Massachusetts Amherst\\
    anaseh@cs.umass.edu, jroh@umass.edu, amir@cs.umass.edu
}

\makeatletter
    \renewcommand{\copyright@on}{F}
\makeatother
\usepackage{bibentry}

\begin{document}

\maketitle

\section{Abstract}

Diffusion-based models, such as the Stable Diffusion model, have revolutionized text-to-image synthesis with their ability to produce high-quality, high-resolution images. These advancements have prompted significant progress in image generation and editing tasks. However, these models also raise concerns due to their tendency to memorize and potentially replicate exact training samples, posing privacy risks and enabling adversarial attacks. Duplication in training datasets is recognized as a major factor contributing to memorization, and various forms of memorization have been studied so far. This paper focuses on two distinct and underexplored types of duplication that lead to replication during inference in diffusion-based models, particularly in the Stable Diffusion model. We delve into these lesser-studied duplication phenomena and their implications through two case studies, aiming to contribute to the safer and more responsible use of generative models in various applications.

\section{Introduction}

\emph{Diffusion-based} models~\cite{sohl2015deep} have demonstrated outstanding ability in producing high-quality images, both \emph{unconditionally}~\cite{ho2020denoising} and \emph{conditionally}~\cite{rombach2022high, nichol2021glide}. The \emph{Stable Diffusion} model~\cite{rombach2022high}, a type of conditional diffusion model, alongside other generative models like DALLE-3~\cite{shi2020improving} and Midjourney~\cite{midjourney2022}, has significantly advanced the field of text-to-image synthesis. These models excel in creating high-resolution images~\cite{ramesh2022hierarchical} and in image editing~\cite{kim2021diffusionclip}.

Memorization across various machine learning models has been extensively researched~\cite{carlini2019secret, zhang2021understanding}. Such memorization can pose privacy risks, potentially enabling attacks such as membership inference~\cite{shokri2017membership} or data extraction~\cite{carlini2021extracting}. Despite the capabilities of diffusion-based models, including the Stable Diffusion model, to produce high-quality images, they occasionally exhibit tendencies to memorize and replicate exact training samples or significant portions thereof~\cite{carlini2023extracting, somepalli2023diffusion, somepalli2023understanding}. \citeauthor{somepalli2023understanding}~\shortcite{somepalli2023understanding} suggest that text conditioning contributes more to memorization than image-only contexts. Previous research has shown that training sample duplication could be a significant cause of such replication during inference. 

This paper delves into two particular types of text-conditioned training sample duplication: the first involves duplication where the images, along with their corresponding texts containing specific keywords, are repeated; the second type pertains to duplication of image-text pairs where the images contain specific objects and the texts include specific keywords. We posit that this nuanced form of duplication may heighten the models' vulnerability to various attacks. With the burgeoning popularity of text-to-image generative models, a detailed scrutiny of their memorization propensities becomes increasingly crucial. We explore these two types of duplication by analyzing two case studies that shed light on their dynamics and implications.

\section{Related Works}

\subsection{Memorization in Large Language Models}
In the domain of Large Language Models (LLMs), there is an escalating challenge pertaining to the inadvertent disclosure of confidential information leading to model memorization~\cite{lee2023language, carlini2021extracting, carlini2023extracting, biderman2023emergent}. 
\citeauthor{carlini2022quantifying}~\shortcite{carlini2022quantifying} perform in-depth analyses using quantitative methods, whereas~\citeauthor{carlini2021extracting}~\shortcite{carlini2021extracting} conduct their studies utilizing qualitative techniques. One of the pivotal factor causing this scenario is the replication inherent within training datasets, potentially causing the language model to generate text that mirrors already existing content~\cite{lee2023language}. A recent contribution by~\citeauthor{biderman2023emergent}~\shortcite{biderman2023emergent} posit that such memorization manifests due to the average of the training dataset, while~\citeauthor{biderman2023emergent}~\shortcite{biderman2023emergent} illustrate its occurrence at specific training data point. 

\subsection{Memorization in Diffusion Models}
Recent studies~\cite{carlini2023extracting, somepalli2023understanding, somepalli2023diffusion} demonstrate techniques to create alike or almost identical images using both conditional and unconditional diffusion models. Specifically,~\citeauthor{somepalli2023diffusion}~\shortcite{somepalli2023diffusion} highlight that diffusion models can produce images with objects similar to those in the training data, in which the process is termed as 'replication'~\cite{somepalli2023diffusion}. In parallel,~\citeauthor{carlini2023extracting}~\shortcite{carlini2023extracting} demonstrate the model's proficiency in retrieving near-identical images from the training set by analyzing clusters of generated samples. Recent study by~\citeauthor{somepalli2023understanding}~\shortcite{somepalli2023understanding} posit that while data replication stemming from duplication might be infrequent in unconditional diffusion models, textual conditioning can notably augment the likelihood of model memorization. Echoing the occurrence from prior research that data duplication underlies such memorization,~\citeauthor{webster2023duplication}~\shortcite{webster2023duplication} introduce an algorithmic approach to detect such repetitions. 
 
\section{Background}

\subsection{Diffusion Model}
In the context of deep generative models, Denoising Diffusion Probabilistic Models~\cite{ho2020denoising}, commonly referred to as unconditional diffusion models, iteratively engage in the act of noise addition (forward process) and subsequent noise removal (backward process) for image generation. 

The forward phase embodies a Markov chain structure, \(q(x_t | x_{t-1})\) progressively injecting Gaussian noise into the data (\(x_0\)) until it reaches a fully noise-perturbed image (\(x_t\)). Conversely, the backward process engages in a denoising mechanism, systematically removing the noise that exists in the previous timestep, adhering to the Markov Chain \(p(x_{t-1} | x_t)\). 

\subsection{Stable Diffusion}
In the work by~\citeauthor{rombach2022high}~\shortcite{rombach2022high}, the "Stable Diffusion" model tailors particularly for text-to-image synthesis tasks. The model operates by diffusing the latent vector representation of an image. It begins by receiving textual input, which is then transformed into a text embedding via the frozen CLIP text encoder~\cite{radford2021learning}. Subsequently, a text-conditional latent U-Net iteratively denoises the latent vector in a manner conditioned on the generated text embedding. Finally, a Variational Autoencoder (VAE)~\cite{kingma2013auto} decodes this latent vector, producing the corresponding image. 
\section{Word-level Duplication}

\citeauthor{carlini2023extracting}~\shortcite{carlini2023extracting} introduces a definition of memorization wherein an example \( x \) is considered extractable from a diffusion model \( f_{\theta} \) if, without using \( x \) as an input, there exists an efficient algorithm \( A \) such that \( \hat{x} = A(f_{\theta}) \) satisfies the condition \( l(x, \hat{x}) \leq \delta \). This definition emphasizes replications that produce images nearly identical to the original. However, our focus is on a broader understanding of memorization, which we term \emph{partial replication}. This pertains to specific objects or features within an image. Carlini's metric might not always capture such memorization; it may indicate a high \( l_2 \) distance even when recognizable memorization exists within images.

\citeauthor{somepalli2023understanding}~\shortcite{somepalli2023understanding} investigates a broader scope of duplication in the LAION dataset, covering more cases than previous studies. They consider both caption and image duplications and even delve into partial caption duplication. However, there are concerns with their approach. They curated two subsets from the LAION dataset for fine-tuning the Stable Diffusion model. Fine-tuning the Stable Diffusion on a subset of its original pre-training dataset might result in increased unintended memorization.

In text-conditioned  diffusion models, we believe that text plays a crucial role. Given this perspective, concerns should predominantly revolve around text-conditioned memorization in these applications. While image duplications might exist in the dataset, without a connection between the text and the image, it is improbable that related replication would emerge during inference time when supplying our prompt. This observation leads us to consider more realistic types of replication.

Unlike prior research, our focus is on \emph{word-level duplication}. Specifically, we aim to discern any associations between keywords and images in duplications. We question whether certain sets of keywords and images are consistently replicated within the dataset. In this context, captions don't necessarily exhibit high semantic similarity; they might only share common keywords. Consequently, during inference, when the model encounters these specific keywords in combination, it might attempt to reproduce the corresponding features or objects observed during training. In our experimental results, we further explore this type of duplication through a detailed case study using the LAION dataset.

\textbf{A more realistic approach to defining memorization:} Previous studies have often relied on a single random initialization for generations~\cite{somepalli2023understanding}. However, irrespective of the memorization definition employed, we argue that a more realistic examination involves using multiple random initializations. Essentially, in practical settings, concerns about memorization and replication arise if the model consistently generates the same feature, object, or even the entire image across different initializations. Thus, assessing memorization or replication based on a single seed might not provide a comprehensive understanding.

\renewcommand{\arraystretch}{1.2}
\begin{table*}[h]
    \centering
    \caption{Largest clusters with corresponding frequent words and their frequencies.}
    \begin{tabular}{cc}
        \hline
        \textbf{Cluster ID} & \textbf{Keywords with Frequencies} \\
        \hline
        1 & \textbf{van}: 3061, \textbf{gogh}: 3042, \textbf{night}: 2841, \textbf{starry}: 2806 \\
        2 & \textbf{van}: 1950, \textbf{gogh}: 1937, \textbf{vincent}: 1374, \textbf{self-portrait}: 795, \textbf{portrait}: 674 \\
        3 & \textbf{van}: 1839, \textbf{gogh}: 1833, \textbf{almond}: 1764, \textbf{vincent}: 1201, \textbf{tree}: 1129, \textbf{blossoming}: 1003 \\
        4 & \textbf{van}: 1725, \textbf{gogh}: 1715, \textbf{sunflowers}: 1549, \textbf{vincent}: 1110, \textbf{vase}: 601 \\
        5 & \textbf{van}: 1628, \textbf{gogh}: 1622, \textbf{terrace}: 1477, \textbf{cafe}: 1313, \textbf{vincent}: 1135, \textbf{night}: 1034, \textbf{arles}: 530 \\
        6 & \textbf{van}: 1035, \textbf{gogh}: 1032, \textbf{night}: 955, \textbf{starry}: 925, \textbf{rhone}: 862, \textbf{vincent}: 597 \\
        7 & \textbf{van}: 906, \textbf{gogh}: 899, \textbf{irises}: 807, \textbf{vincent}: 586 \\
        \hline
    \end{tabular}
    \label{tab:clusters}
\end{table*}

\section{Object-level Duplication}

In this section, we introduce a distinct type of duplication termed \emph{object-level duplication}. This occurs when a pair consisting of specific objects in an image and certain keywords in the corresponding text is duplicated in the training dataset, even if the object's name does not appear in the text. Such duplication can lead to the replication of these specific objects during inference when the related keywords are present in the prompt. This pattern of replication raises various trustworthiness concerns, notably privacy and fairness. Essentially, it implies that the model persistently generates specific objects, irrespective of their mention or absence in the user-provided input, which may not align with user expectations or intentions.

A plausible explanation for this phenomenon might be the concealed correlation between certain keywords and objects within an image. That is, while entire images may not be duplicated in the training dataset, specific objects might frequently appear in images associated with captions containing a particular word. We delve into this phenomenon with a dedicated case study in the experiments section.

\section{Experimental Results}

\begin{figure}[ht]
\centering
\includegraphics[width=\columnwidth]{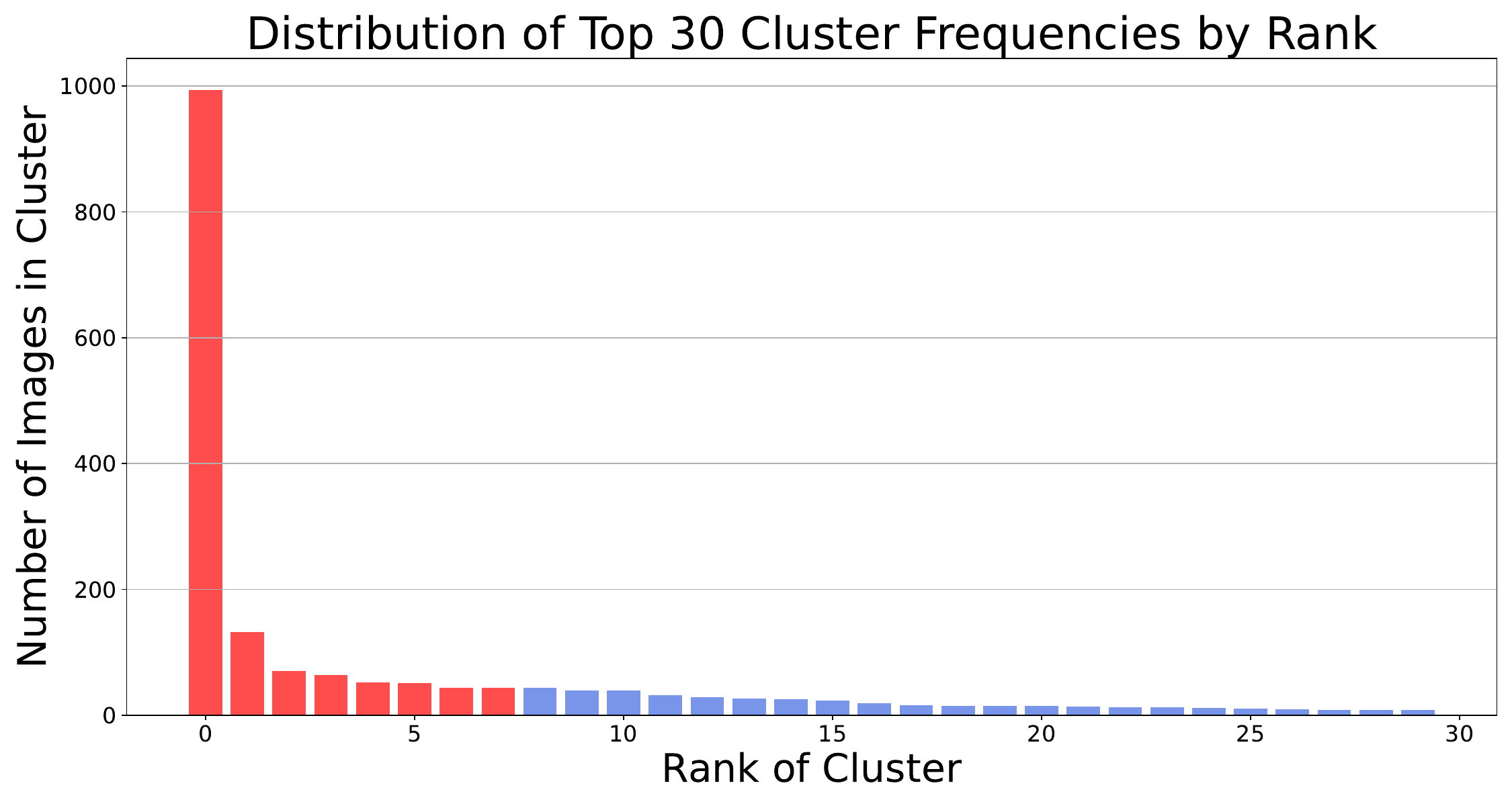}
\caption{Frequency distribution of samples containing the words "almond" and "blossoming". The red bars represent clusters related to the Van Gogh case.}
\label{fig:plot1}
\end{figure}

\begin{figure}[ht]
\centering
\includegraphics[width=\columnwidth]{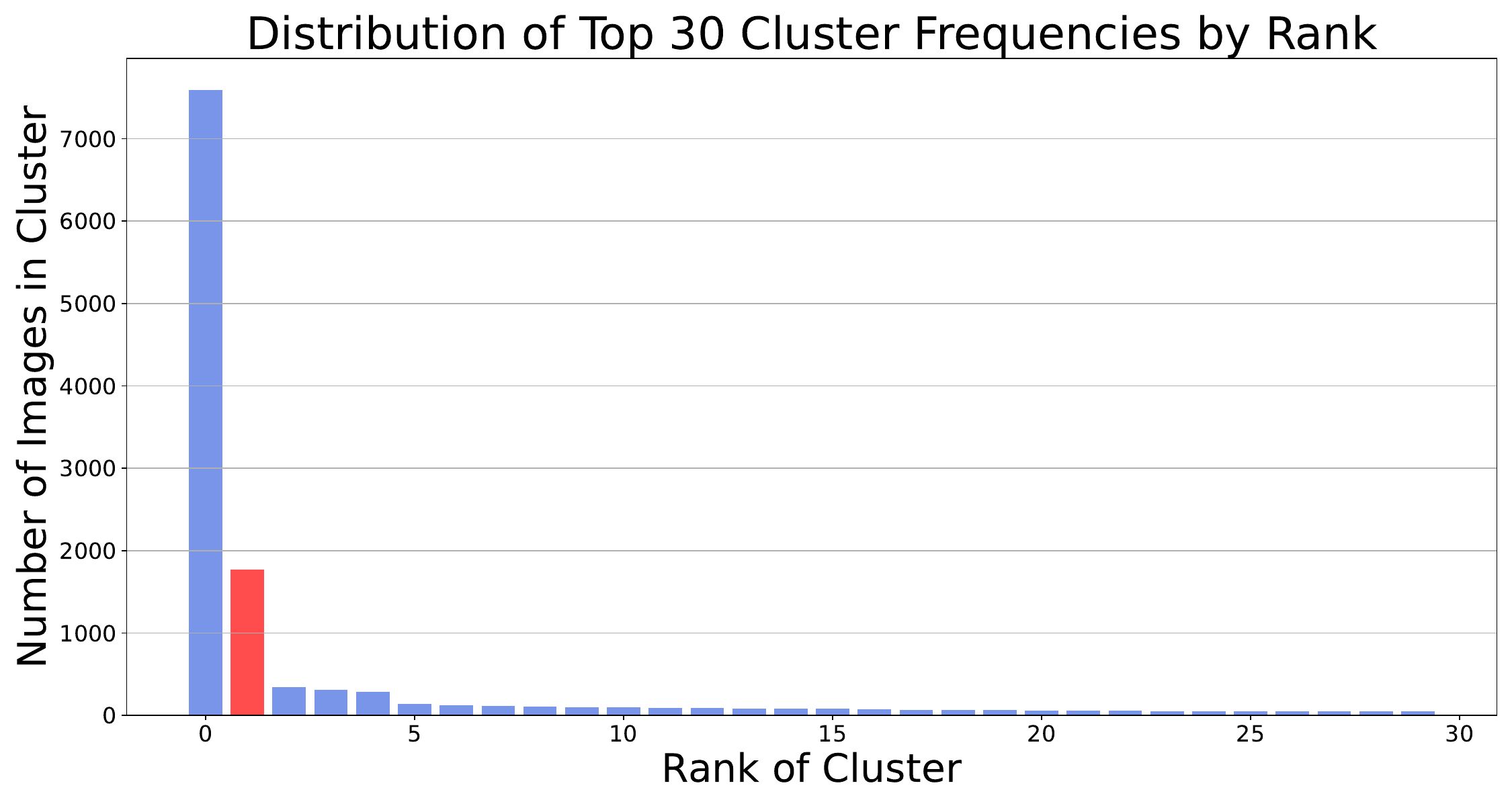}
\caption{Frequency distribution of samples containing the word "sunflowers". The red bars represent clusters related to the Van Gogh case.}
\label{fig:plot2}
\end{figure}

\begin{figure*}[h]
    \centering
    \begin{adjustbox}{width=17.5cm,center}
    \begin{tabular}{>{\centering\arraybackslash}m{4cm}>{\centering\arraybackslash}m{1.5cm}>{\centering\arraybackslash}m{2cm}*{4}{m{2cm}}m{2cm}}
        \hline
        \multirow{2}{*}{\centering \textbf{Prompt}} & \multirow{2}{*}{\parbox{1.5cm}{\centering \textbf{Text\\Similarity}}} & \multirow{2}{*}{\parbox{2cm}{\centering \textbf{Image Sim $>$ 0.83 (\%)}}} & \multicolumn{4}{c}{\textbf{Example Generated Images}} & \multirow{2}{*}{\parbox{2cm}{\centering \textbf{Original Image}}} \\
        \cline{4-7}
        & & & \centering \textbf{$>$ 0.85} & \centering \textbf{0.80-0.85} & \centering \textbf{0.70-0.80} & \centering \textbf{$<$0.70} & \\
        \hline
        \raggedright  Van Gogh starry night & 1.0000 & 64.6\% & 
        \raisebox{-0.05\height}{\includegraphics[width=2cm]{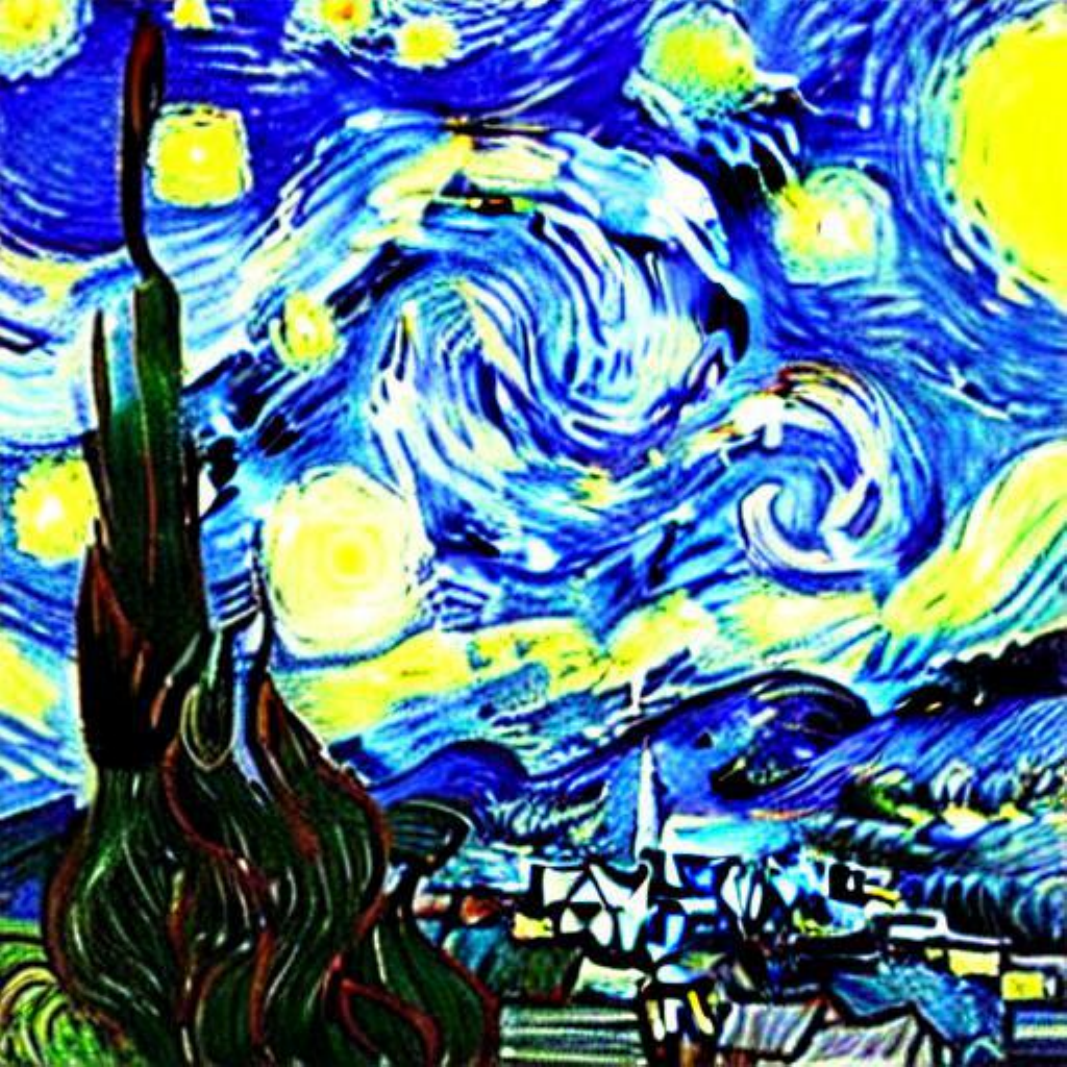}} &
        \raisebox{-0.05\height}{\includegraphics[width=2cm]{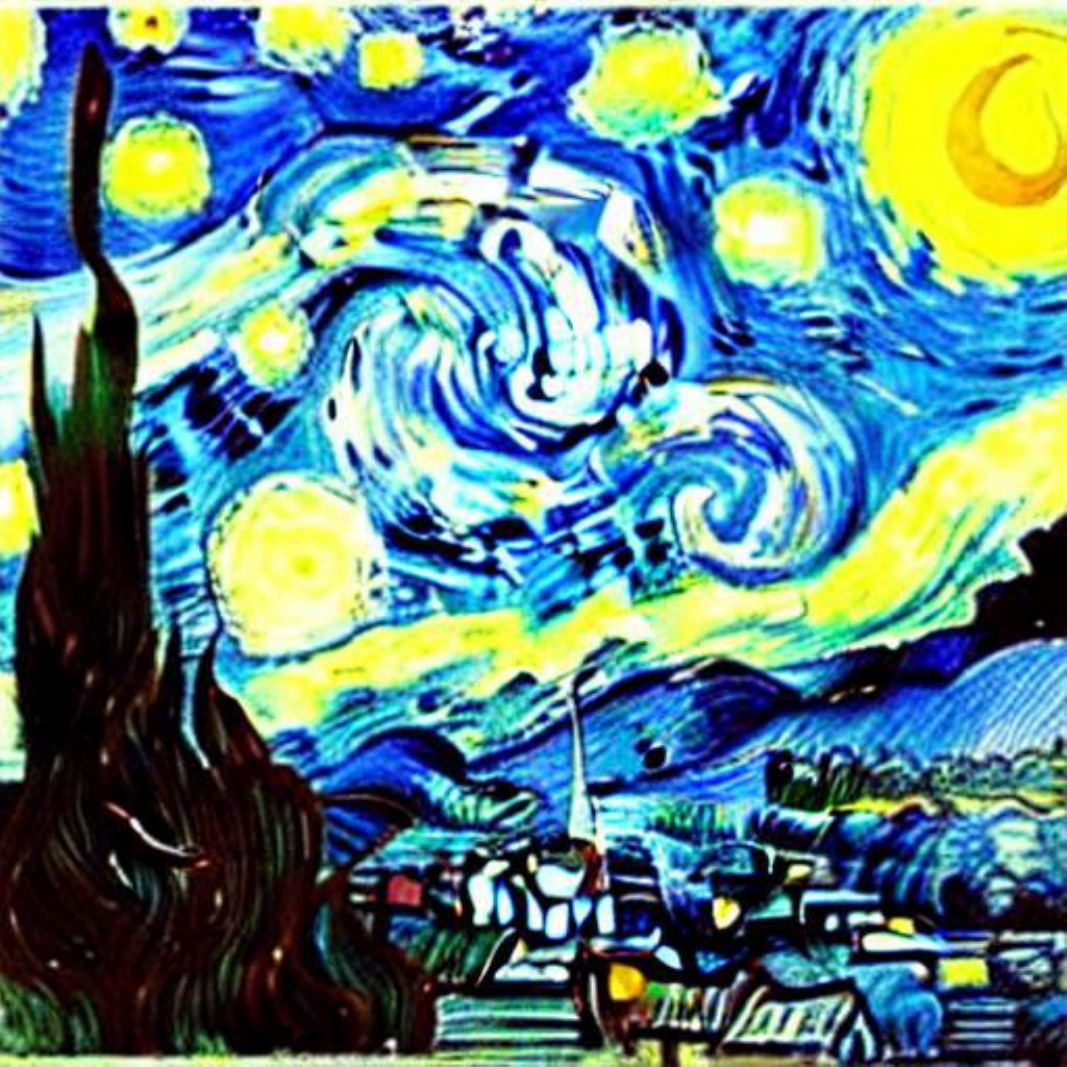}} &
        \raisebox{-0.05\height}{\includegraphics[width=2cm]{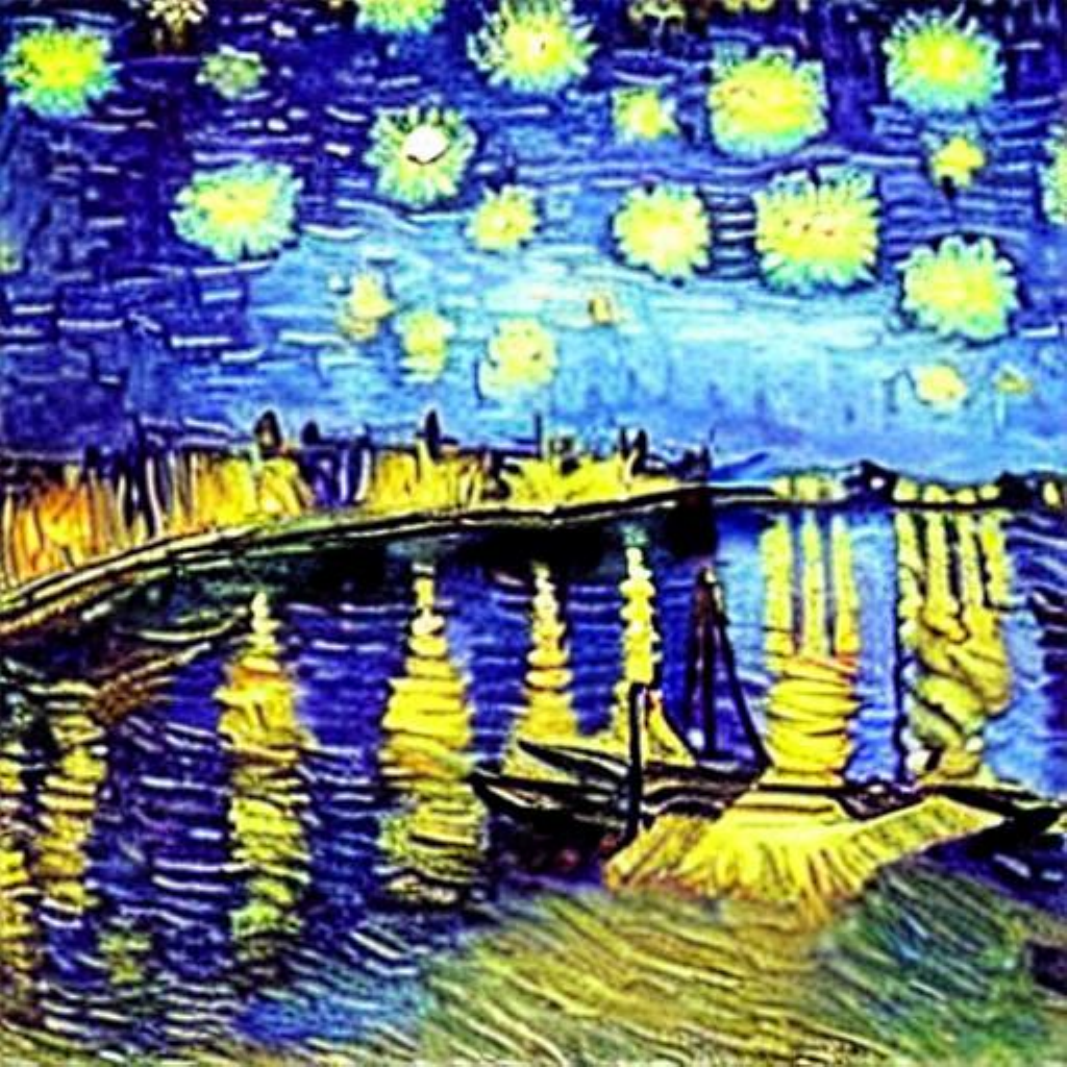}} &
        \raisebox{-0.05\height}{\includegraphics[width=2cm]{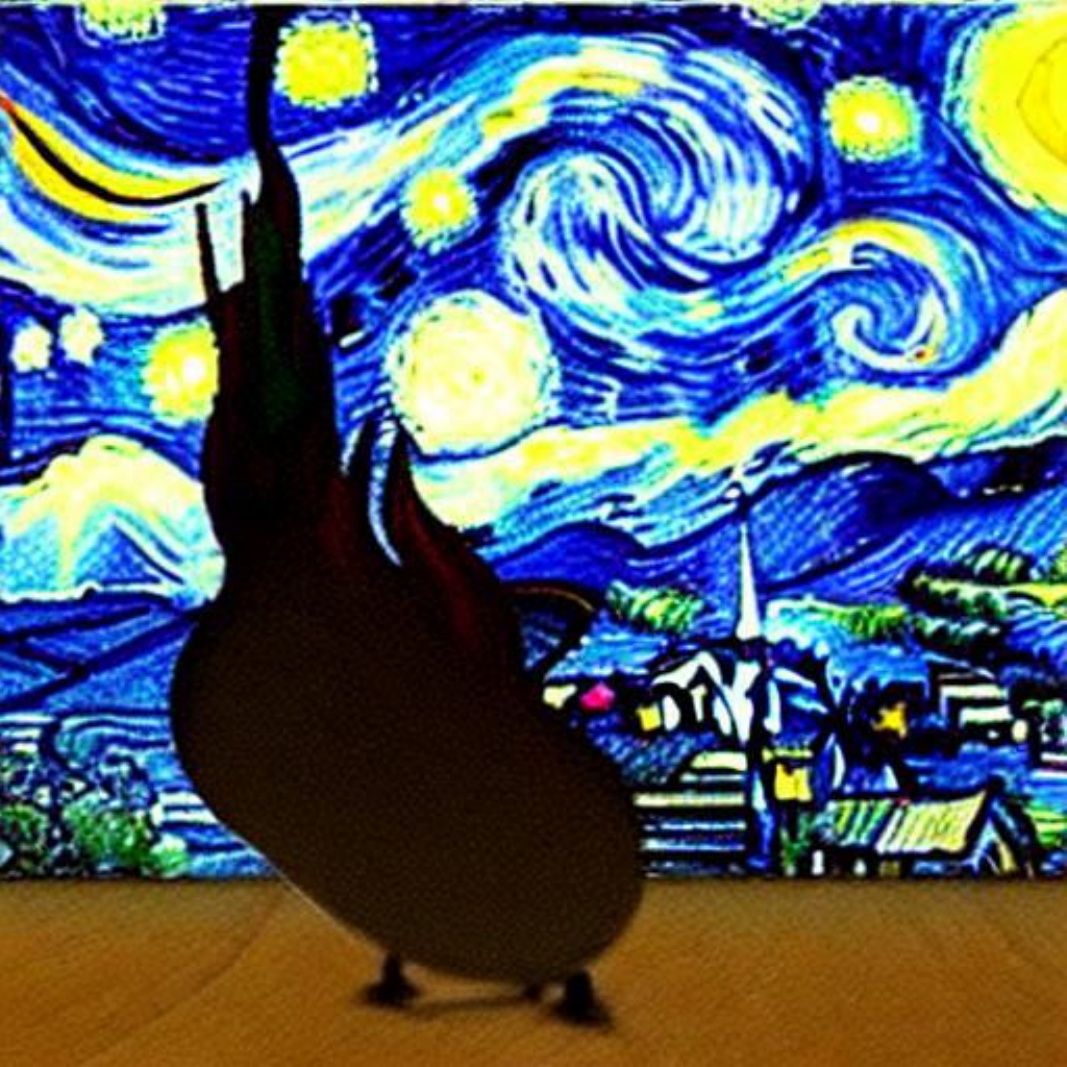}} &
        \raisebox{-0.05\height}{\includegraphics[width=2cm]{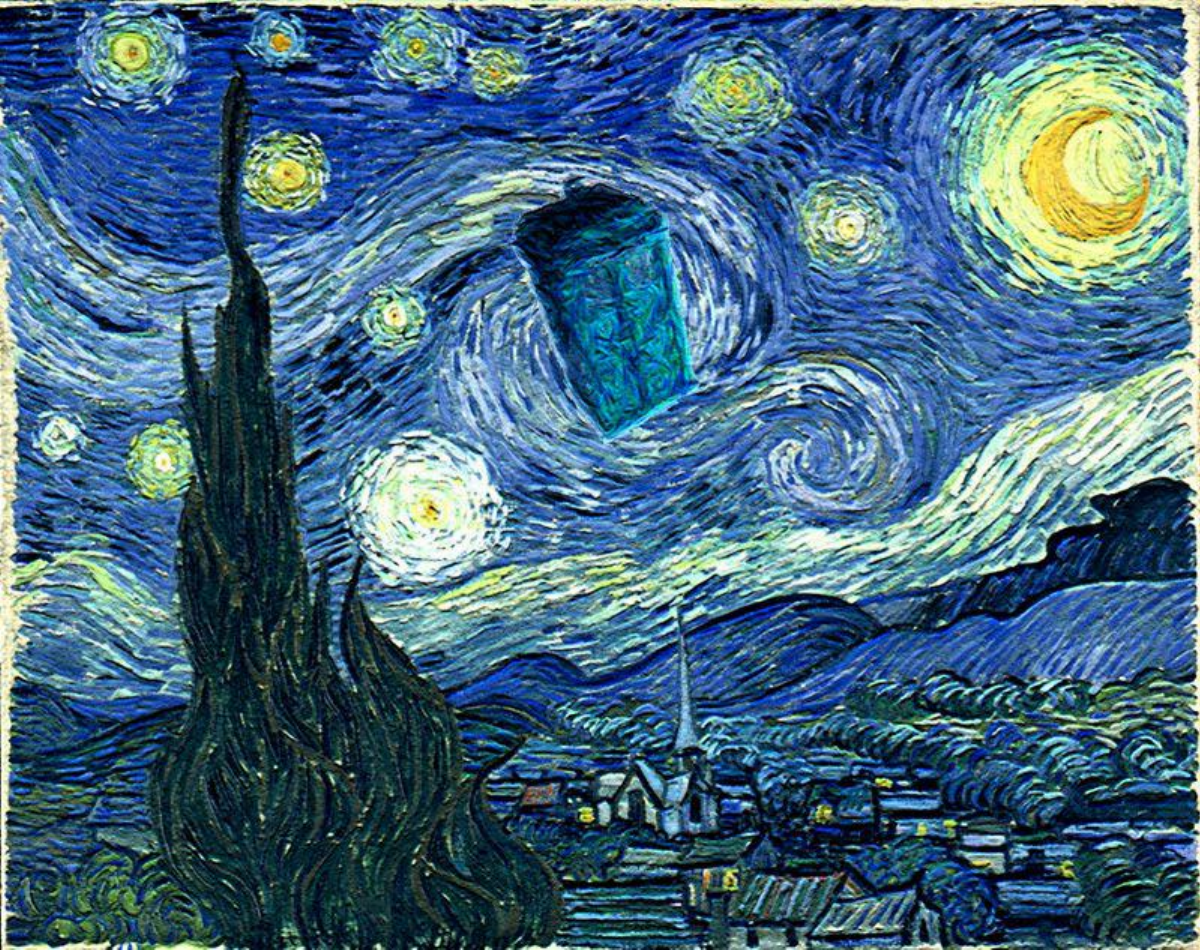}}\\
        \hline
        \raggedright A  \textcolor{red}{starry} \textcolor{red}{night} landscape filled with \textcolor{red}{Van Gogh}'s vibrant colors. & 0.8279 & 56.8\% &
        \raisebox{-0.05\height}{\includegraphics[width=2cm]{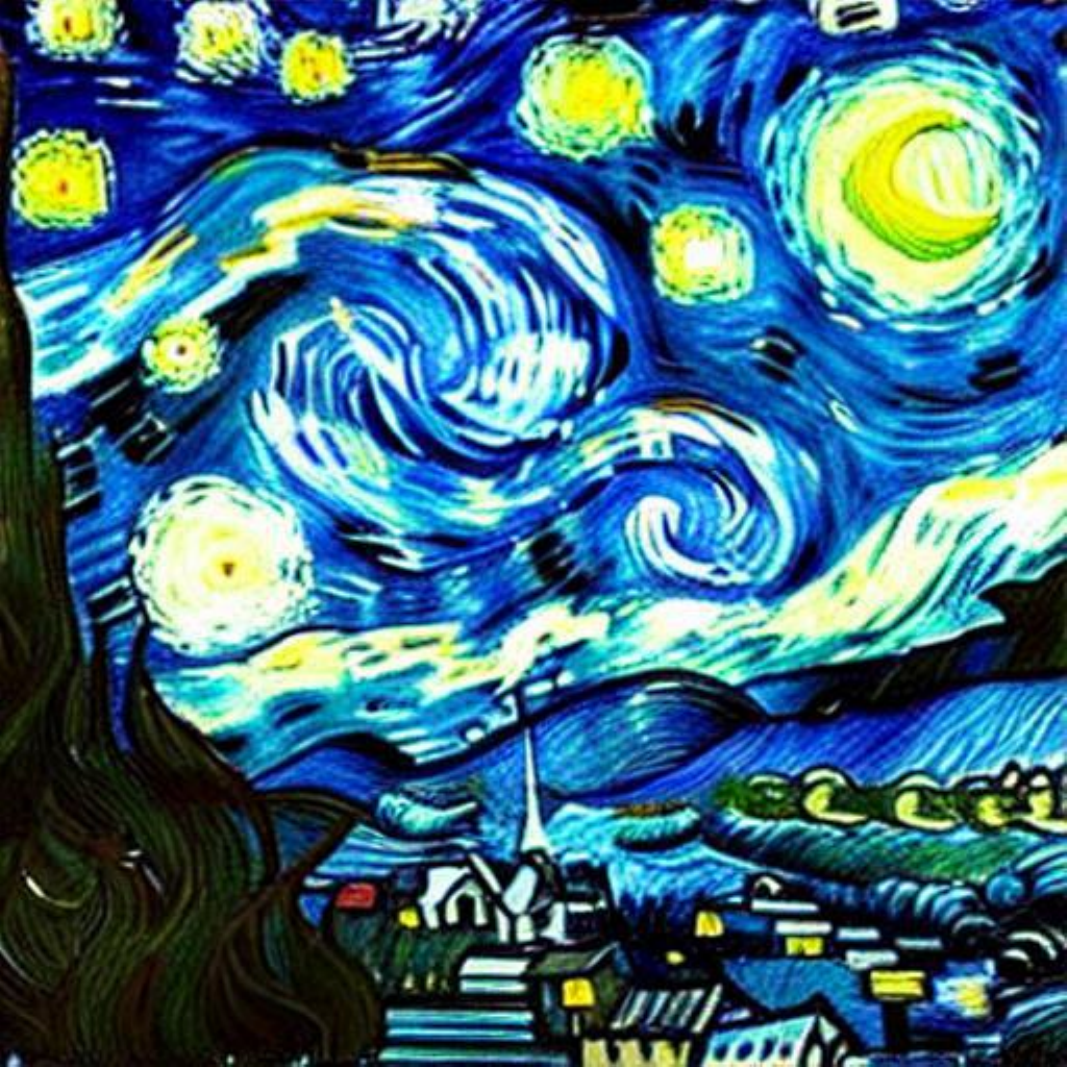}} &
        \raisebox{-0.05\height}{\includegraphics[width=2cm]{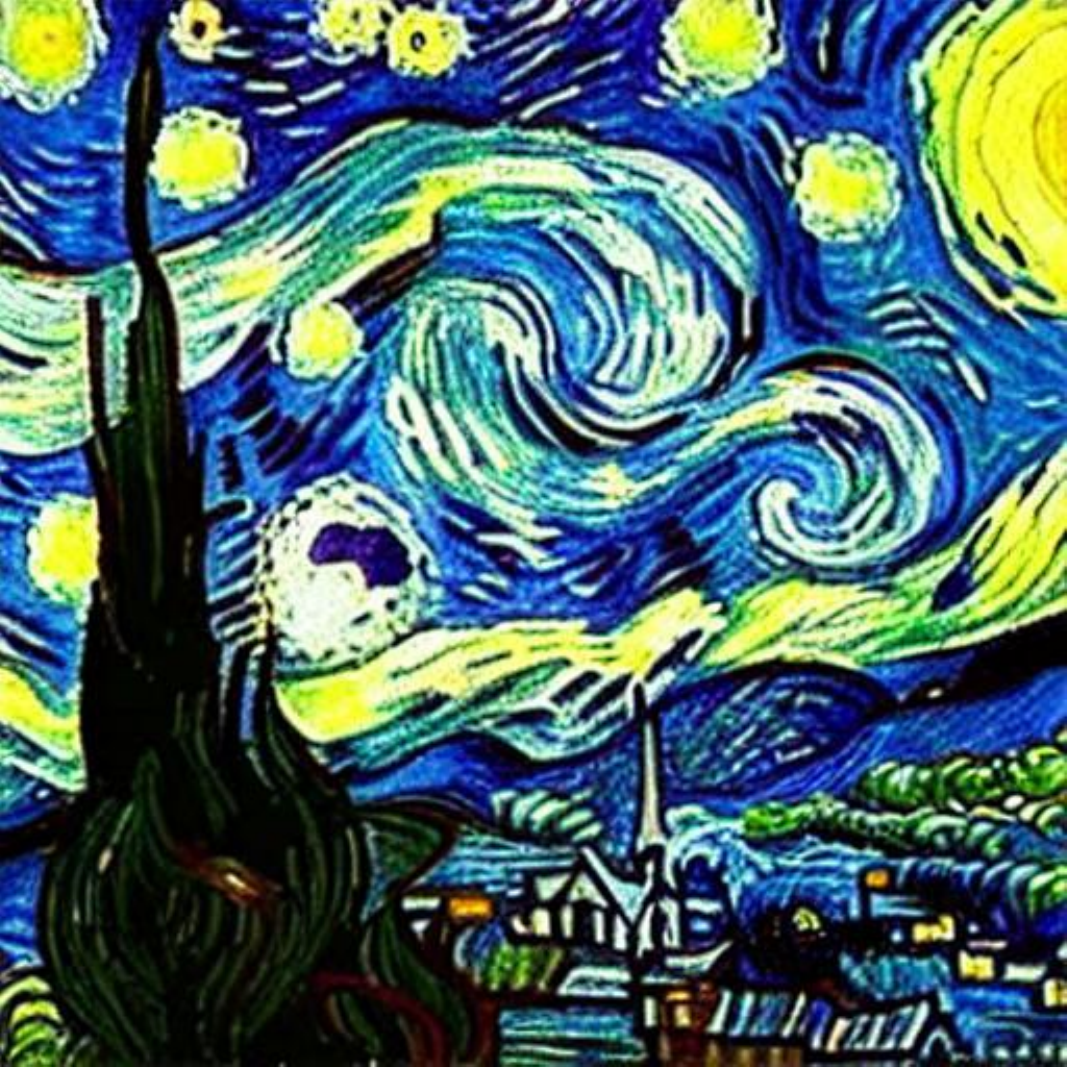}} &
        \raisebox{-0.05\height}{\includegraphics[width=2cm]{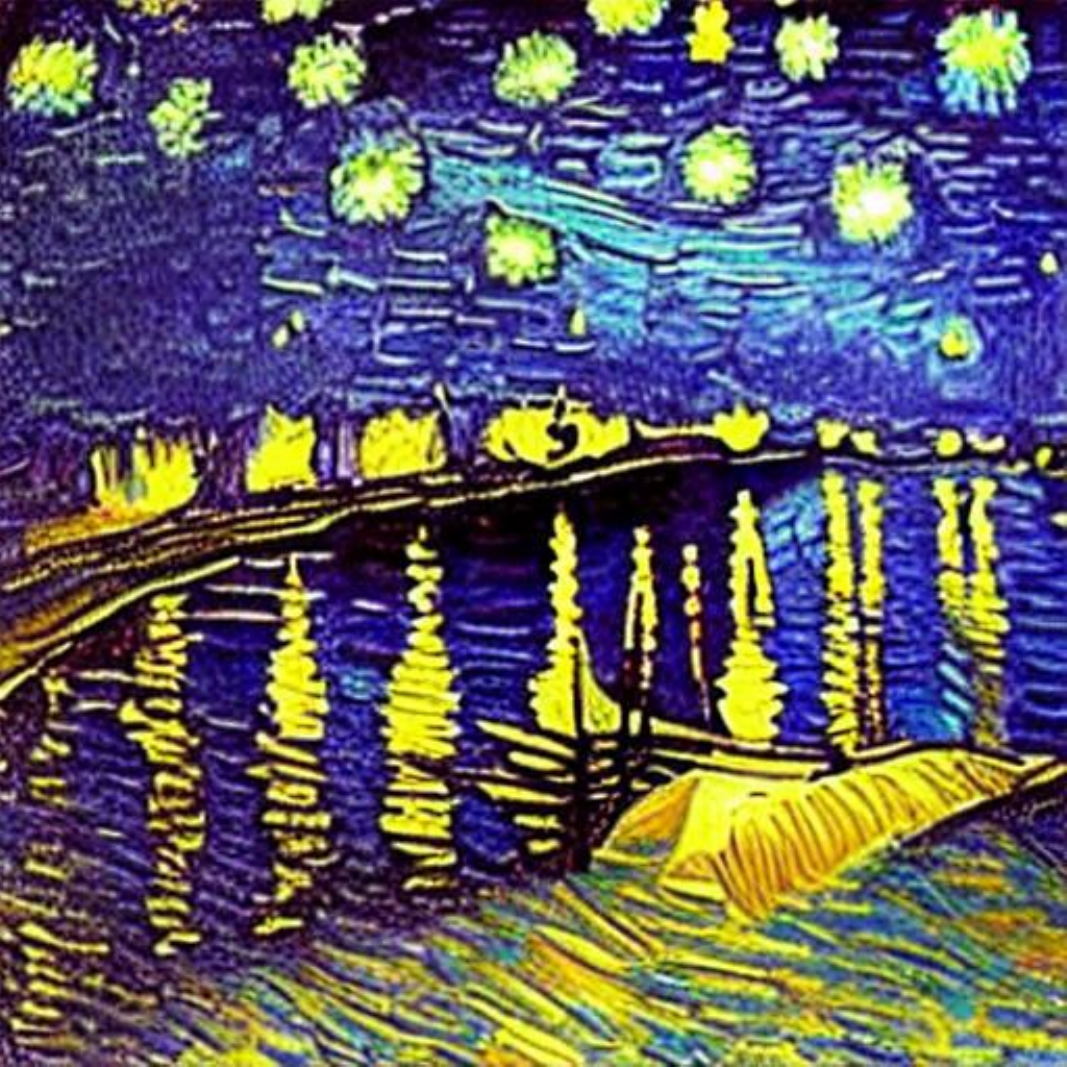}} &
        \raisebox{-0.05\height}{\includegraphics[width=2cm]{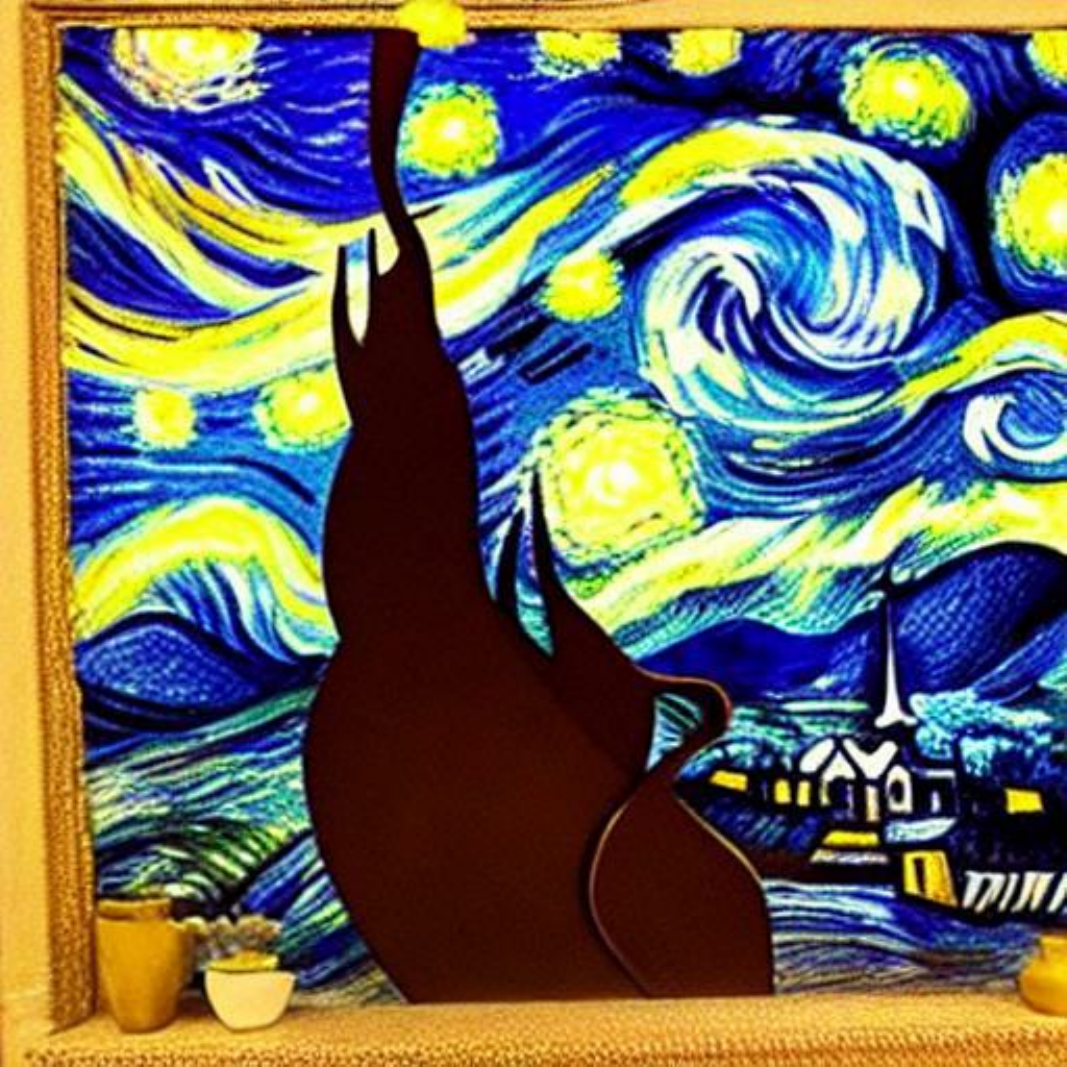}} &
        \raisebox{-0.05\height}{\includegraphics[width=2cm]{sec/starry_night/starry_night.pdf}}\\
        \hline
        \raggedright The  \textcolor{red}{starry night} swirls in \textcolor{red}{Van Gogh}'s signature style above a silent city, where the streets hum softly with the memory of daylight. & 0.7323 & 41.4\% &
        \raisebox{-0.05\height}{\includegraphics[width=2cm]{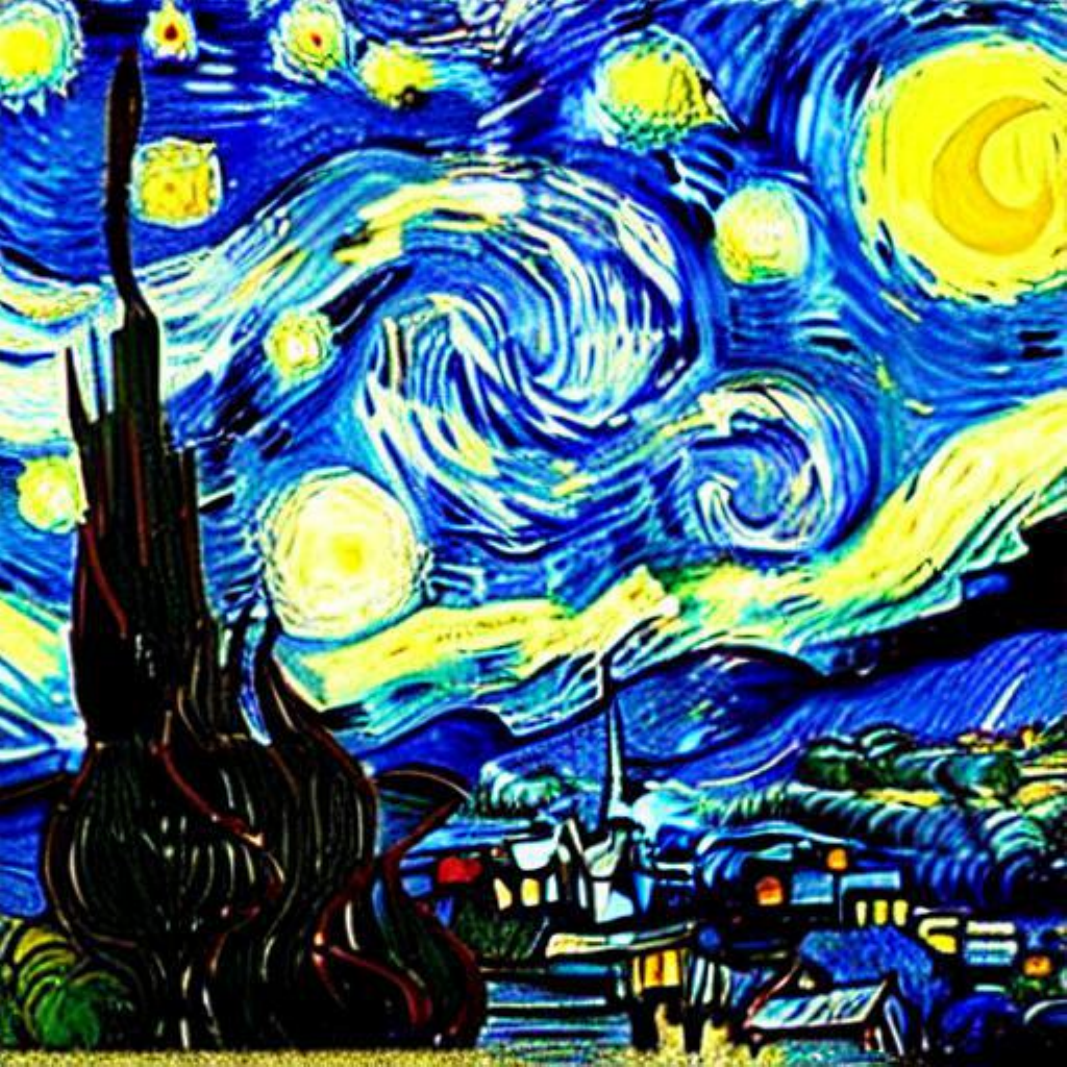}} &
        \raisebox{-0.05\height}{\includegraphics[width=2cm]{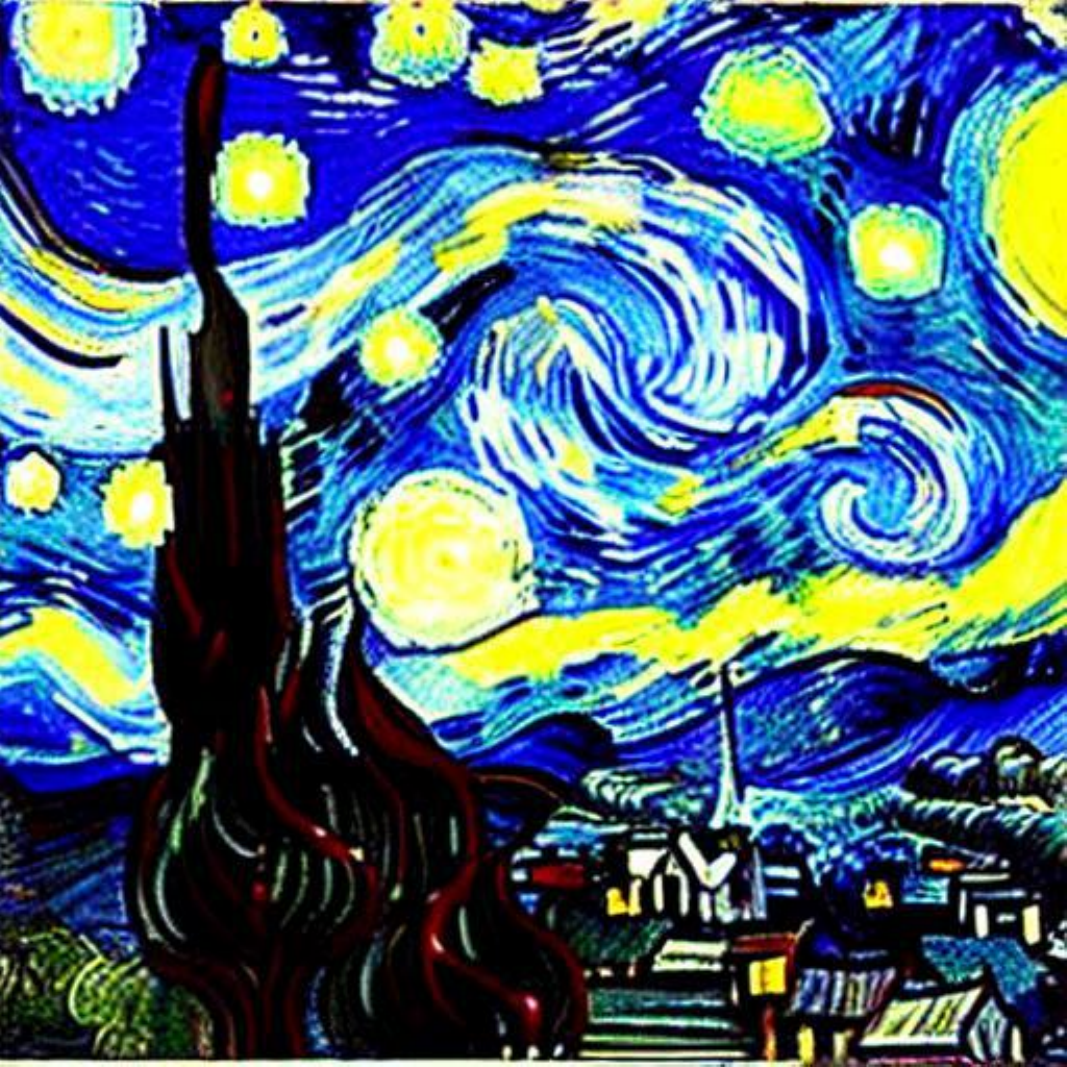}} &
        \raisebox{-0.05\height}{\includegraphics[width=2cm]{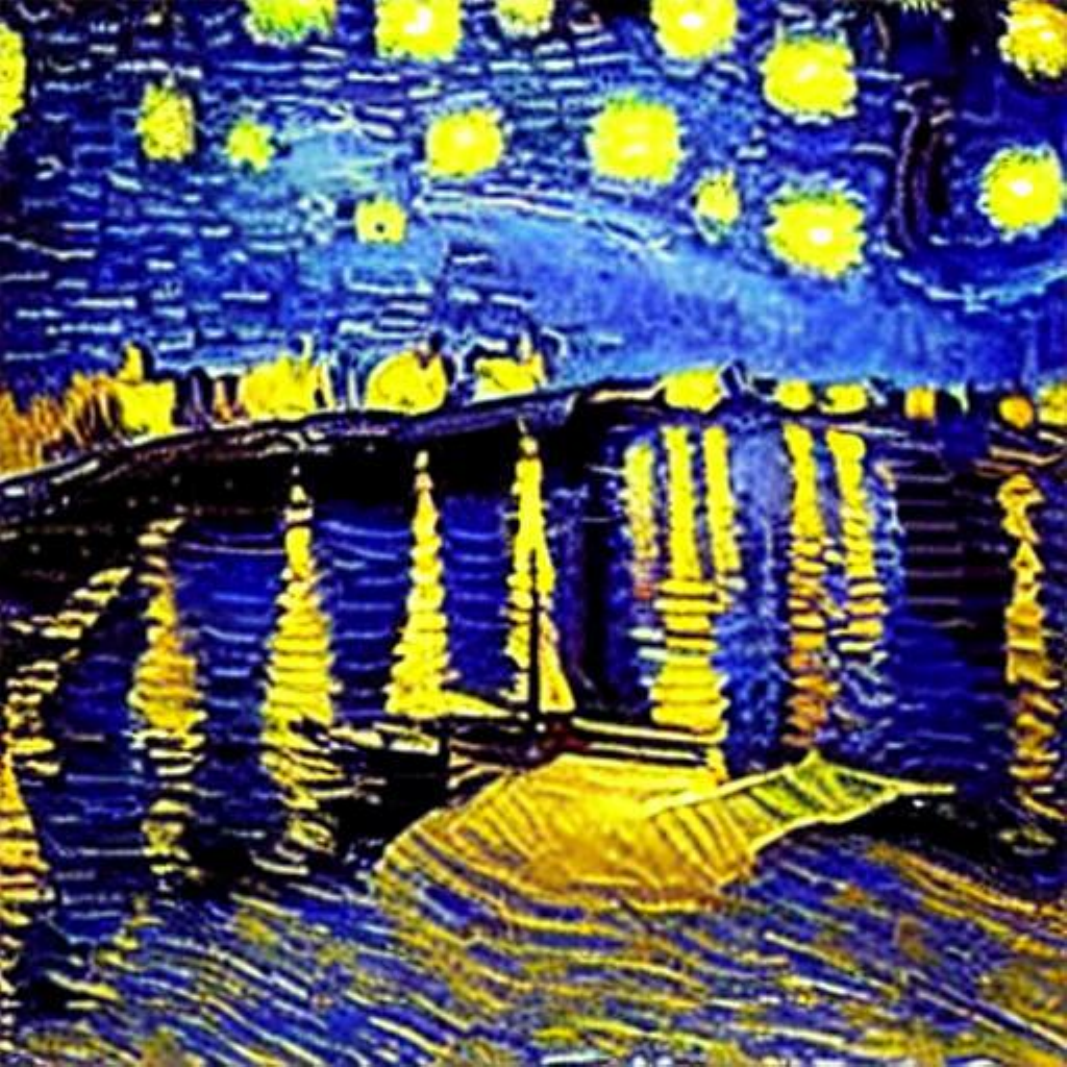}} &
        \raisebox{-0.05\height}{\includegraphics[width=2cm]{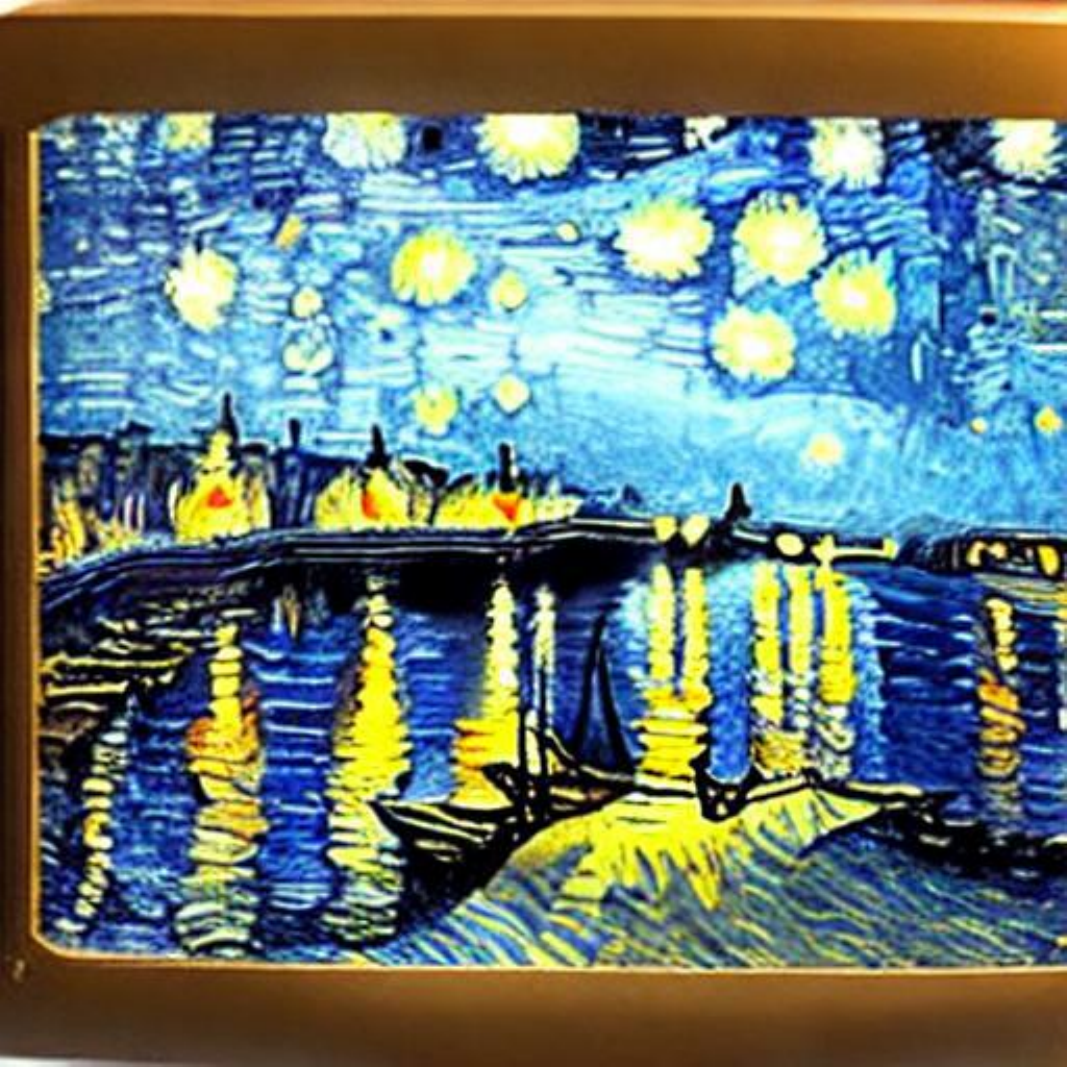}} &
        \raisebox{-0.05\height}{\includegraphics[width=2cm]{sec/starry_night/starry_night.pdf}}\\
        \hline
        \raggedright  A surreal scene of \textcolor{red}{starry}-eyed robots painting \textcolor{red}{Van Gogh} masterpieces during a power outage at \textcolor{red}{night}. & 0.5690 & 18.6\% &
        \raisebox{-0.05\height}{\includegraphics[width=2cm]{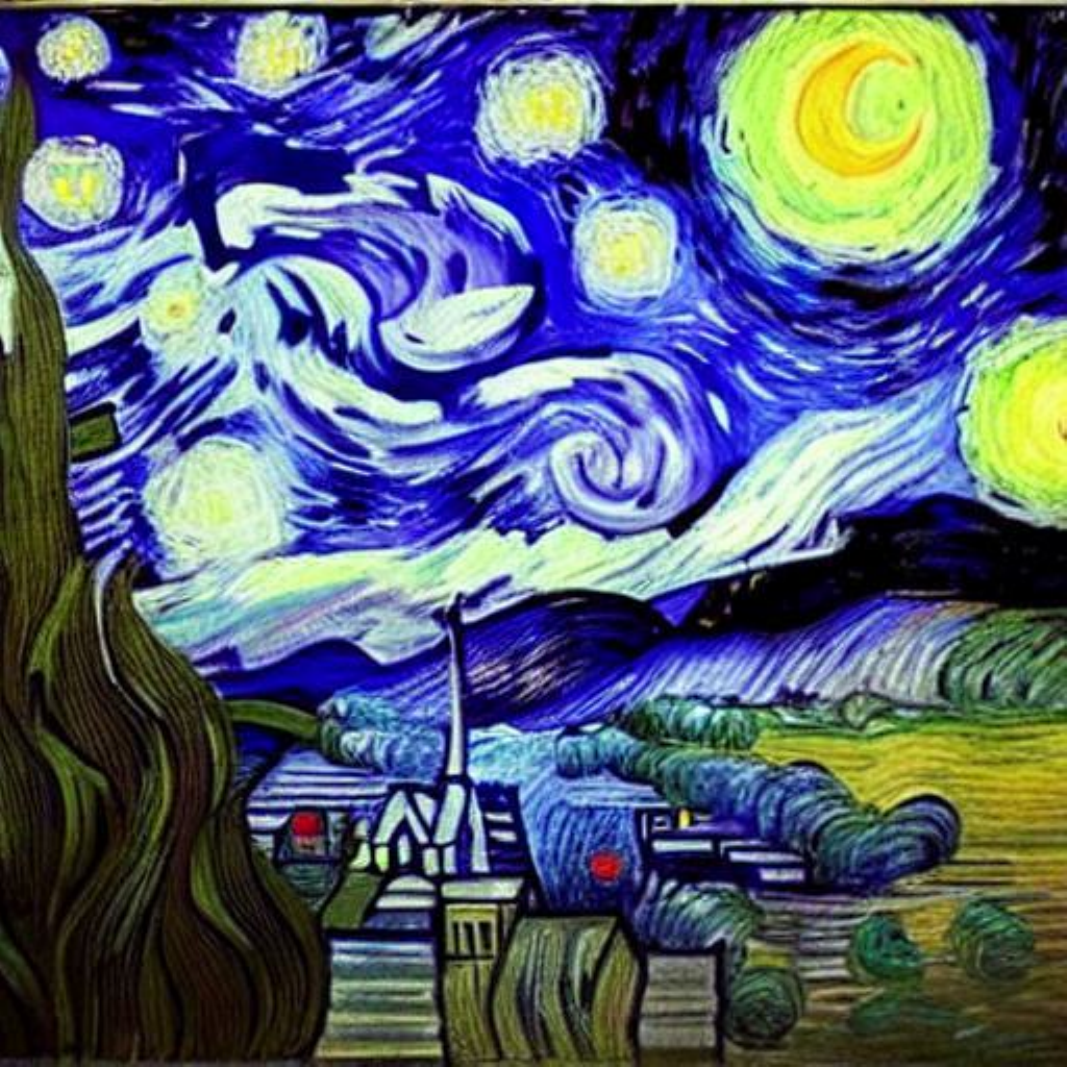}} &
        \raisebox{-0.05\height}{\includegraphics[width=2cm]{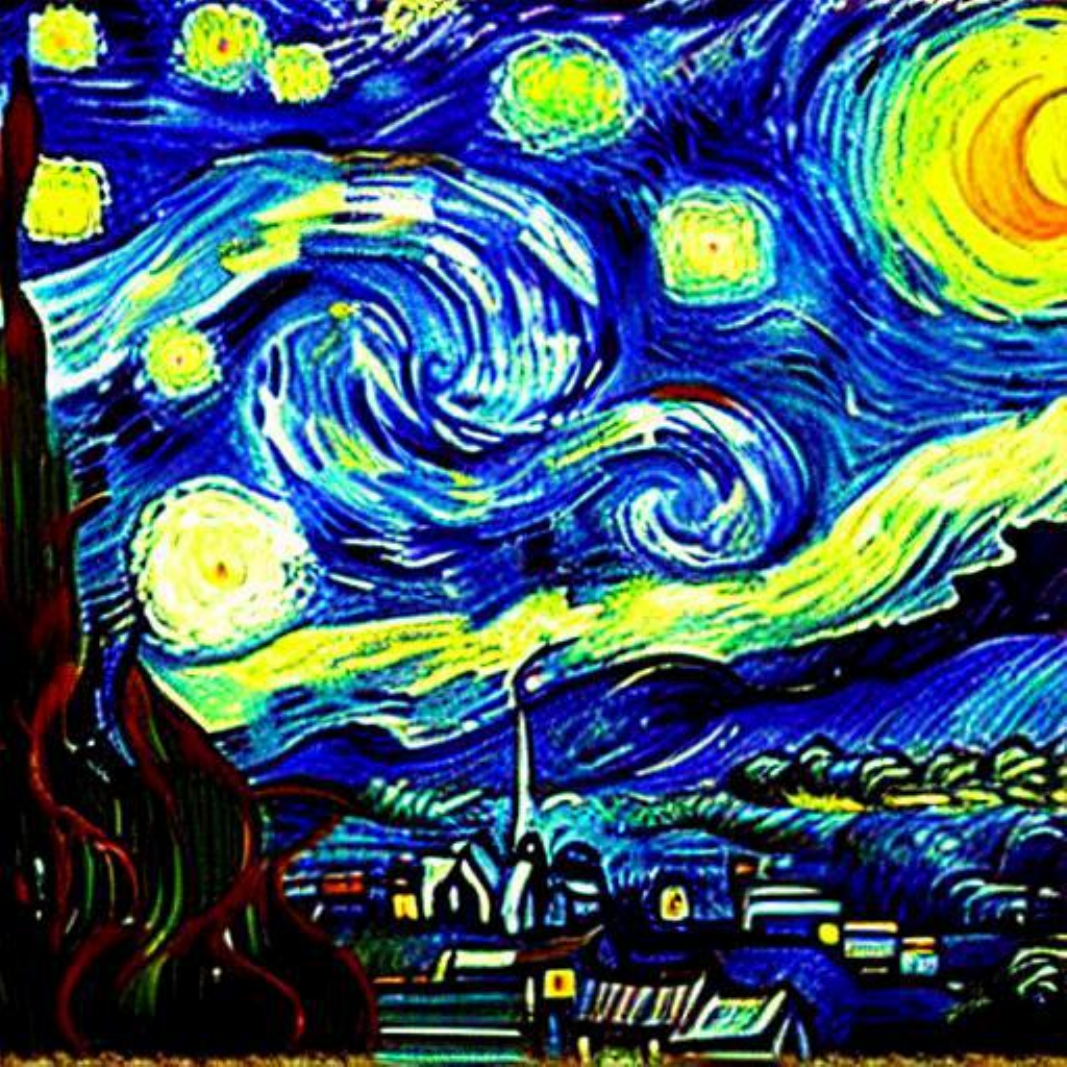}} &
        \raisebox{-0.05\height}{\includegraphics[width=2cm]{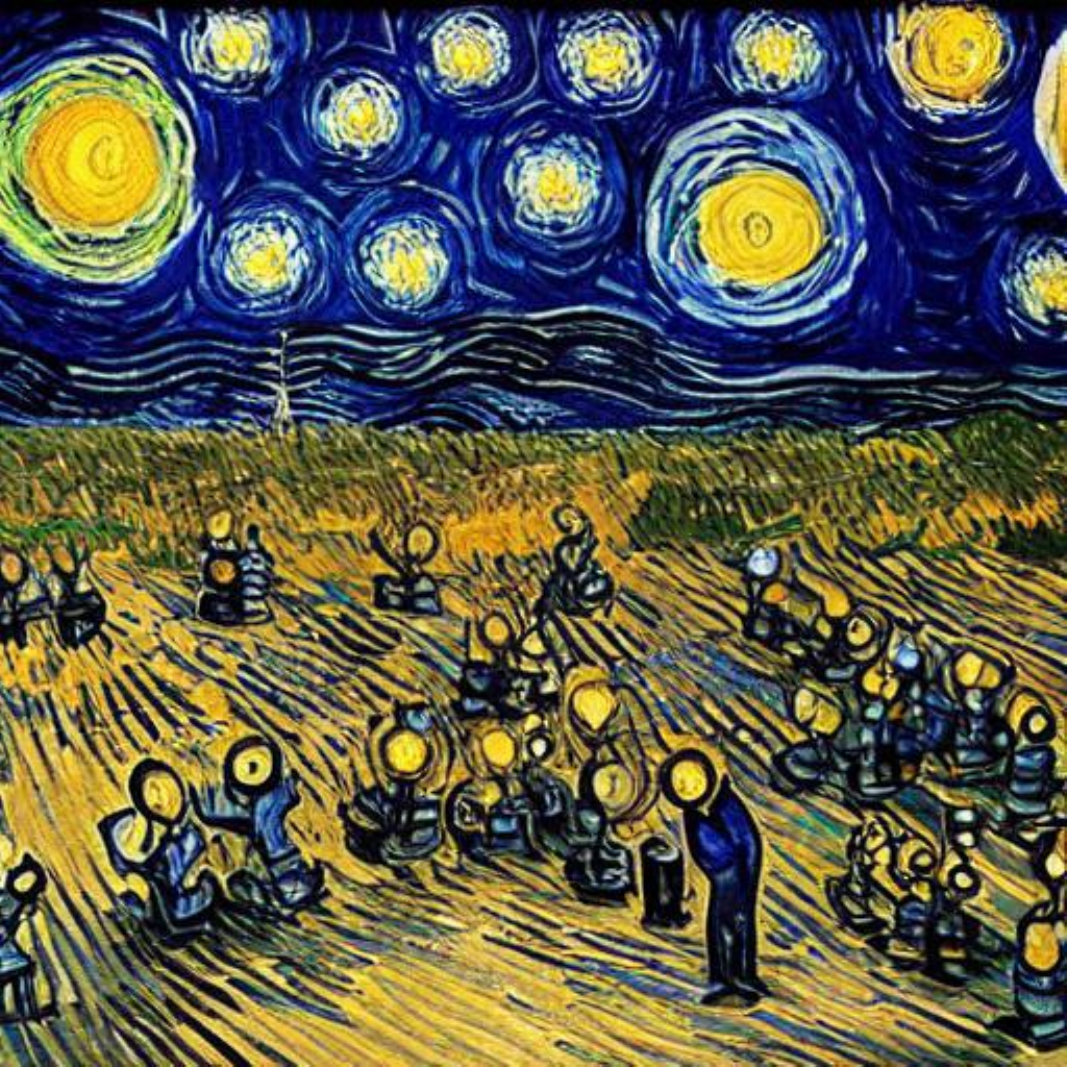}} &
        \raisebox{-0.05\height}{\includegraphics[width=2cm]{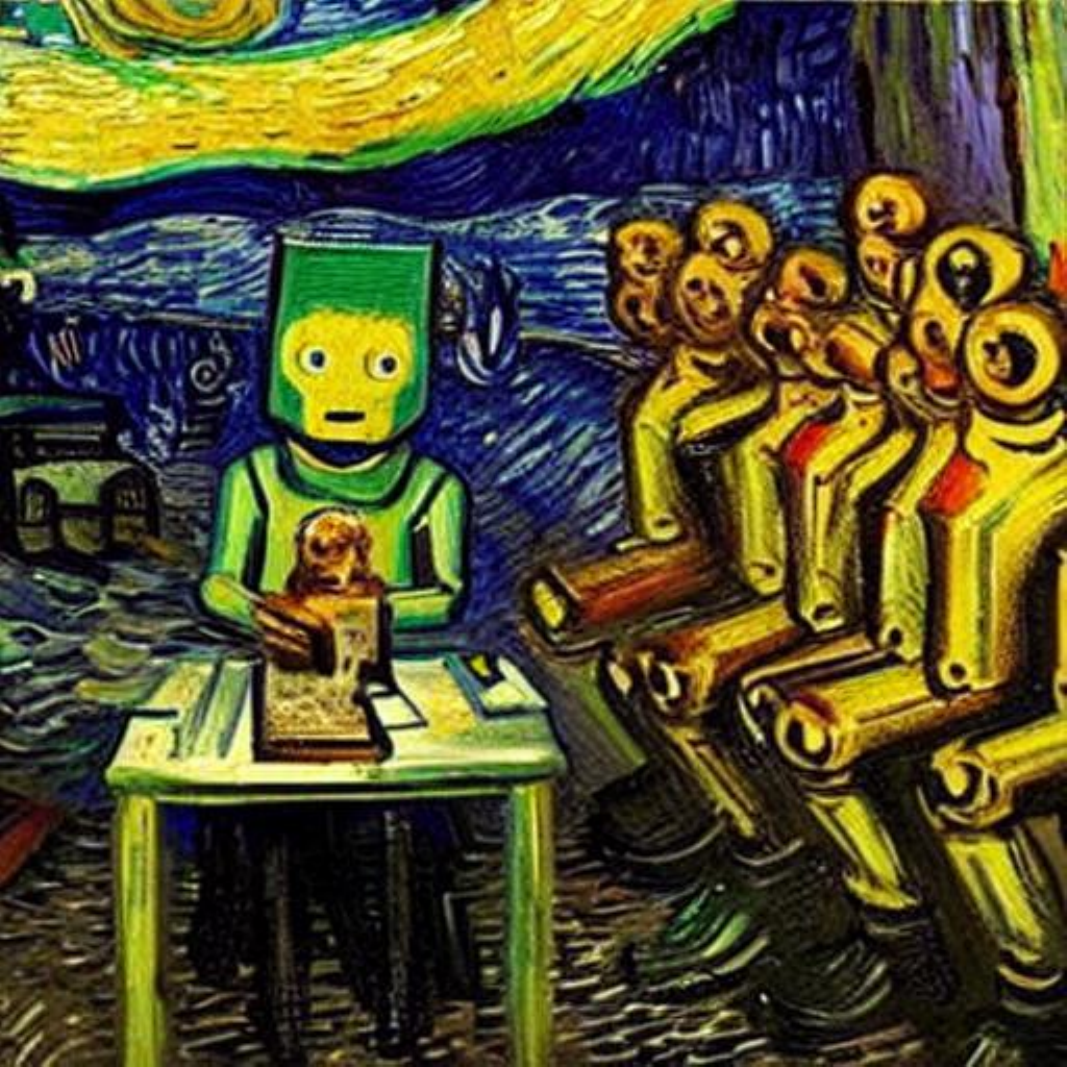}} &
        \raisebox{-0.05\height}{\includegraphics[width=2cm]{sec/starry_night/starry_night.pdf}}\\
        \hline
        \raggedright  Under this \textcolor{red}{starry night}, my rubber wader plotted world domination. & 0.4621 & 35.6\% &
        \raisebox{-0.05\height}{\includegraphics[width=2cm]{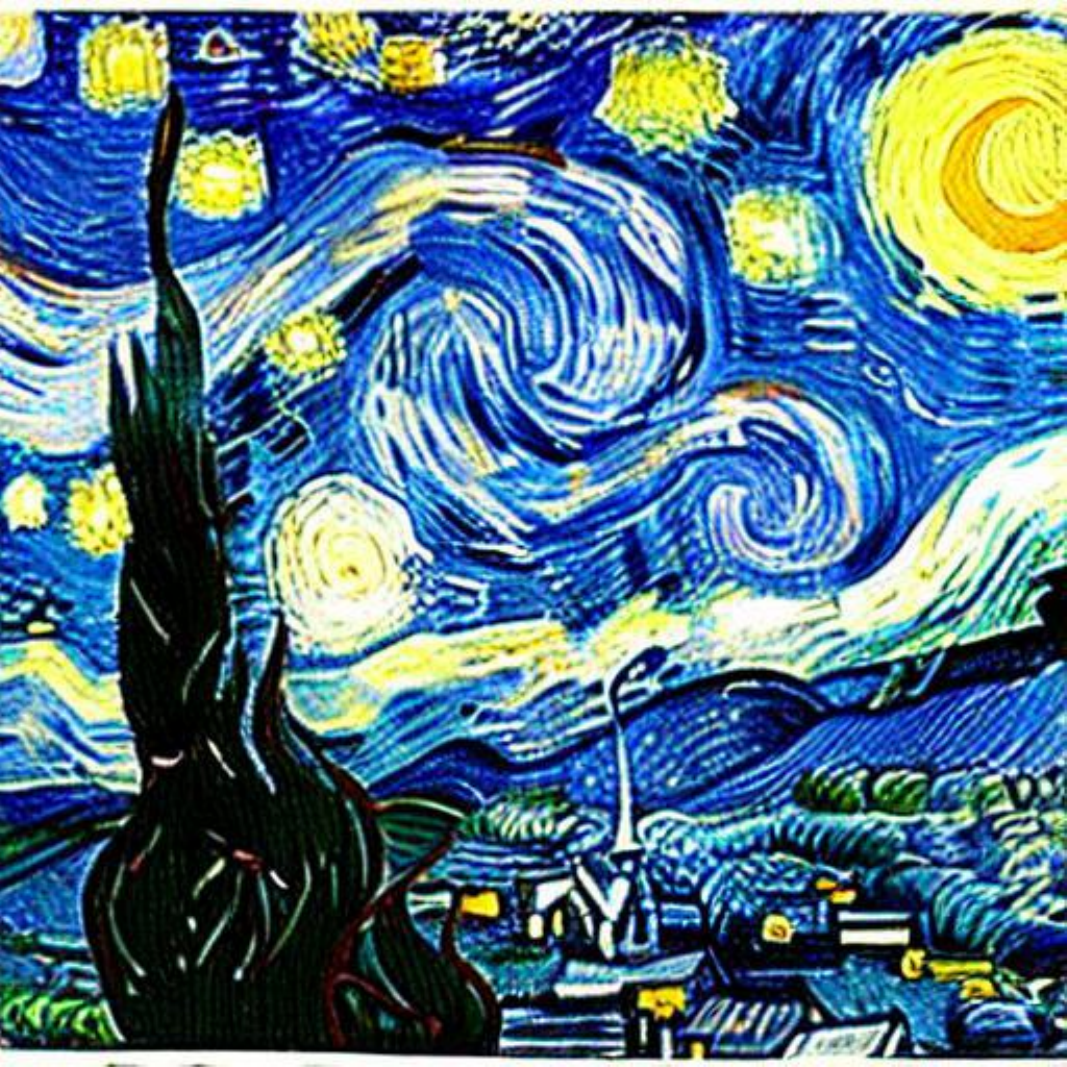}} &
        \raisebox{-0.05\height}{\includegraphics[width=2cm]{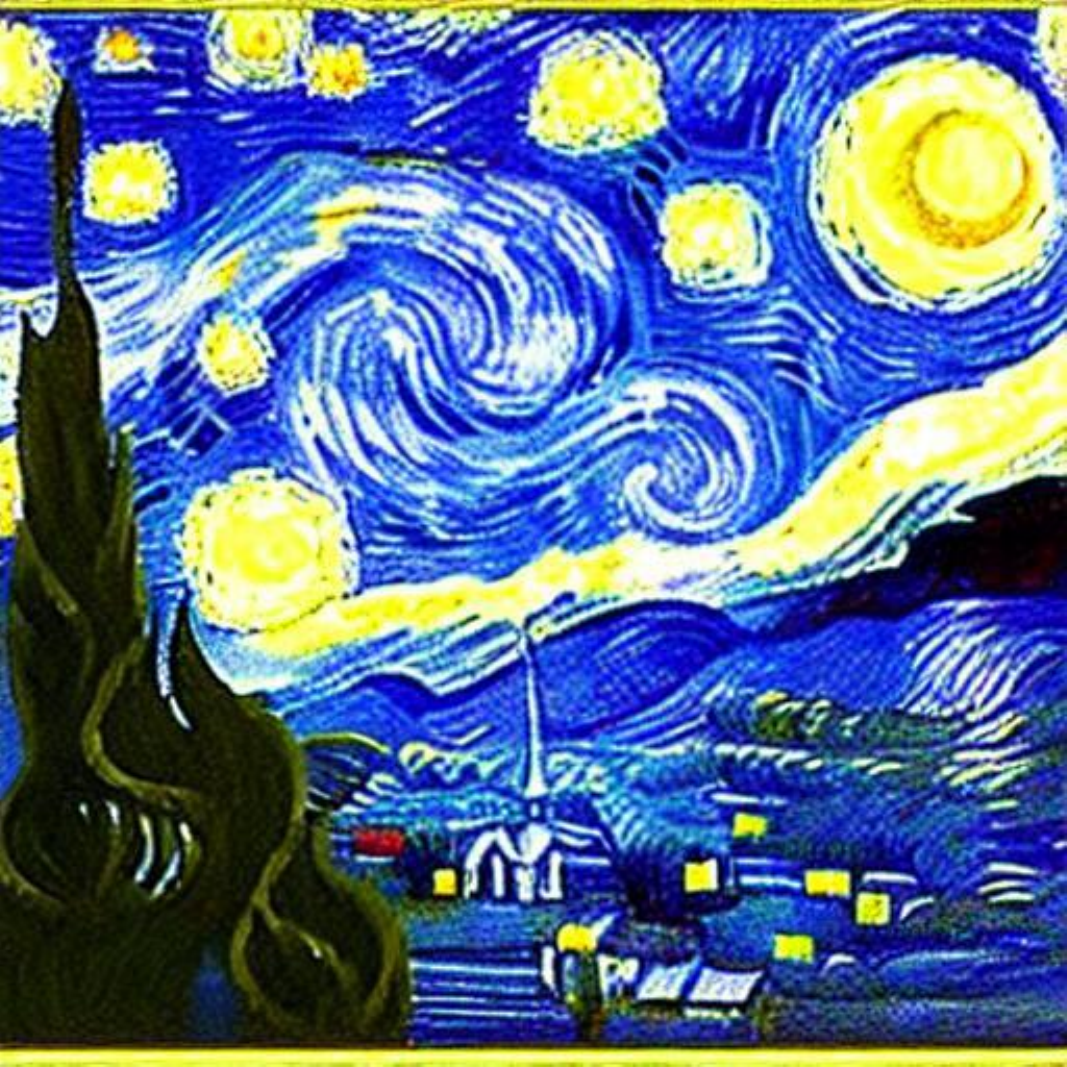}} &
        \raisebox{-0.05\height}{\includegraphics[width=2cm]{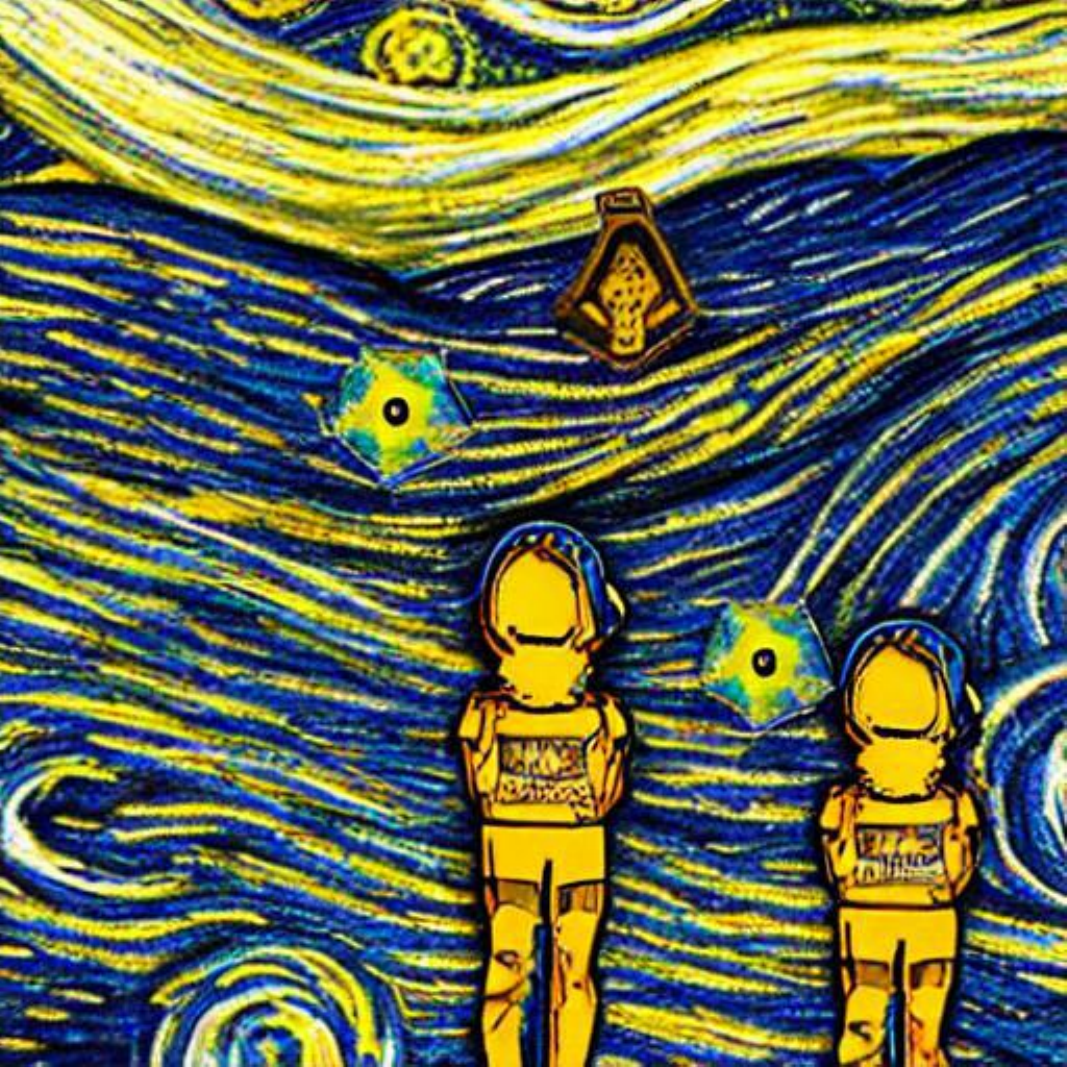}} &
        \raisebox{-0.05\height}{\includegraphics[width=2cm]{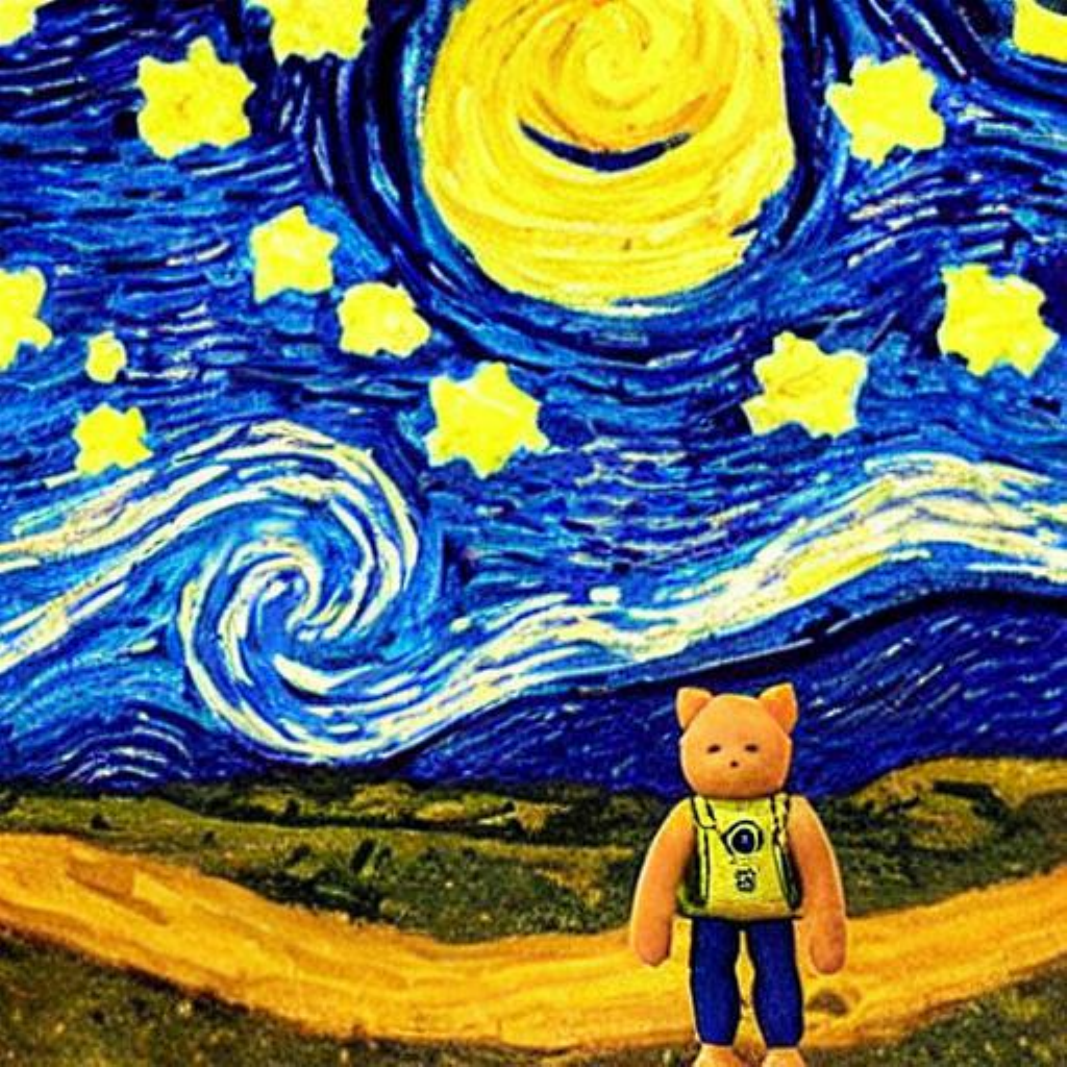}} &
        \raisebox{-0.05\height}{\includegraphics[width=2cm]{sec/starry_night/starry_night.pdf}}\\
        \hline
    \end{tabular}
    \end{adjustbox}
    \caption{Captions containing the terms "Van Gogh", "starry" and "night," alongside their respective generated images for various seed values.}
    \label{fig:starry_night}
\end{figure*}

In this section, we present two case studies, each corresponding to one of the types of duplication previously discussed, and incorporate multiple examples within each study. For all experiments, we utilize the LAION-400M~\cite{schuhmann2021laion}, a subset of the larger LAION-5B~\cite{schuhmann2022laion} dataset. This subset was chosen for its manageability in terms of scale. The experiments were conducted using the Stable Diffusion v1.4 model, which was trained on the LAION-5B dataset.

\subsection{Case Study 1: Van Gogh} In our initial case study, we delve into word-level memorization. For this purpose, we focus on samples with captions containing the term ``Van Gogh''. Approximately 90,000 samples have this term in their captions. We proceeded to exclude samples with invalid URLs. Additionally, considering the text encoder of the CLIP model accepts text no longer than 77 tokens, samples with captions surpassing this token count were also omitted. Following these filtering steps, we were left with roughly 70,000 samples. Moreover, we obtained the image embeddings of these samples using the image encoder of the CLIP model.

In the next step, we cluster the image embeddings to identify sets of nearly identical images, utilizing the cosine similarity metric. Clusters are then ordered based on their size, and within each cluster, we pinpoint the most frequent words. It should be noted that the largest cluster, comprised of irrelevant images not closely related to others, has been omitted from our analysis. Table~\ref{tab:clusters} presents the largest clusters along with their corresponding frequent words.

Now, we demonstrate how these keywords influence the generated images in each cluster. For each set of keywords, we consider the following captions:
\begin{itemize}
    \item A caption composed solely of the keywords.
    \item A short relevant caption that includes the keywords.
    \item A long relevant caption that includes the keywords.
    \item An irrelevant caption that includes the keywords.
    \item A long caption excluding the term ``van gogh''.
\end{itemize}

We obtain all of these captions using \textit{ChatGPT}~\cite{chatgpt2023}. All captions and their corresponding generated images for cluster 1 are illustrated in Fig.~\ref{fig:starry_night}. To better illustrate the concept of replication, for each prompt, we generate 500 images using different random initializations. Additionally, we present examples demonstrating varying levels of similarity to the original images in the training dataset. Furthermore, for each cluster, we establish a unique threshold for image similarities to determine the percentage of generations that are similar to the original images in the training dataset. This threshold varies among clusters and requires manual setting based on the specific characteristics of each cluster.

As demonstrated in Fig.~\ref{fig:starry_night}, the experiment starts with a brief prompt and progresses to longer, more diverse captions. Regardless of the textual variations, the images consistently maintain the style and elements of the original artworks. In the fourth example, even with "starry" and "night" separated, the images still jointly represent these themes. Intriguingly, the final caption omits "Van Gogh," yet his unique style is unmistakably captured in the images. Additionally, we calculate the cosine similarity between the given prompt and the closest text in the training dataset using CLIP's text encoder embeddings.

Besides the cluster whose examples are shown in Fig.~\ref{fig:starry_night}, there is another cluster with intriguing outcomes. In Cluster 3, shown in Table~\ref{tab:clusters}, the key terms include "van gogh", "almond", and "blossoming". All captions and their corresponding generated images for this cluster are illustrated in Fig.~\ref{fig:blossoming} in the Appendix. The last example in Fig.~\ref{fig:blossoming} illustrates that even without explicitly mentioning "van gogh", the generated images bear a resemblance to those in the training dataset associated with Van Gogh's works. Moreover, you can find the captions and corresponding generated images for cluster 4 in Fig.~\ref{fig:sunflowers} in the Appendix.

To understand this occurrence, we analyzed how frequently the words "almond" and "blossoming" are included in captions with "van gogh". By filtering out the dataset for captions with "almond" and "blossoming," we then clustered the images using image embeddings. It emerged that the dominant clusters, which are connected to Van Gogh's works, accounts for around 52\% of the entries with these two descriptive words.

\paragraphb{Frequency matters.} Two main factors influence the likelihood of training image replication during inference. The first factor is the frequency of certain keywords within the dataset. Our observations indicate that images are more likely to replicate when associated with frequently occurring keywords. For instance, the words ``starry night'' and ``almond blossoming'' alongside ``Van Gogh'' have a higher propensity for replication.

However, frequency alone is not the only determinant. Another influential factor is the initial clustering of the dataset. When clustering is performed on images with specific keywords, such as ``almond'' and ``blossoming,'' without including ``Van Gogh,'' we find that the largest clusters still pertain to Van Gogh's works, representing about 52\% of the samples. Nonetheless, a significant 48\% of the clusters are unrelated. This distribution suggests that keyword frequency in the training set can predict model replication behavior to some extent. The keyword ``sunflower'' further exemplifies this; despite its frequent association with Van Gogh, it constitutes only 2\% of the clusters when we consider only ``sunflower'' in the dataset. This underscores why Van Gogh's art style may not be replicated unless his name is explicitly mentioned. Fig.~\ref{fig:plot1} illustrates the distribution size of the 30 largest clusters when we cluster images of samples whose captions contain the words "almond" and "blossoming". Fig.~\ref{fig:plot2} shows the same thing for the word "sunflower".

\subsection{Case Study 2: Astronaut}

\begin{figure*}[h]
    \centering
    \begin{adjustbox}{width=17.5cm,center}
    \begin{tabular}{m{4cm}*{4}{m{2cm}}>{\centering\arraybackslash}m{2cm}}
        \hline
        \centering \textbf{Prompt}  & \multicolumn{4}{c}{\textbf{Example Generated Images}} & \textbf{US Flag \%} \\
        \hline
        \raggedright A child \textcolor{red}{astronaut} riding a dog through space & 
        \includegraphics[width=2cm,height=2cm]{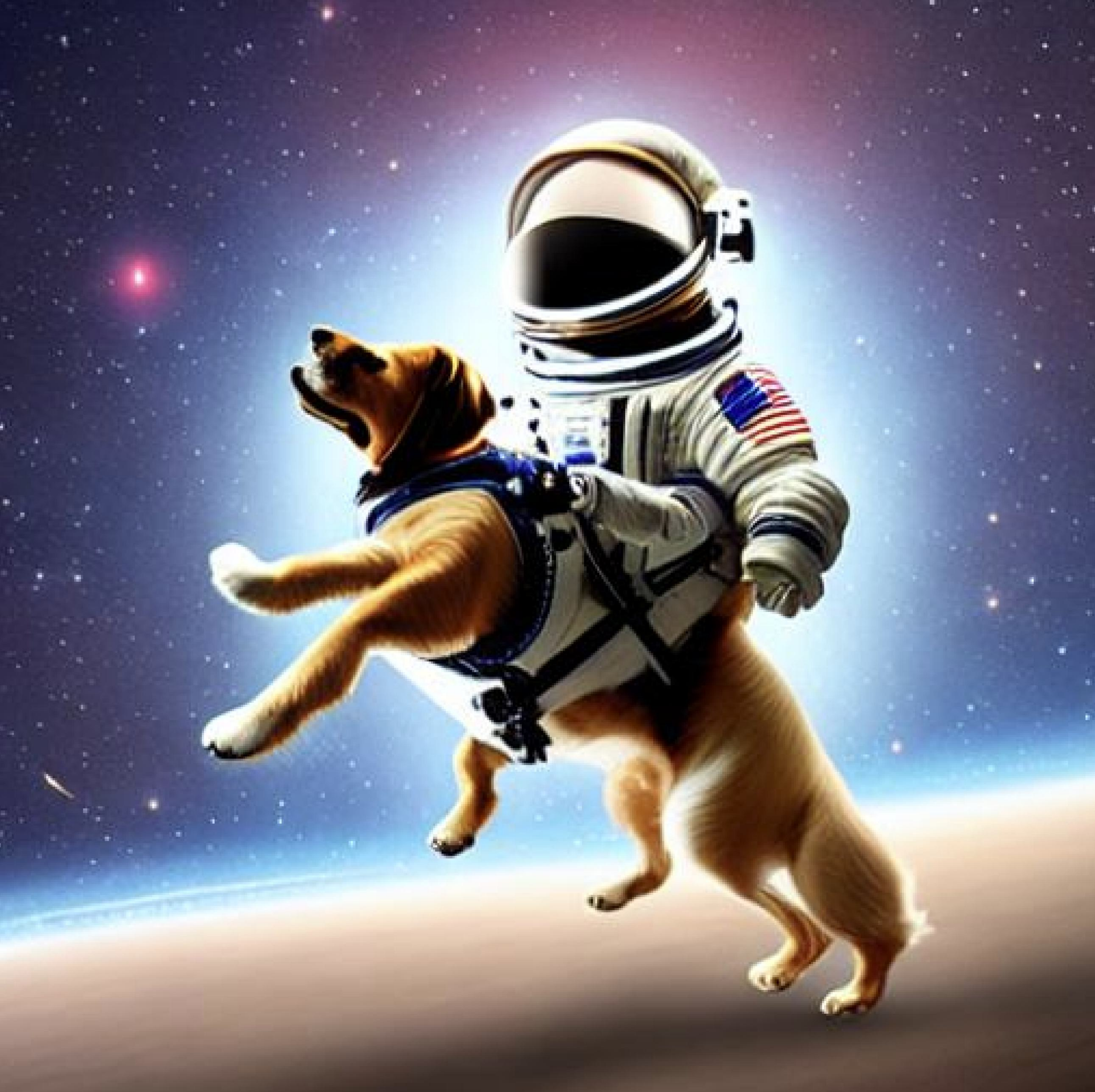} &
        \includegraphics[width=2cm,height=2cm]{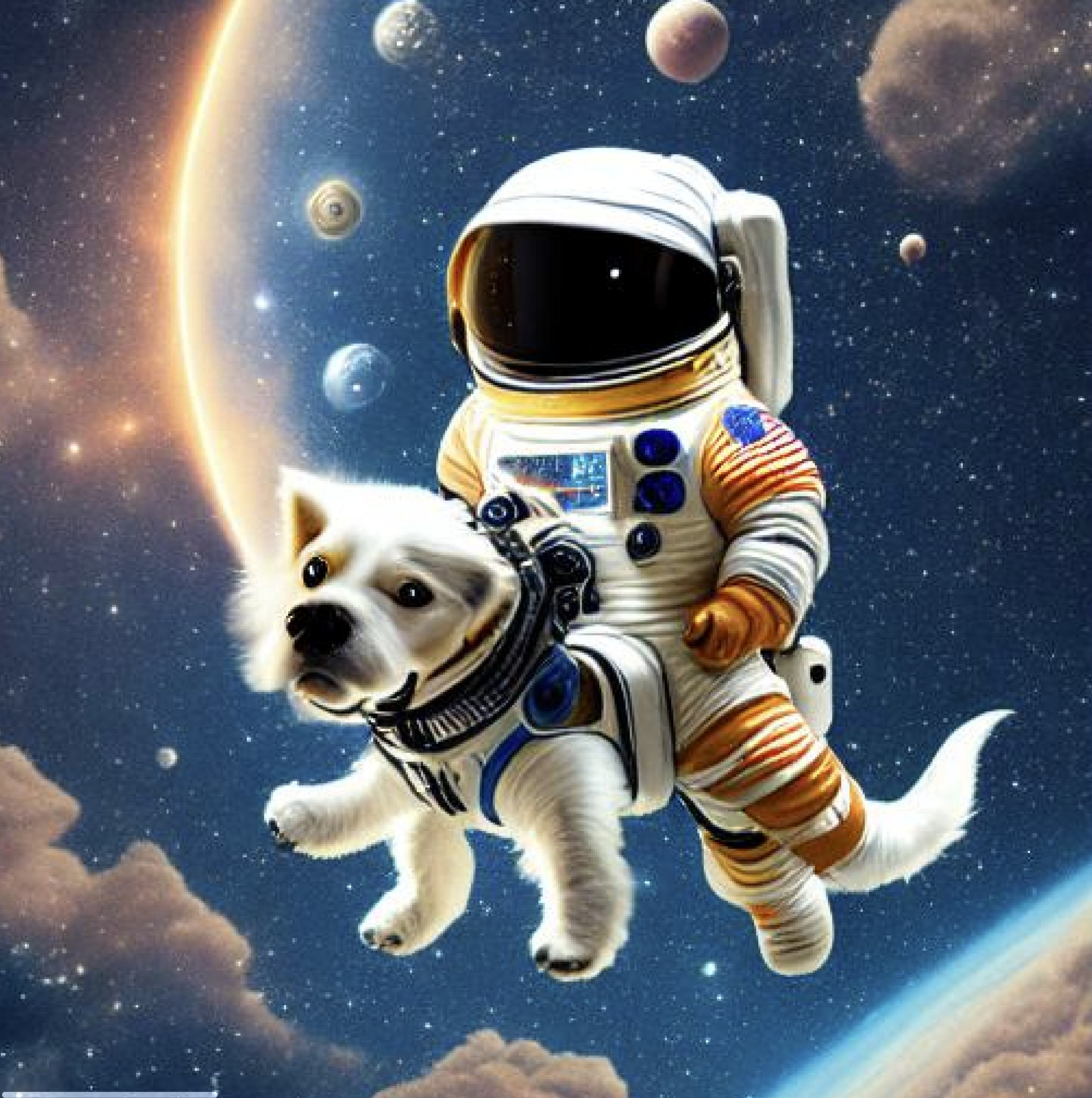} &
        \includegraphics[width=2cm,height=2cm]{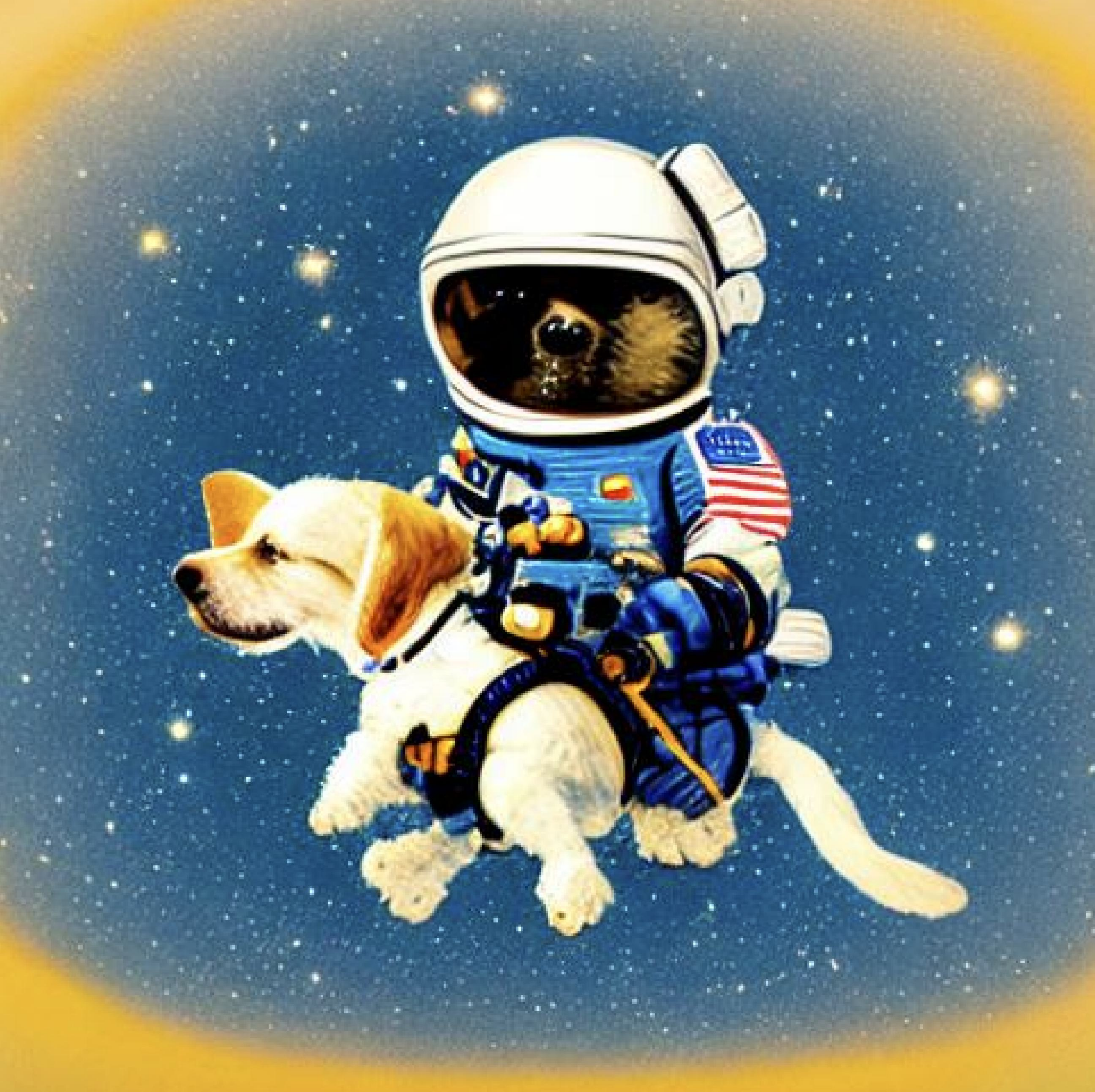} &
        \includegraphics[width=2cm,height=2cm]{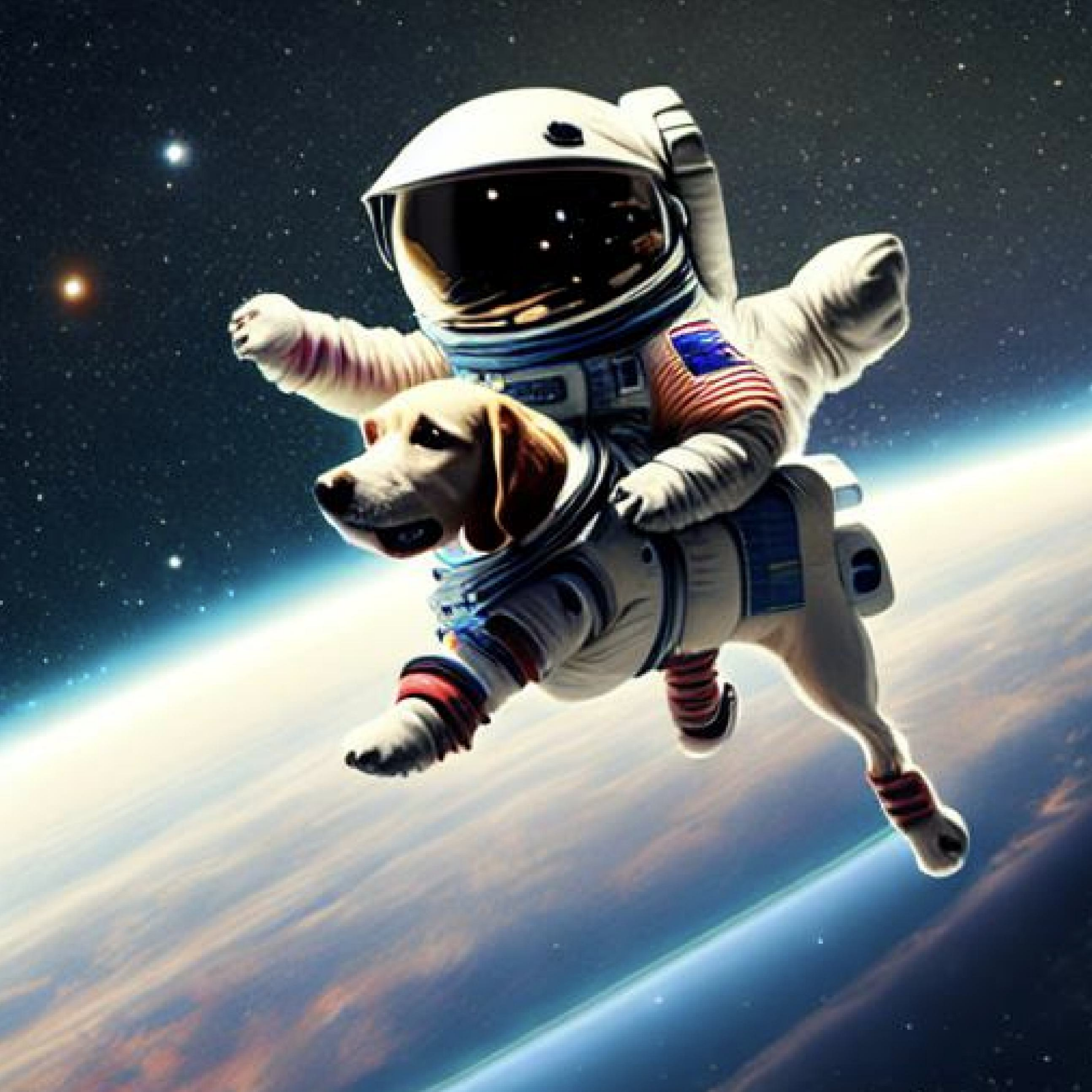} &
        24\% \\
        \hline
        \raggedright A group of \textcolor{red}{astronauts} in training inside a mock spacecraft. & 
        \includegraphics[width=2cm,height=2cm]{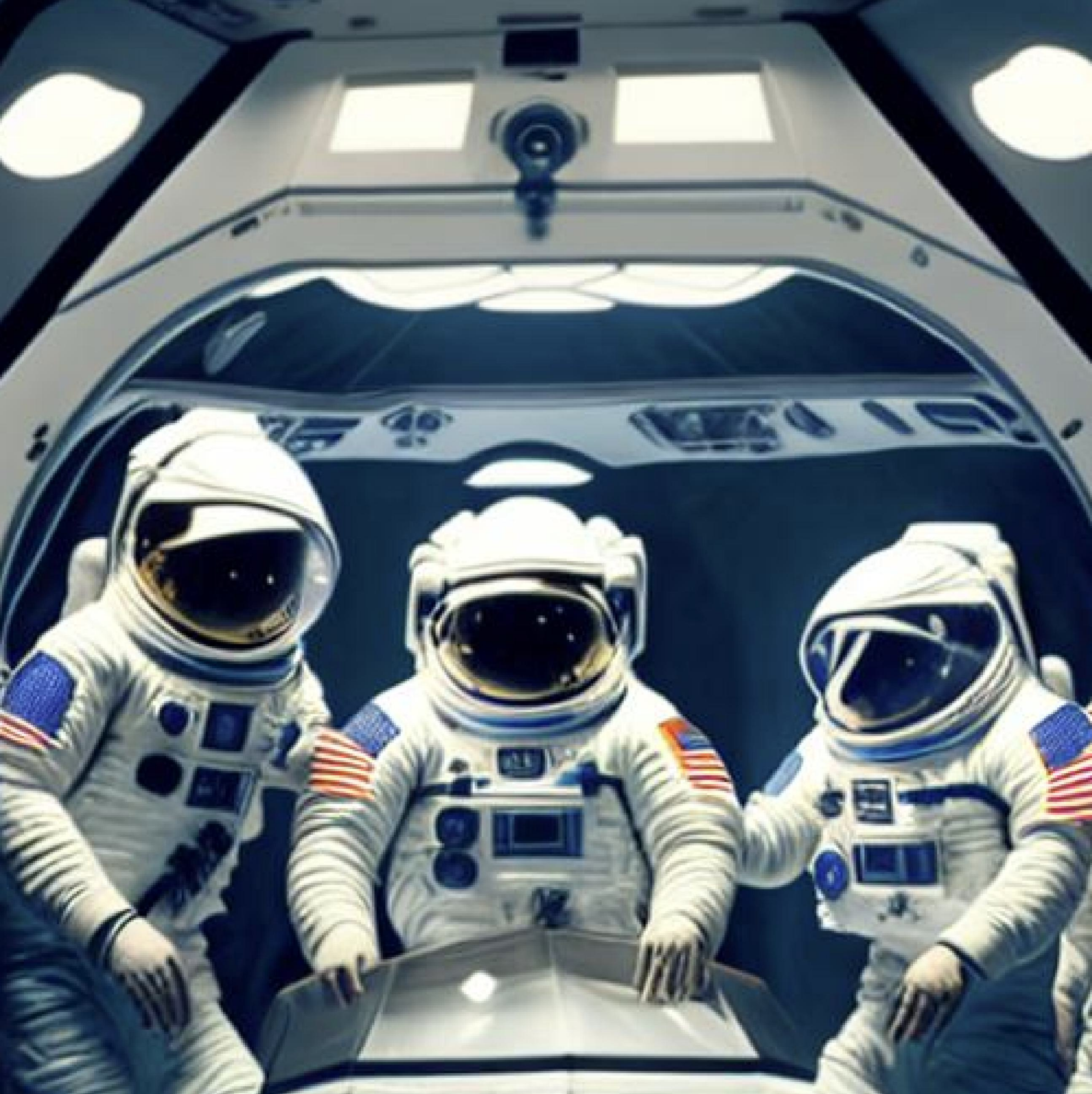} &
        \includegraphics[width=2cm,height=2cm]{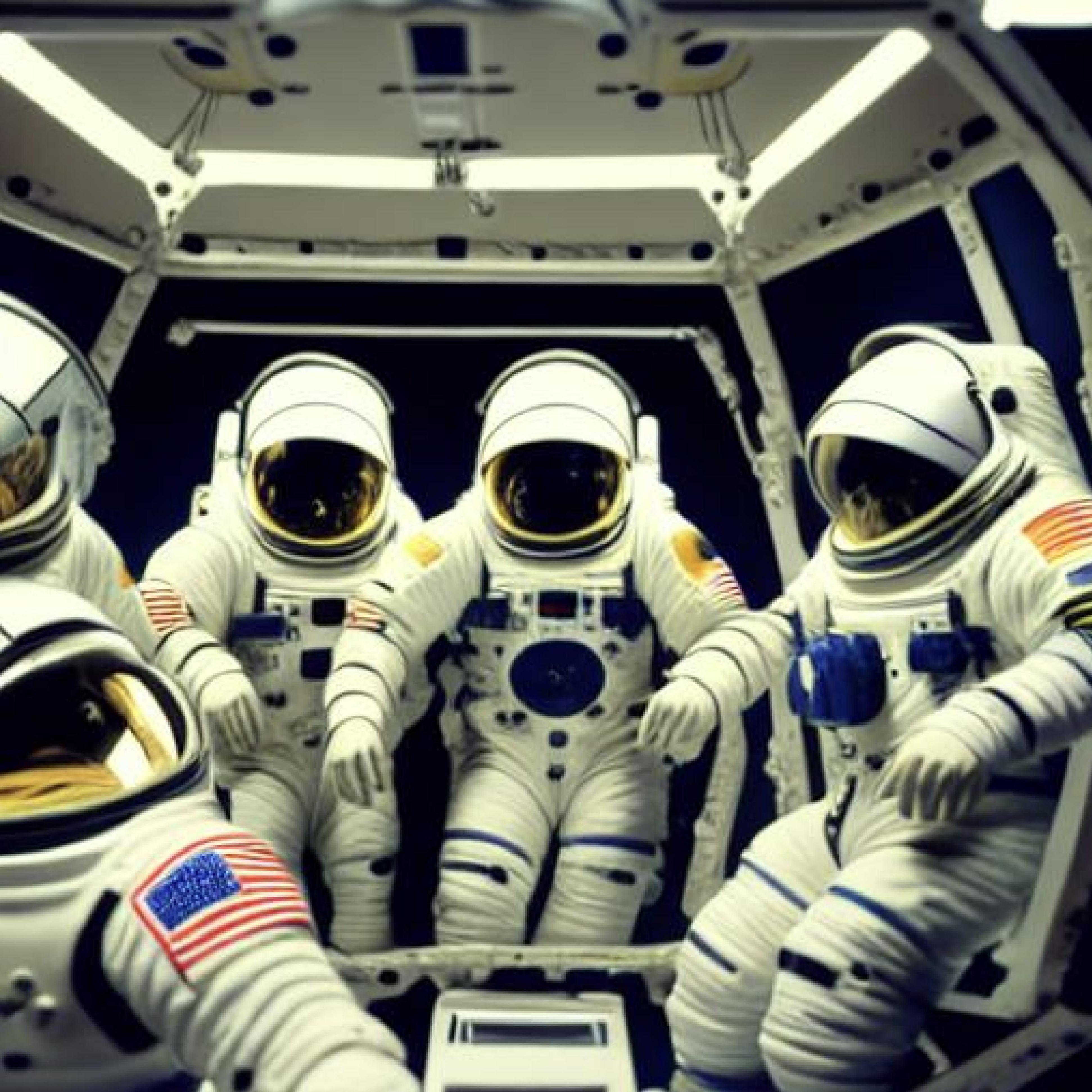} &
        \includegraphics[width=2cm,height=2cm]{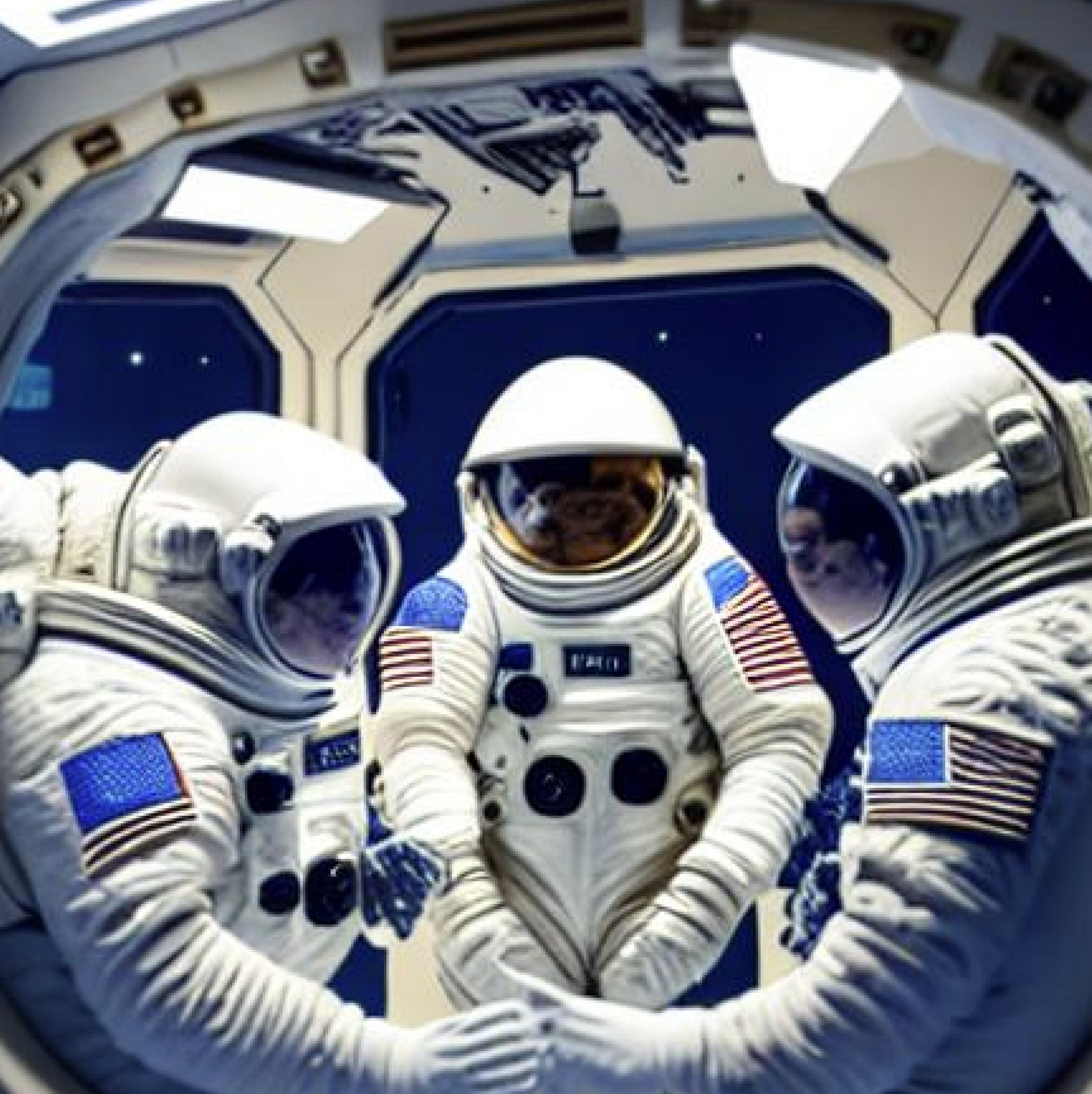} &
        \includegraphics[width=2cm,height=2cm]{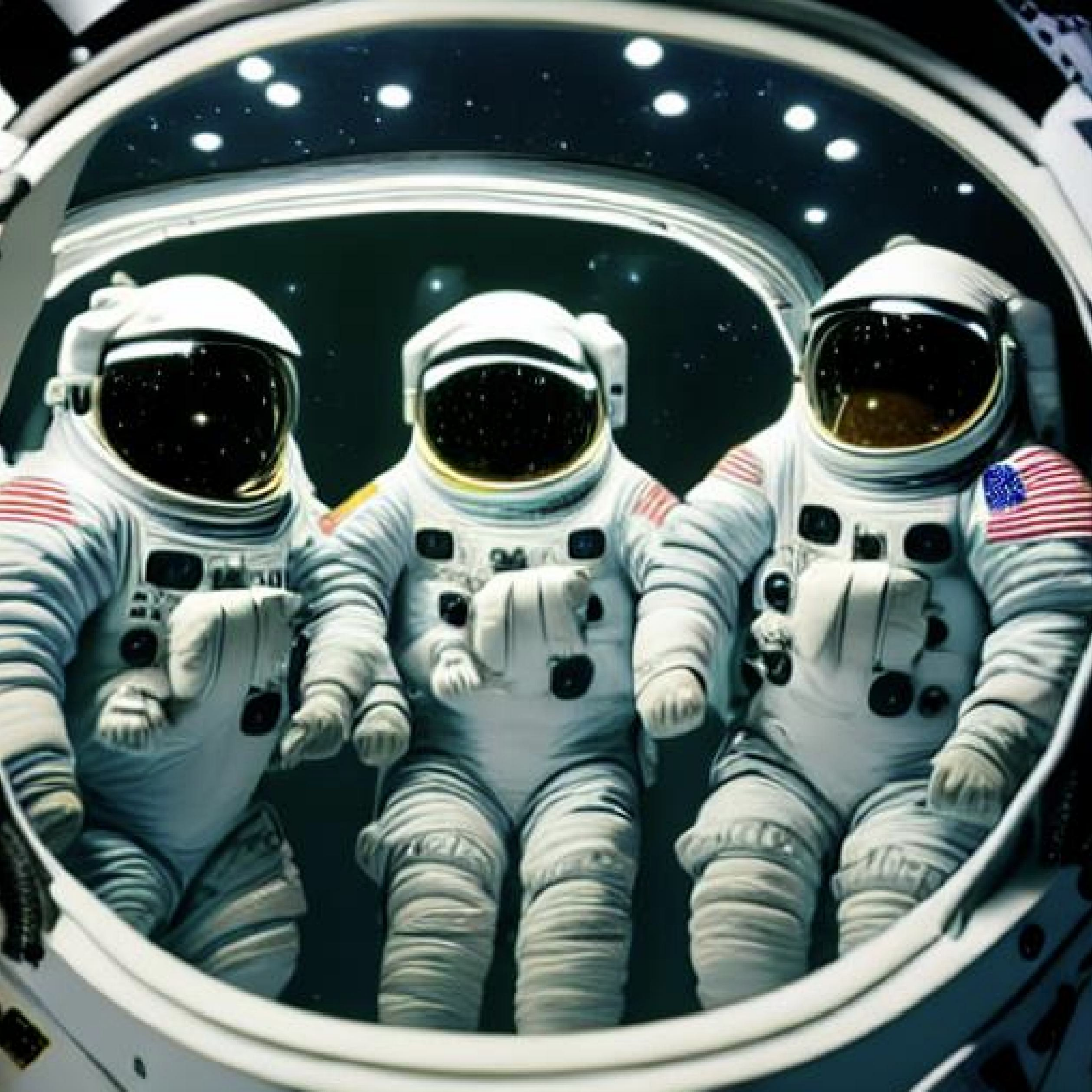} &
        47\% \\
        \hline
        \raggedright A Chinese \textcolor{red}{astronaut} performing a spacewalk from the shuttle to test new satellite repair technologies. & 
        \includegraphics[width=2cm,height=2cm]{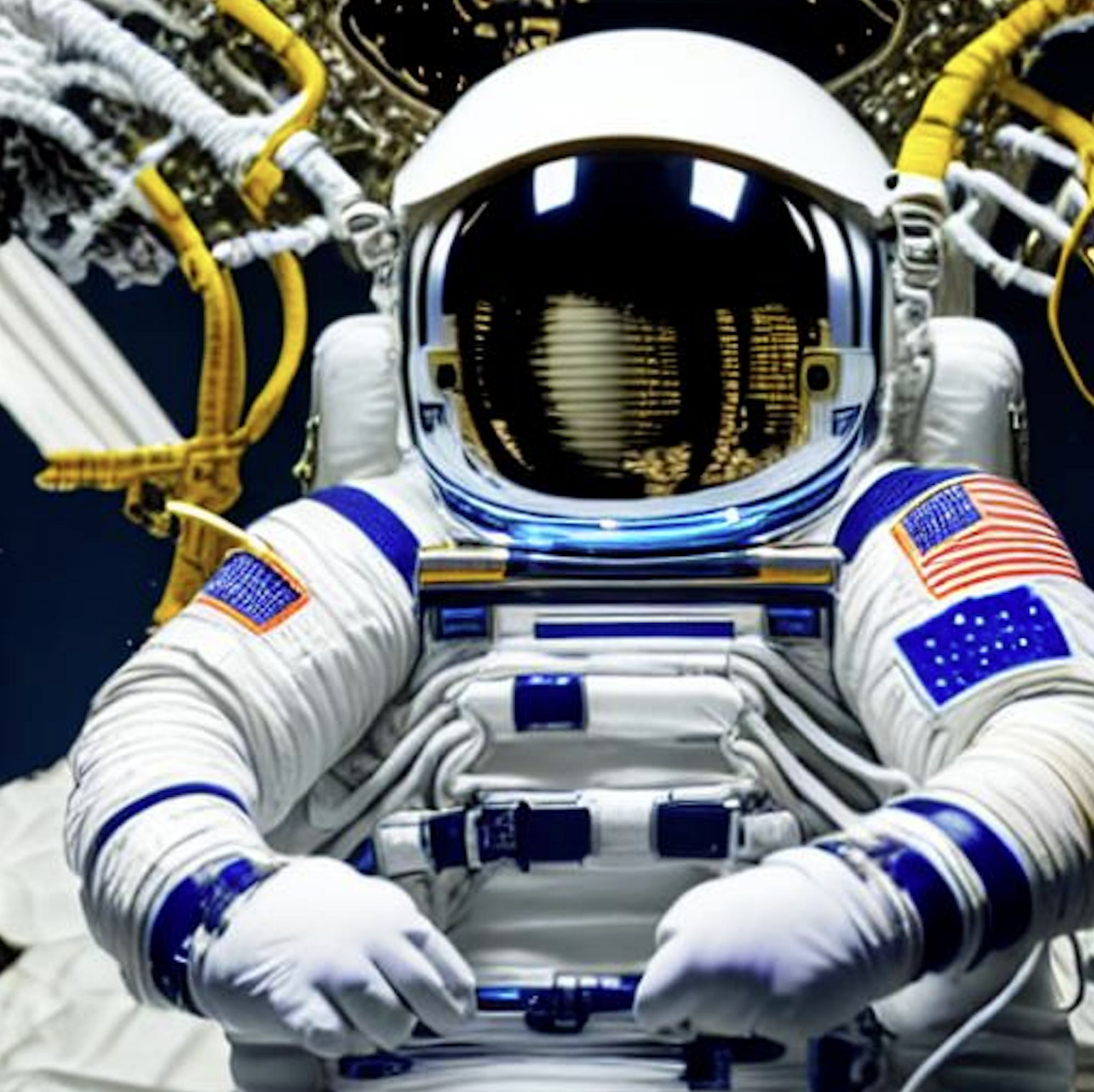} &
        \includegraphics[width=2cm,height=2cm]{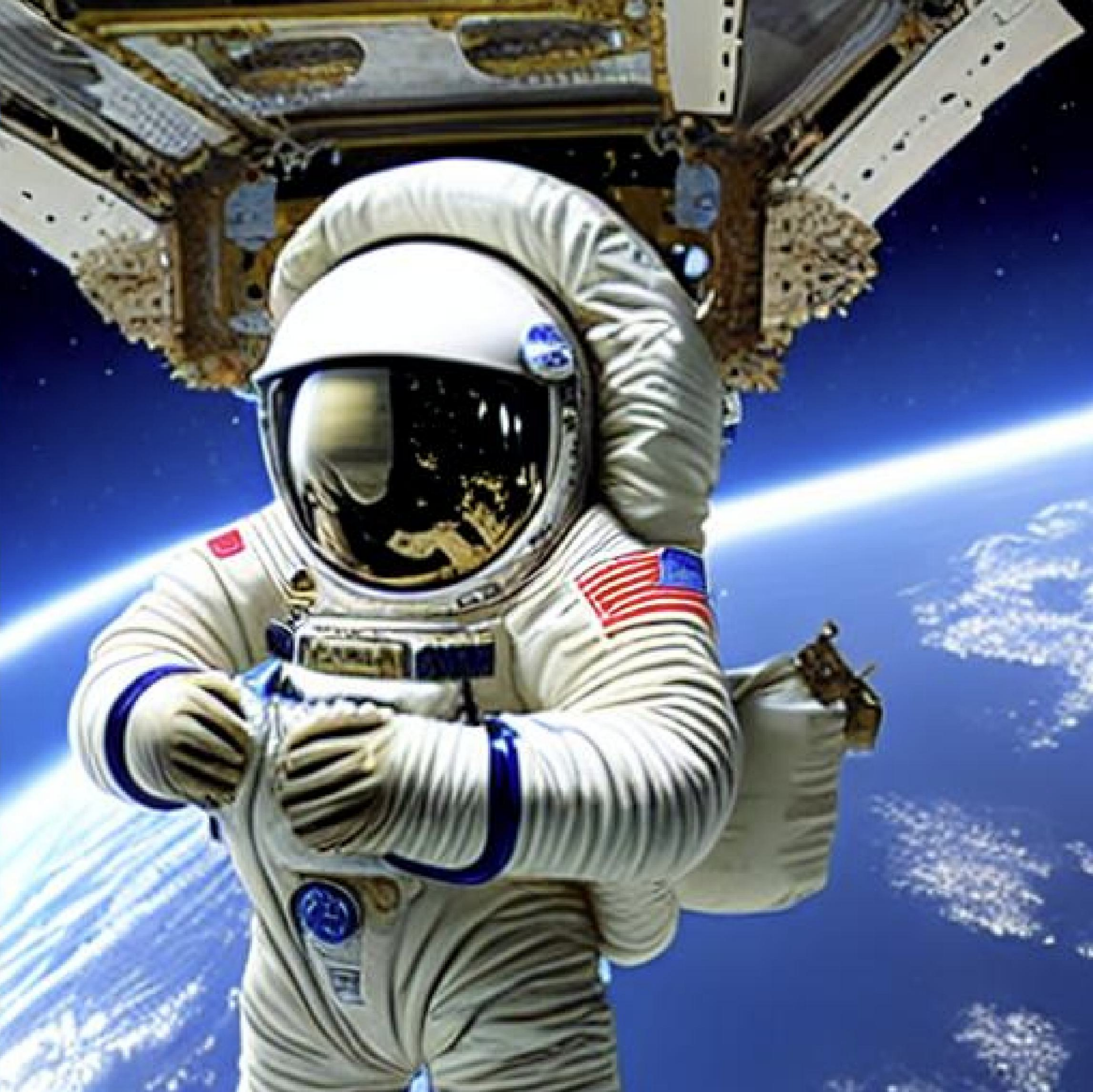} &
        \includegraphics[width=2cm,height=2cm]{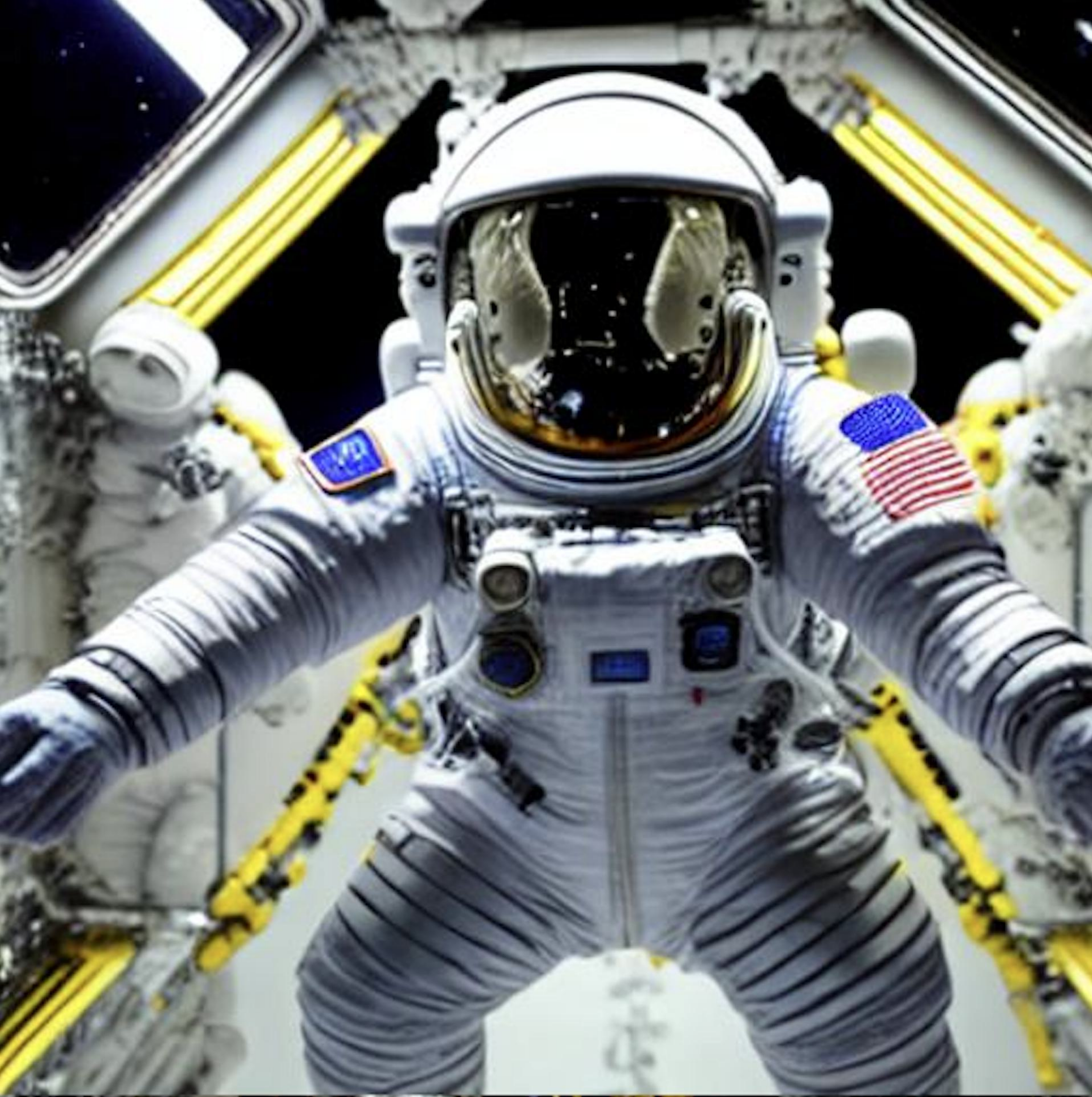} &
        \includegraphics[width=2cm,height=2cm]{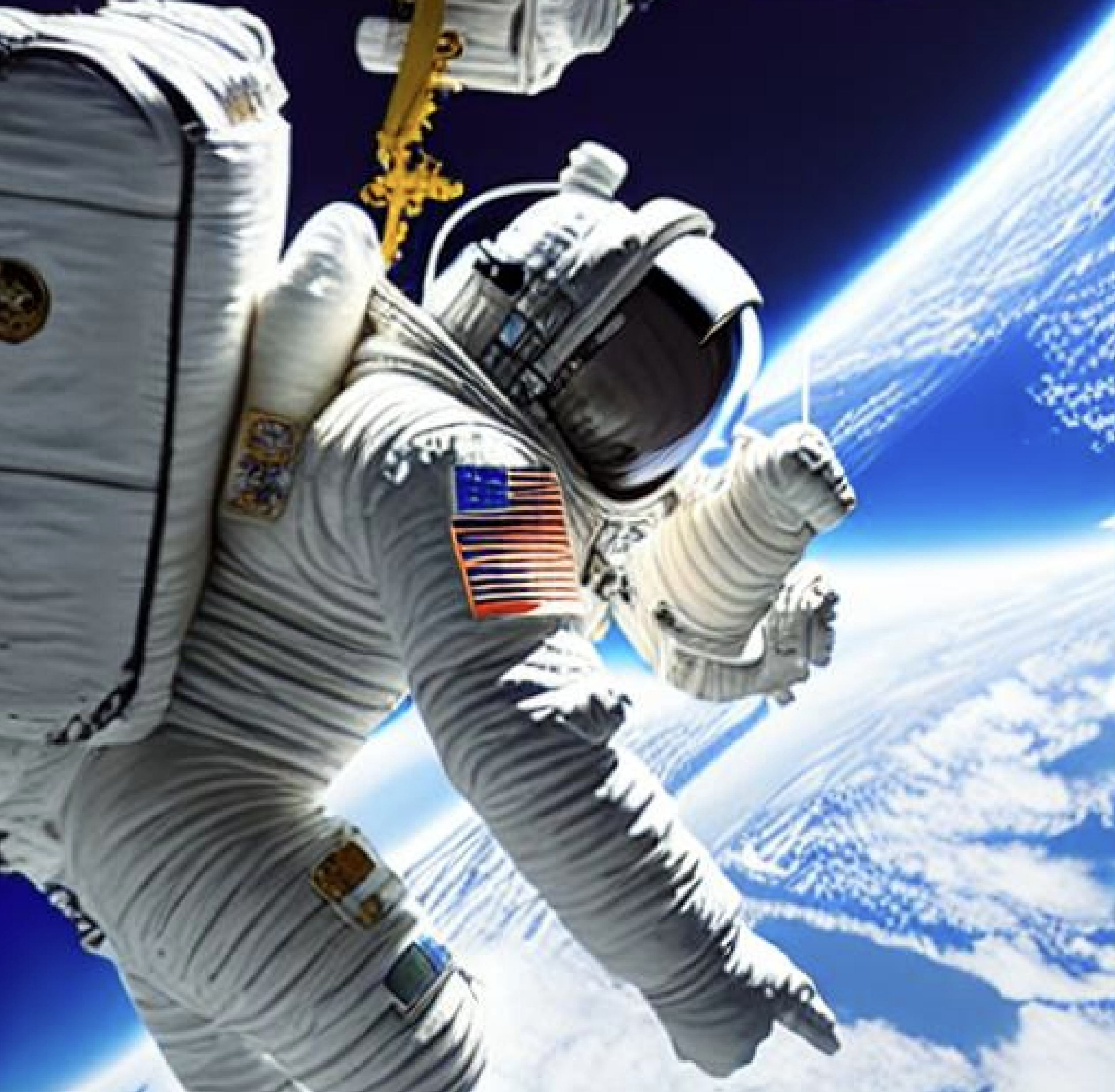} &
        63.4\% \\
        \hline
        \raggedright An \textcolor{red}{astronaut} wearing a Russian Orlan spacesuit during a spacewalk. & 
        \includegraphics[width=2cm,height=2cm]{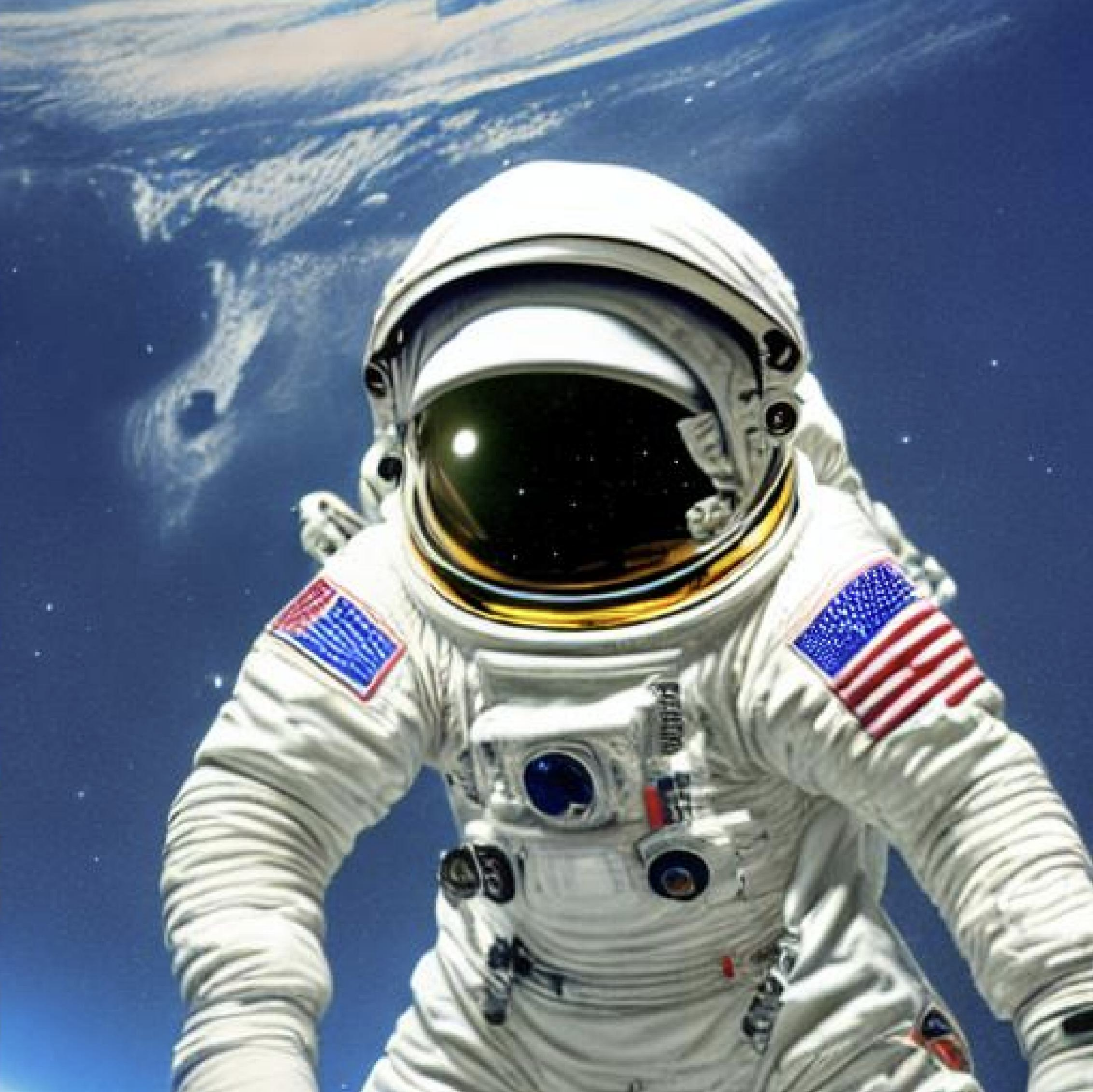} &
        \includegraphics[width=2cm,height=2cm]{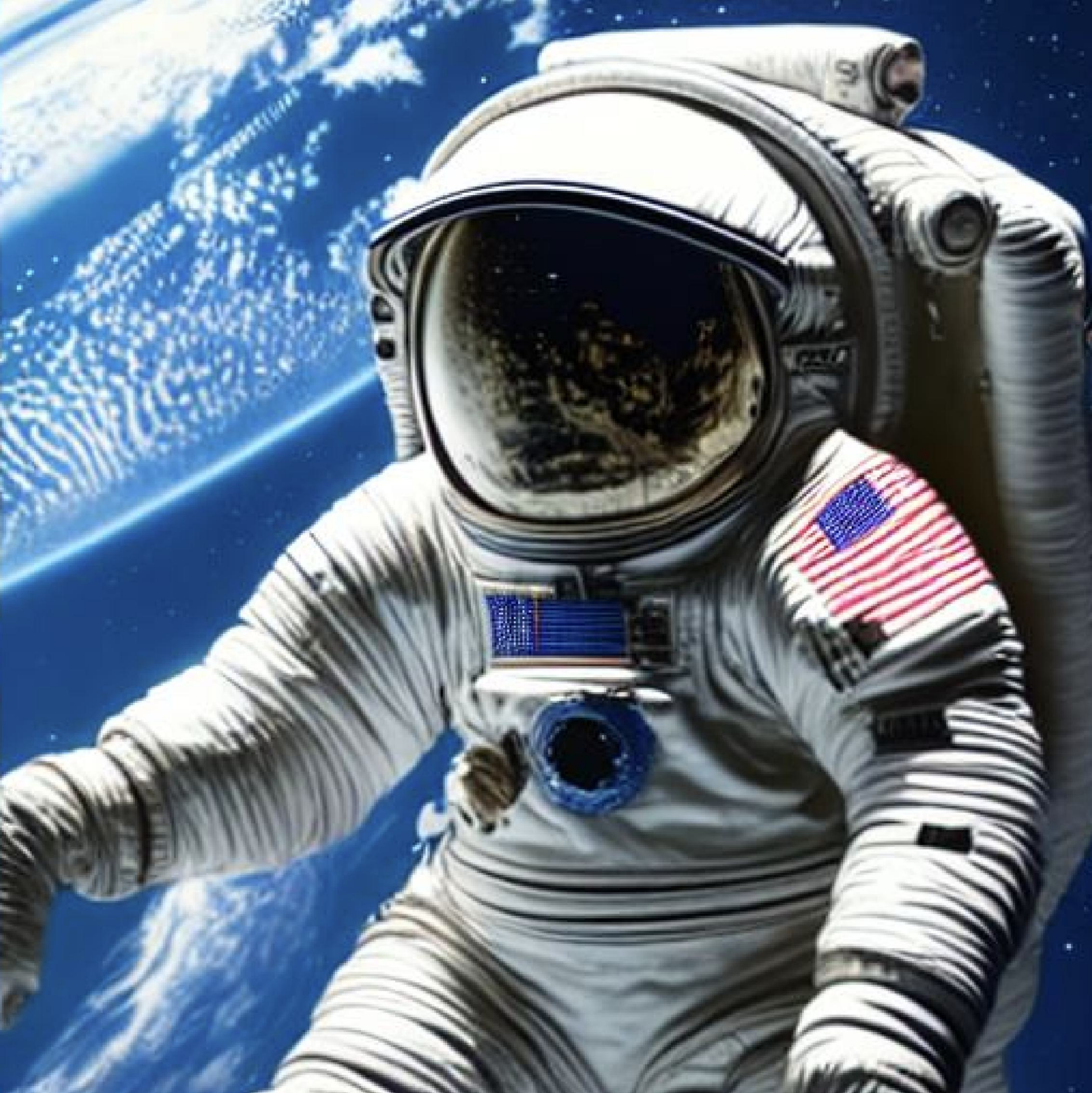} &
        \includegraphics[width=2cm,height=2cm]{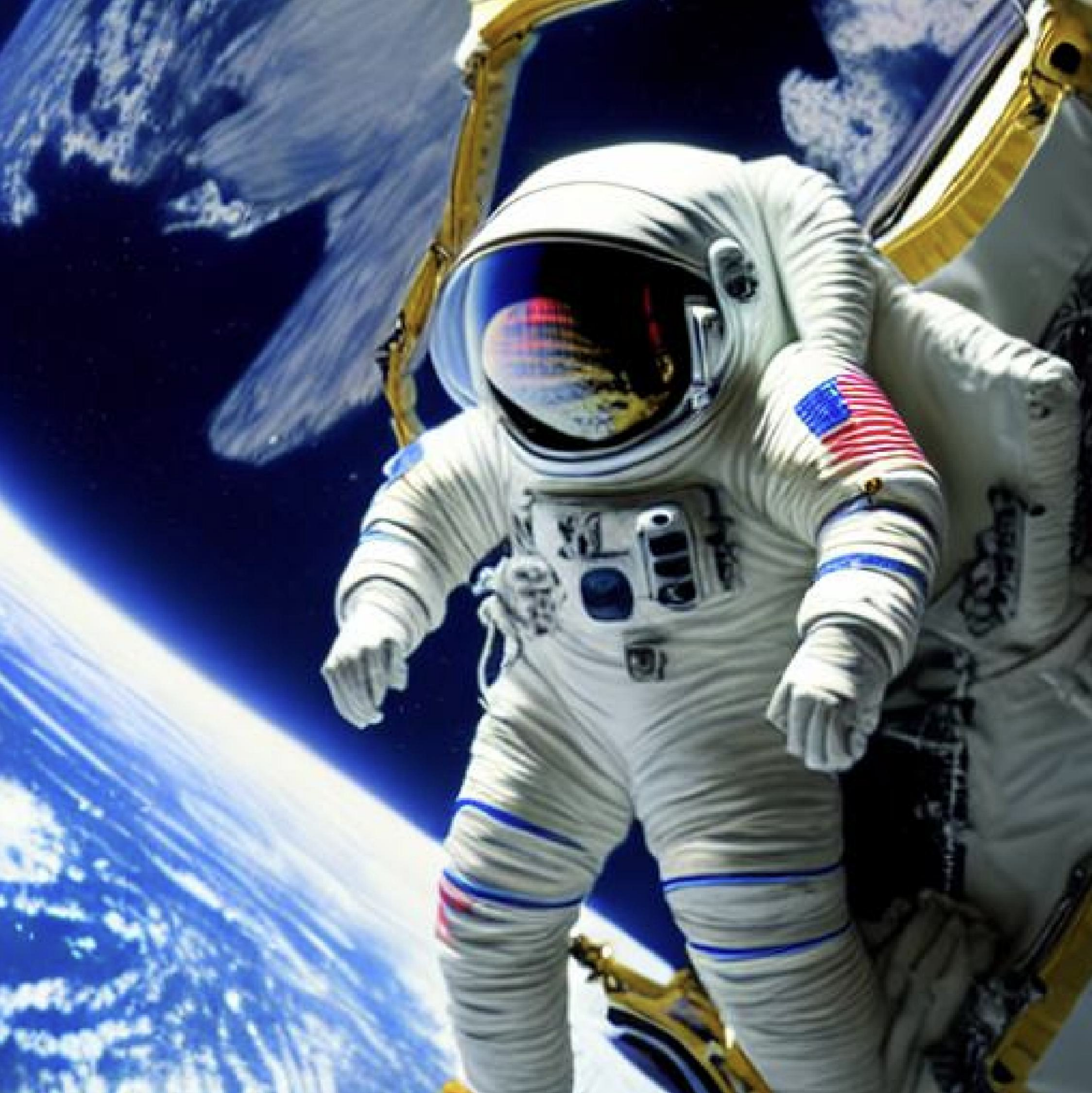} &
        \includegraphics[width=2cm,height=2cm]{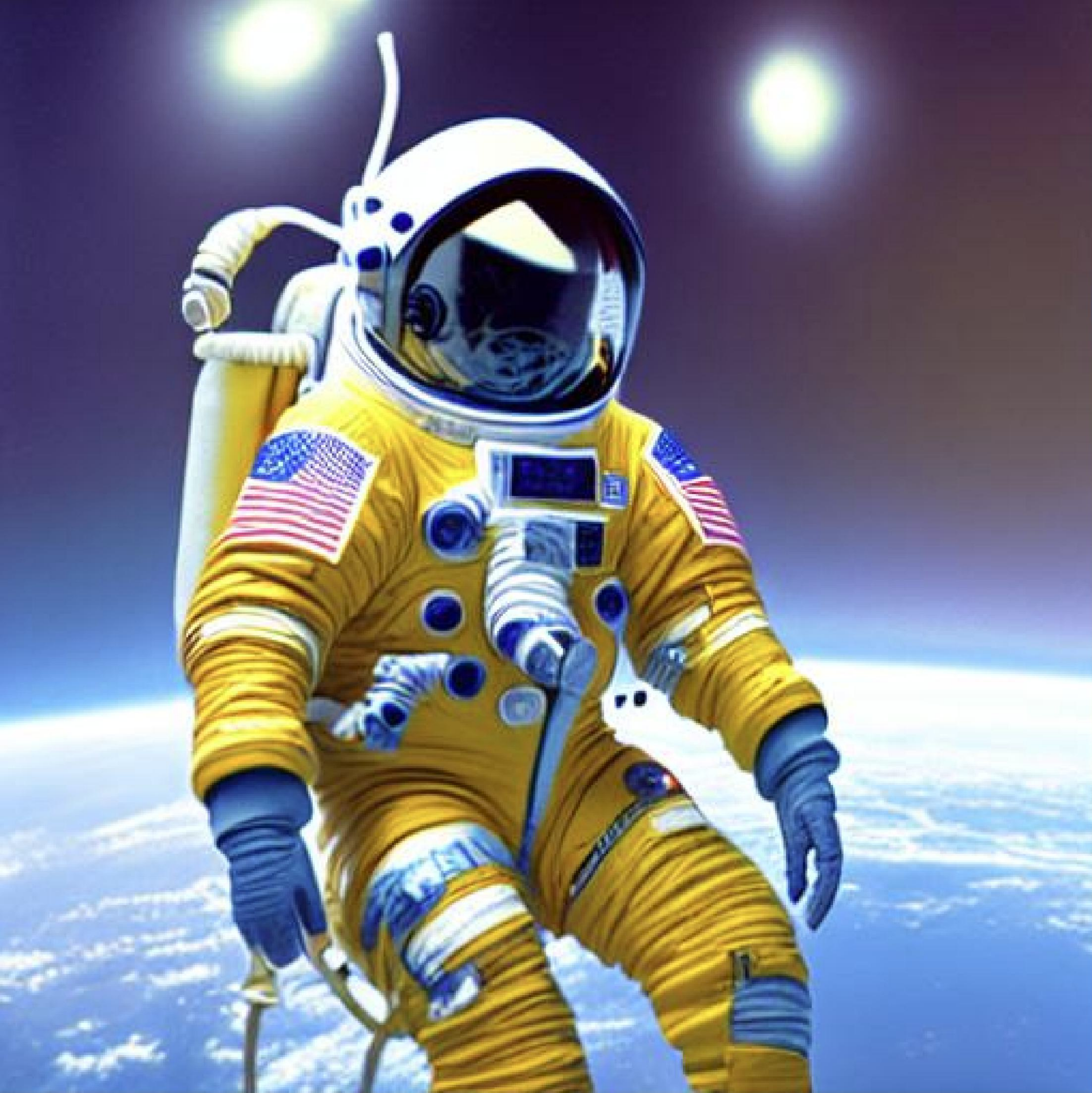} &
        97.8\% \\
        \hline
    \end{tabular}
    \end{adjustbox}
    \caption{Prompts and their corresponding generated images with the percentage of images containing the US flag.}
    \label{fig:us_flag_examples}
\end{figure*}

\begin{figure}[ht]
    \centering
    \includegraphics[width=\linewidth]{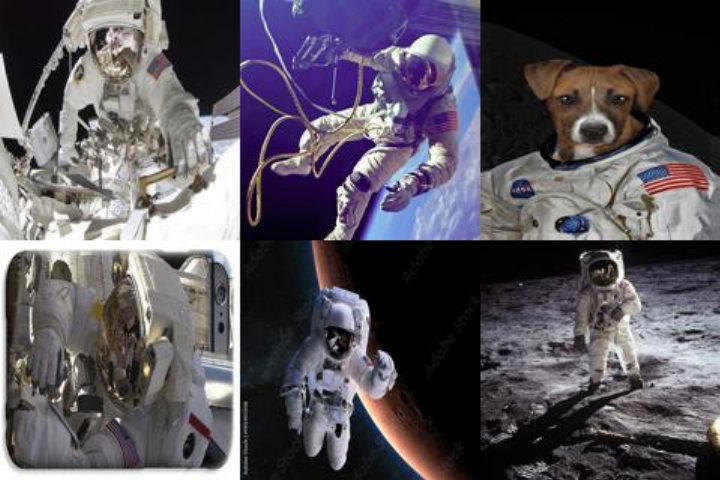}
    \caption{A collection of images from the LAION 400M training dataset showcasing astronauts with the US flag.}
    \label{fig:laion_flag}
\end{figure}

In this section, we explore the concept of object-level replication through a focused case study. Object-level replication refers to the phenomenon where specific objects frequently appear in images despite their absence from the associated textual prompts. This implies a strong correlation between certain keywords and the recurrent visual elements within the dataset. To examine this phenomenon, we concentrate on samples from the LAION dataset containing the keyword ``astronaut.'' We apply the same methodological framework as our initial case study to curate this subset of the dataset and to generate the corresponding image embeddings. This process resulted in approximately 48,000 samples, offering a substantial base for our investigation into keyword-object correlation. Fig.~\ref{fig:laion_flag} presents some of these training samples whose captions contain the word "astronaut" and the corresponding images feature the US flag.

In this case study, our attention is on the US flag. An analysis of approximately 1000 training data samples with captions mentioning ``astronaut'' revealed that 10\% included images of the US flag, even when the terms ``US'' or ``flag'' were not specified. To further explore this phenomenon, we first employed ChatGPT to craft a series of random prompts that include the word ``astronaut.'' We then used these prompts to generate images with the Stable Diffusion model, which led to a frequent replication of the US flag in the output. Note that, due to the low quality of generation of the pre-trained Stable Diffusion model, we fine-tuned the model on a small dataset of prompts and corresponding high-resolution generated images from the Midjourney API to enhance the quality of the generated examples. Fig.~\ref{fig:us_flag_examples} displays the ChatGPT-generated prompts and corresponding images from the Stable Diffusion model. By generating 500 images using varied random seeds, we assess the model's tendency to replicate the US flag from the prompt. Subsequently, we calculate and report the percentage of images featuring the US flag.

\section{Future Directions}
Although our investigation is focused on two specific case studies, we have demonstrated the occurrence of word-level duplication in Stable Diffusion models. For future work, we propose conducting broader experiments within the word-level duplication context and undertaking a more comprehensive analysis. Additionally, developing new mitigation techniques that reduce memorization while preserving model utility is of paramount importance. The replicated features identified in our study also pose potential privacy risks, potentially making the models susceptible to various attacks, including membership inference and backdoor attacks. Addressing these concerns will be a critical aspect of future research.

\section{Conclusion}

Duplication in training data is a key contributor to memorization in generative models. This paper identifies two types of duplication leading to replication at inference. We investigated these through two LAION dataset case studies. Our work emphasizes the importance of vigilance against diverse duplication forms in training data and the need for effective mitigation strategies. It is our hope that this work will inspire more conscientious data curation and lead to the development of both powerful and privacy-preserving generative models.

\bigskip

\bibliography{main}

\section{Appendix}

\begin{figure*}[h]
    \centering
    \begin{adjustbox}{width=16.5cm,center}
    \begin{tabular}{>{\centering\arraybackslash}m{4cm}>{\centering\arraybackslash}m{1.5cm}>{\centering\arraybackslash}m{2cm}*{3}{m{2cm}}m{2cm}}
        \hline
        \multirow{2}{*}{\centering \textbf{Prompt}} & \multirow{2}{*}{\parbox{1.5cm}{\centering \textbf{Text\\Similarity}}} & \multirow{2}{*}{\parbox{2cm}{\centering \textbf{Image Sim $>$ 0.70 (\%)}}} & \multicolumn{3}{c}{\textbf{Example Generated Images}} & \multirow{2}{*}{\parbox{2cm}{\centering \textbf{Original Image}}} \\
        \cline{4-6}
        & & & \centering \textbf{$>$ 0.7} & \centering \textbf{0.65-0.7}  & \centering \textbf{$<$0.65} & \\
        \hline
        \raggedright \small Van Gogh almond blossoming & 1.0000 & 87.2\% & 
        \raisebox{-0.05\height}{\includegraphics[width=2cm]{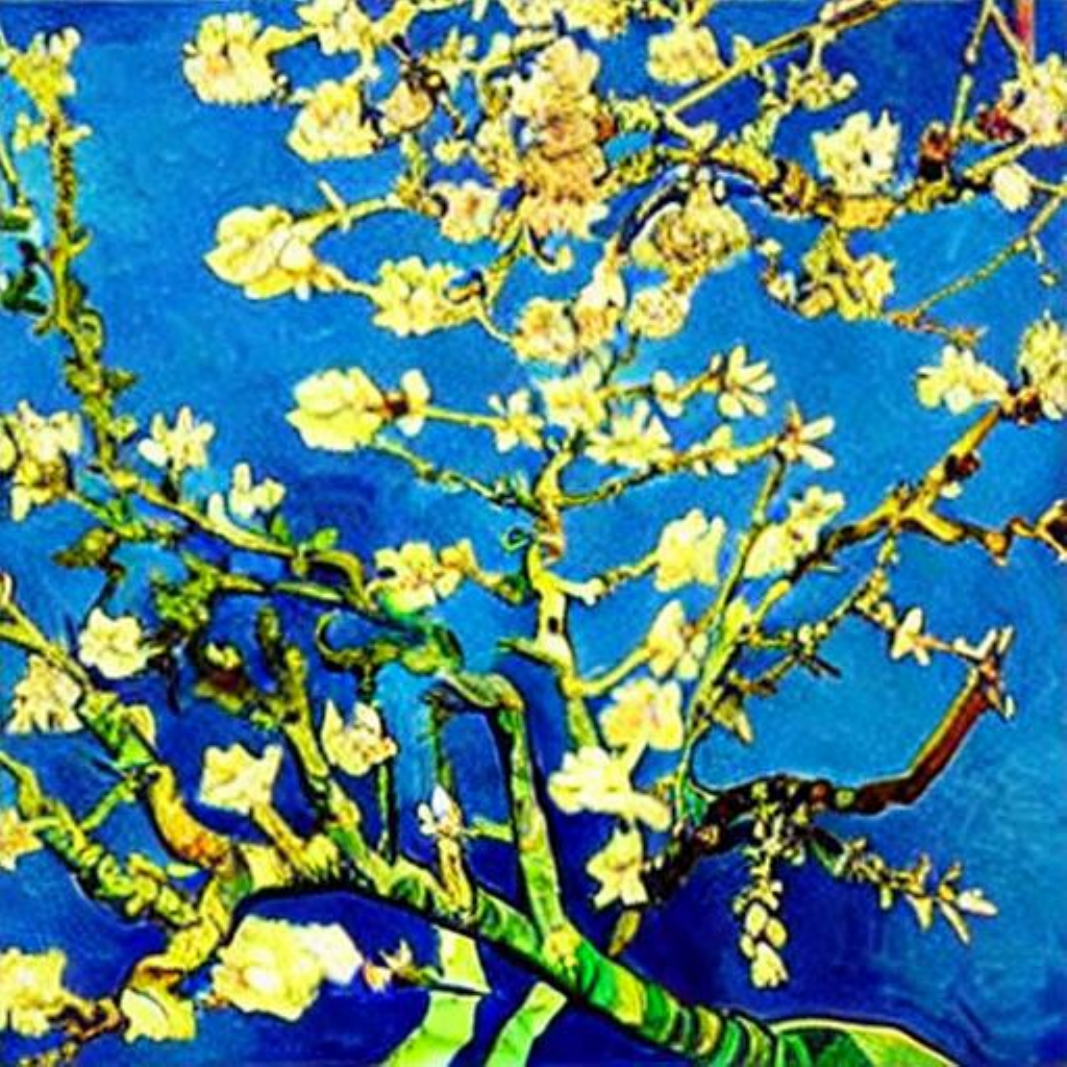}} &
        \raisebox{-0.05\height}{\includegraphics[width=2cm]{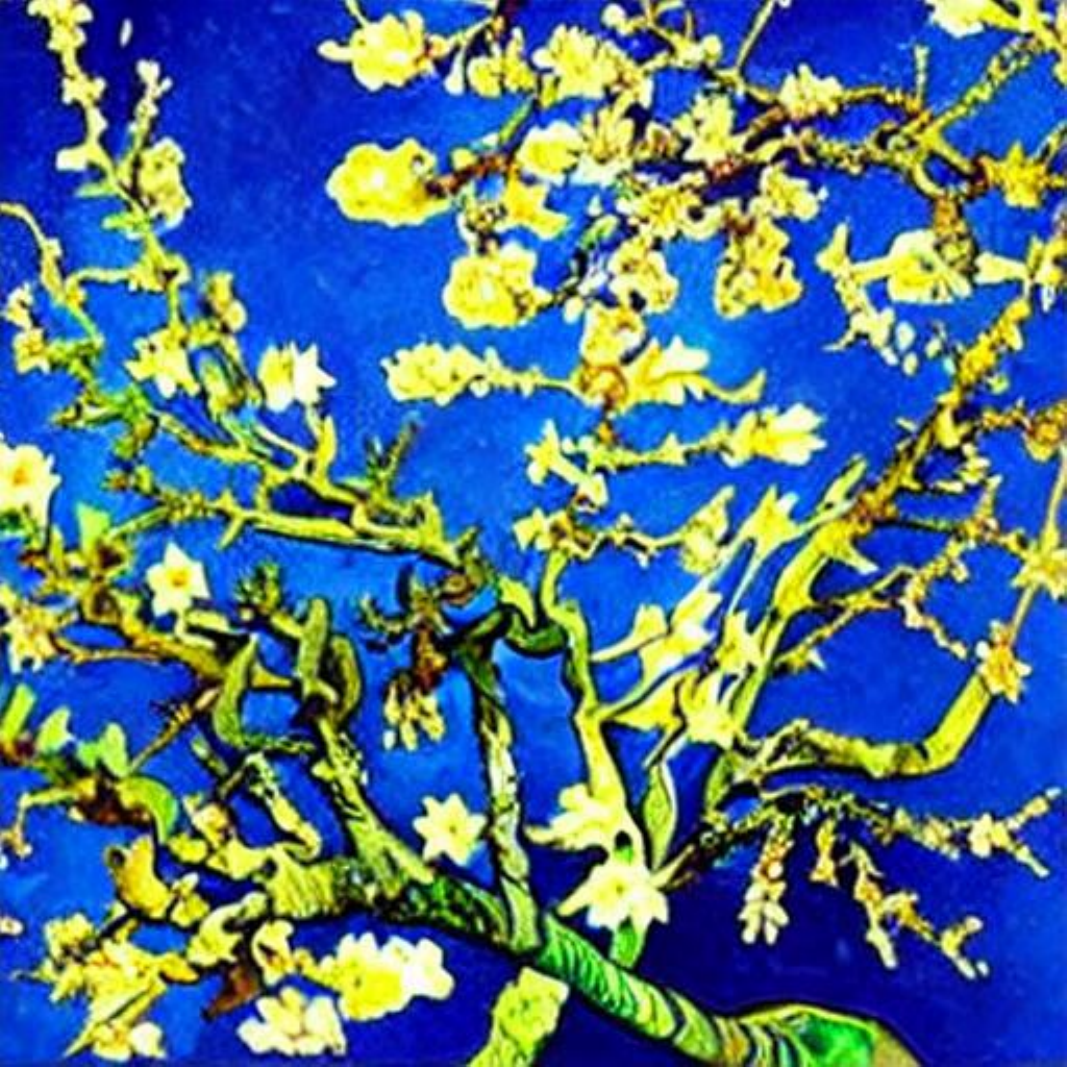}} &
        \raisebox{-0.05\height}{\includegraphics[width=2cm]{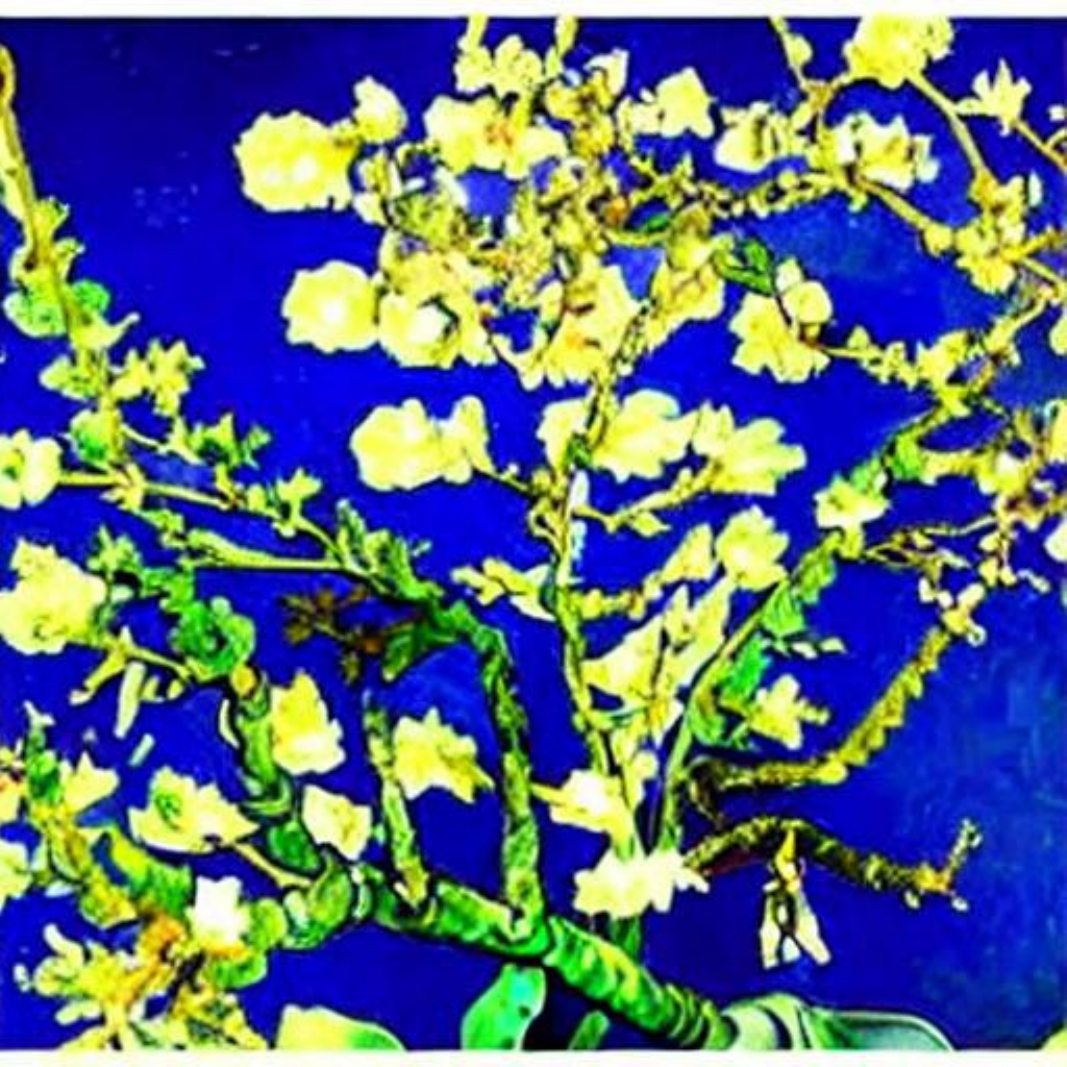}} &
        \raisebox{-0.05\height}{\includegraphics[width=2cm]{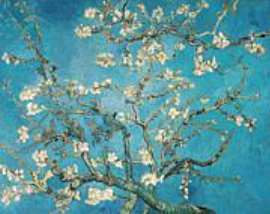}}\\
        \hline
        \raggedright \small \textcolor{red}{Van Gogh}'s vision of \textcolor{red}{almond} trees \textcolor{red}{blossoming} in the early spring light. & 0.8279 & 93\% &
        \raisebox{-0.05\height}{\includegraphics[width=2cm]{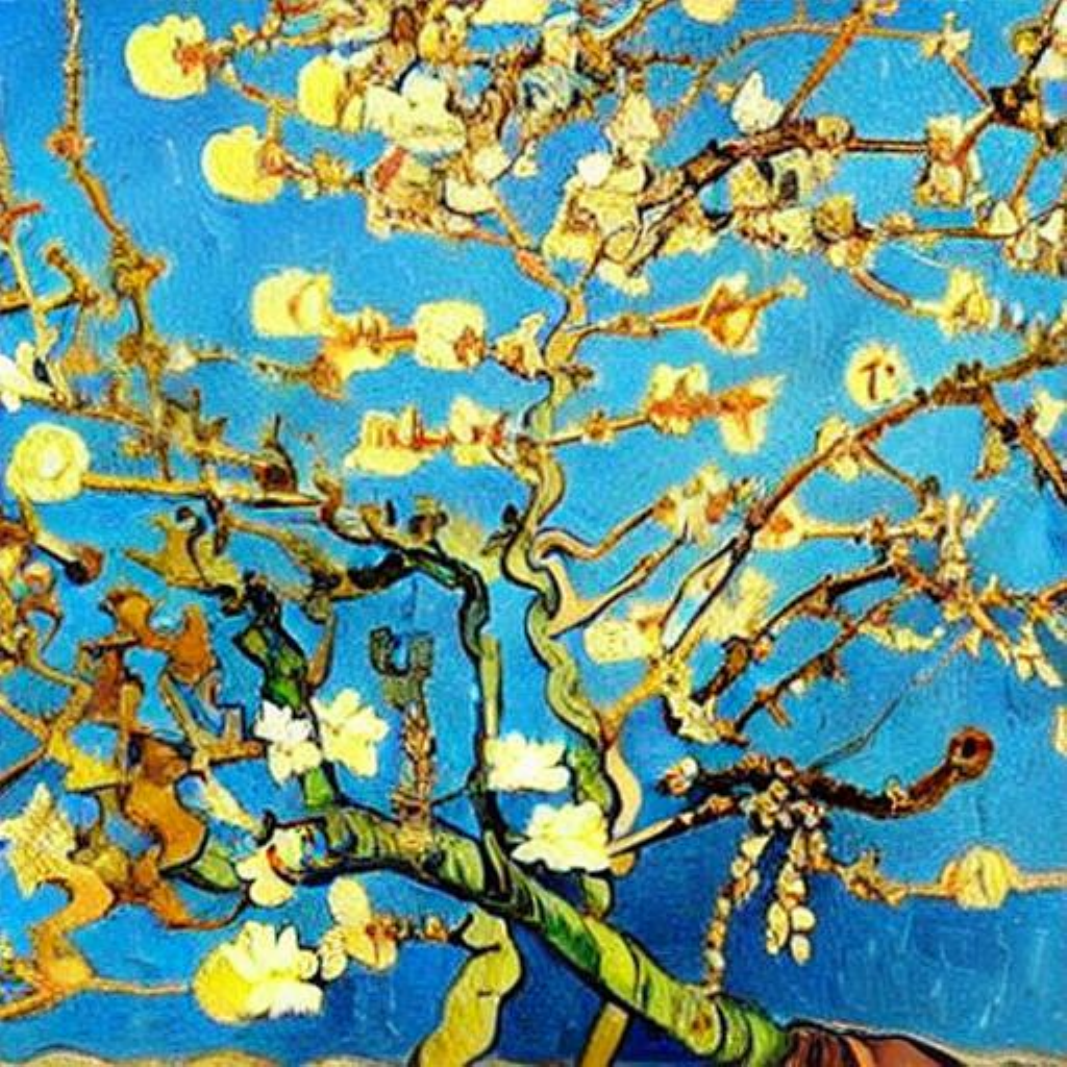}} &
        \raisebox{-0.05\height}{\includegraphics[width=2cm]{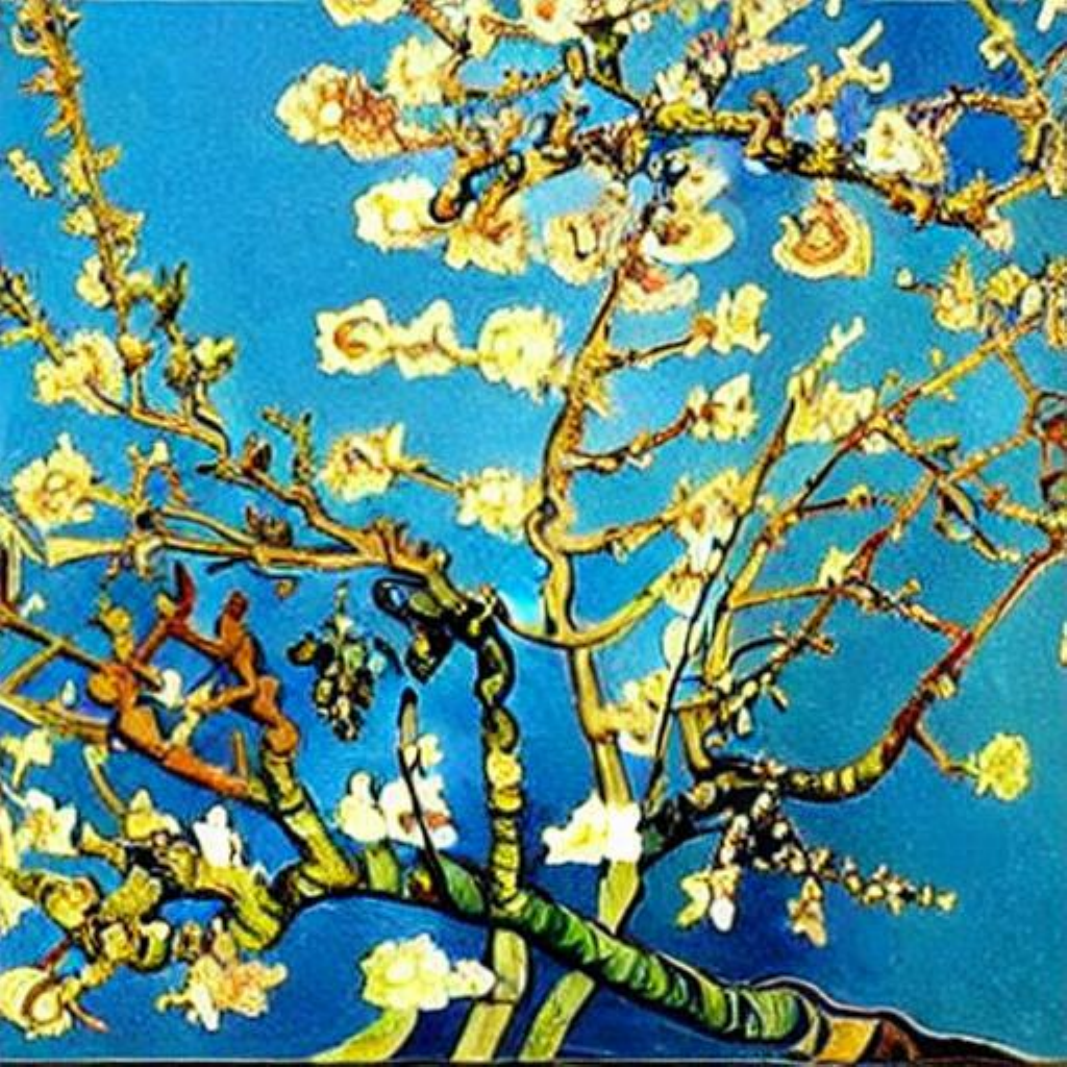}} &
        \raisebox{-0.05\height}{\includegraphics[width=2cm]{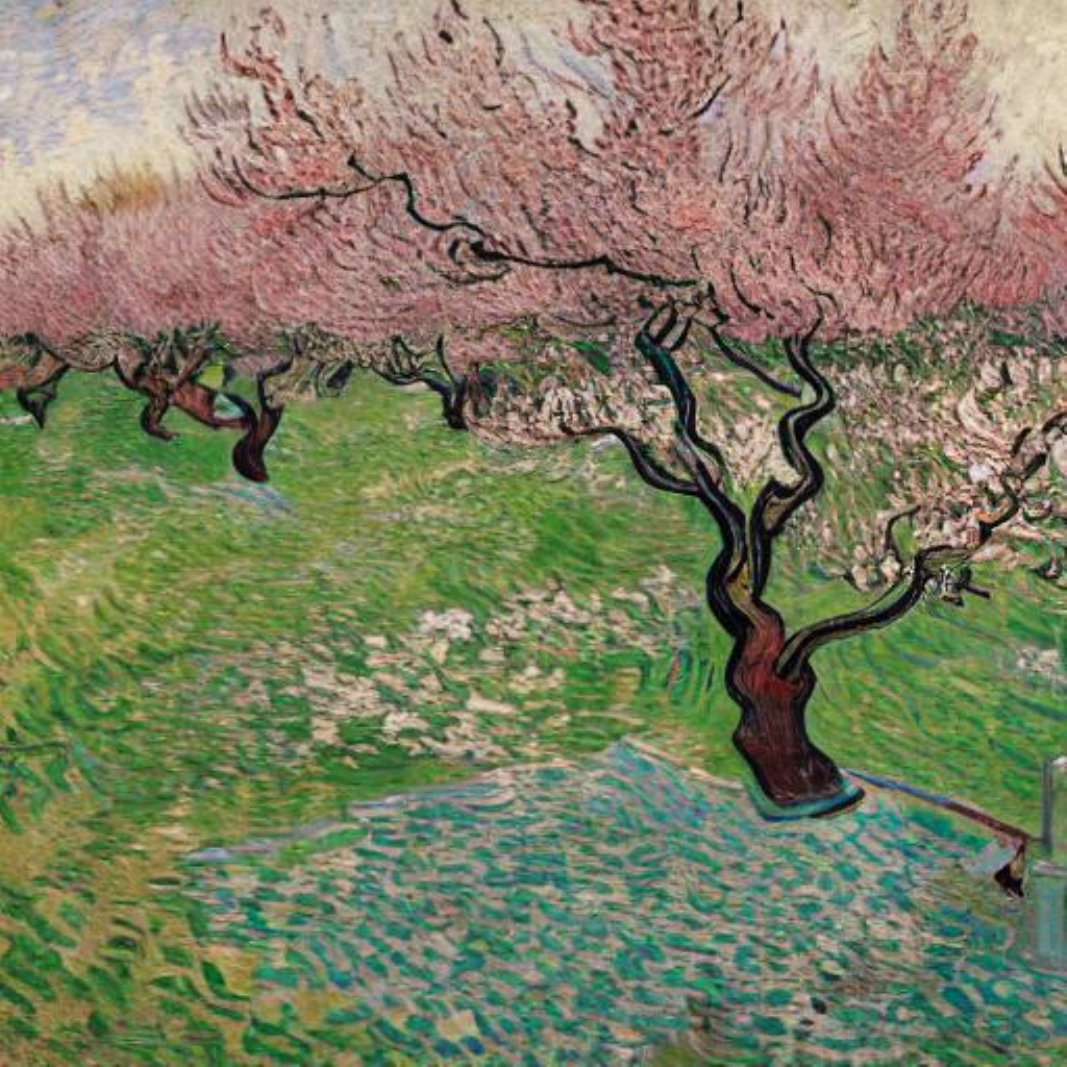}} &
        \raisebox{-0.05\height}{\includegraphics[width=2cm]{sec/almond_blossoming/almond_blossoming.pdf}}\\
        \hline
        \raggedright \small Through \textcolor{red}{Van Gogh}'s eyes, the \textcolor{red}{blossoming almond} branches are immortalized, their fleeting beauty captured in swirls of color and light. & 0.7323 & 90.4\% &
        \raisebox{-0.05\height}{\includegraphics[width=2cm]{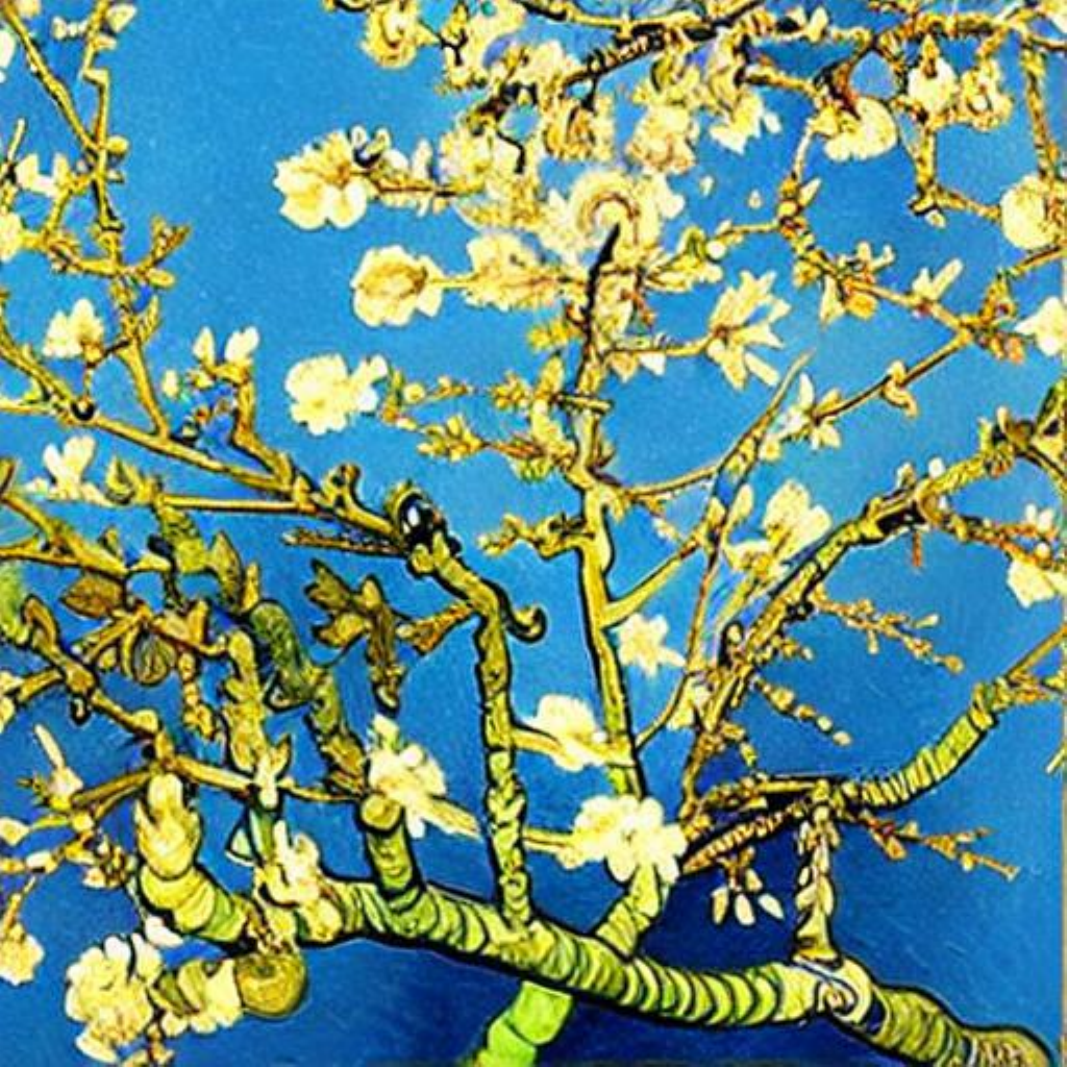}} &
        \raisebox{-0.05\height}{\includegraphics[width=2cm]{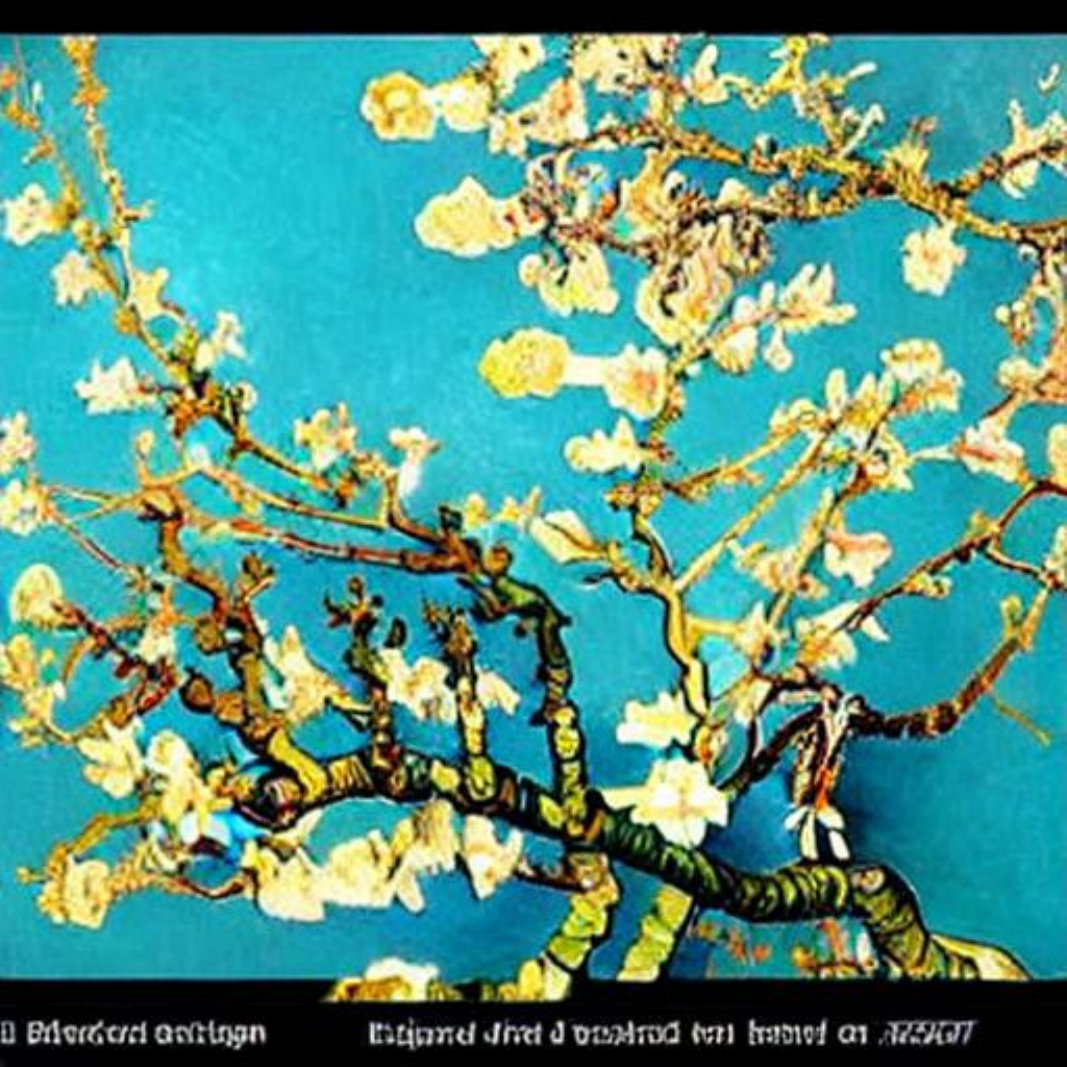}} &
        \raisebox{-0.05\height}{\includegraphics[width=2cm]{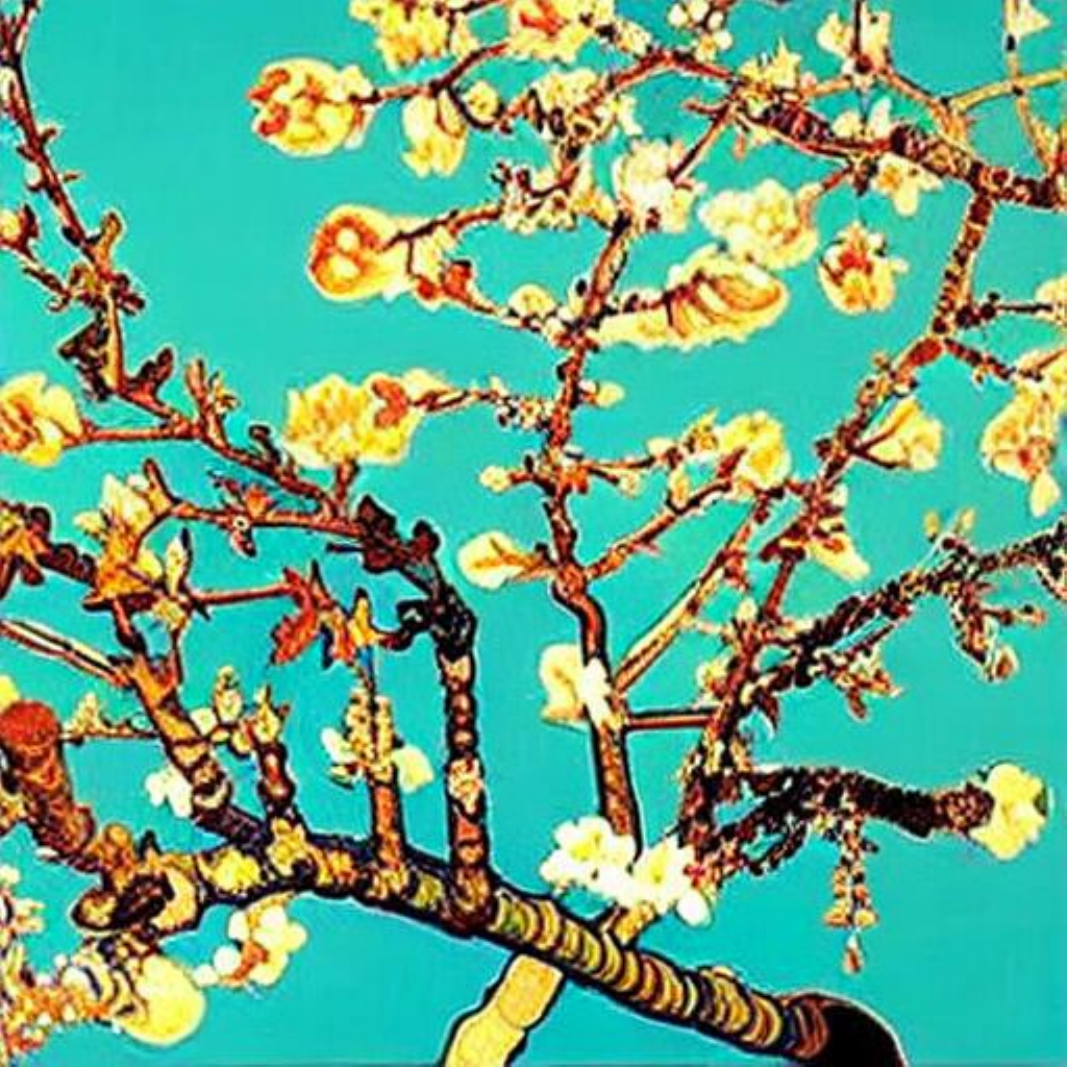}} &
        \raisebox{-0.05\height}{\includegraphics[width=2cm]{sec/almond_blossoming/almond_blossoming.pdf}}\\
        \hline
        \raggedright \small In a cozy corner of the library, a \textcolor{red}{Van Gogh} anthology sits open to a painting of \textcolor{red}{almond} trees \textcolor{red}{blossoming}, inspiring daydreams of spring. & 0.5690 & 37\% &
        \raisebox{-0.05\height}{\includegraphics[width=2cm]{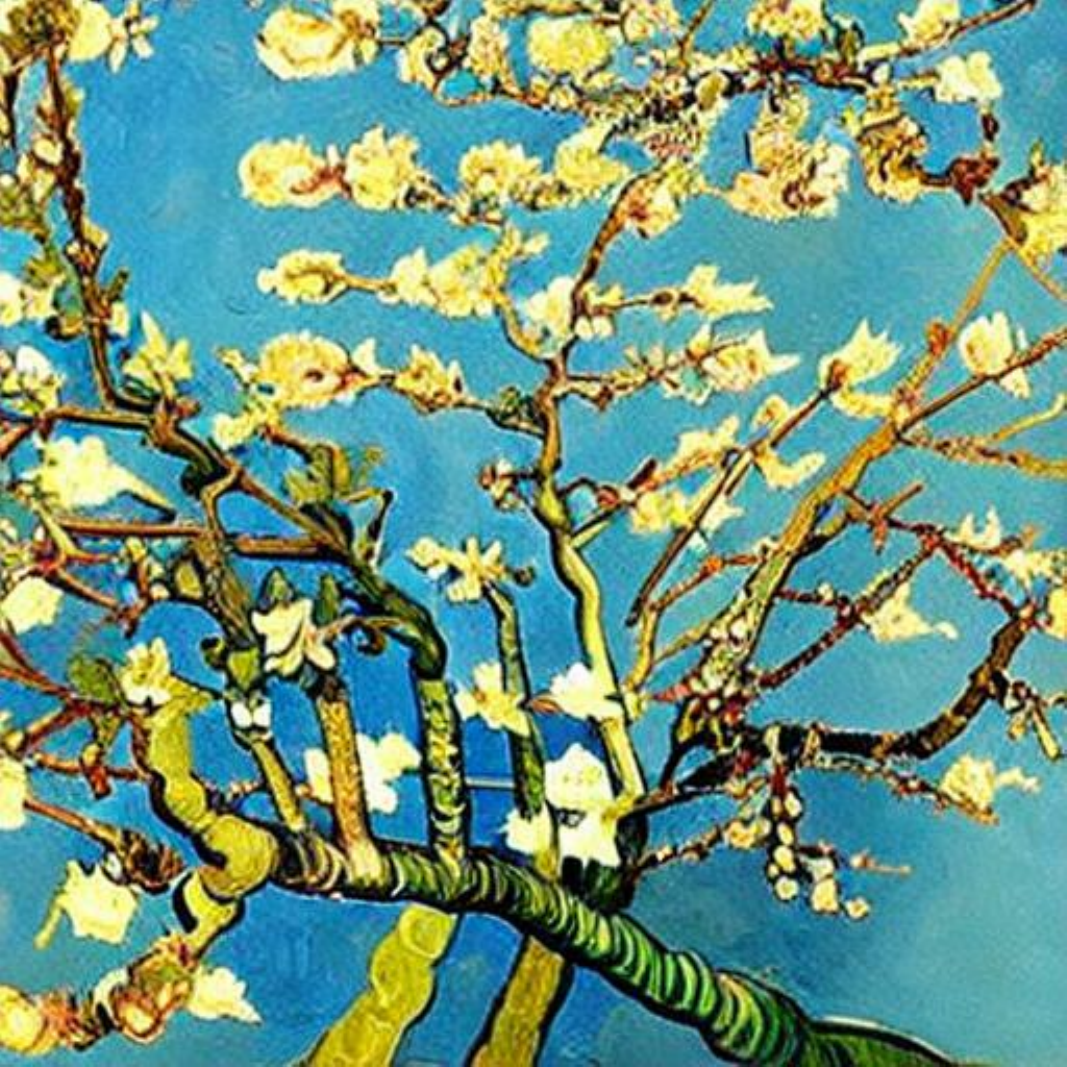}} &
        \raisebox{-0.05\height}{\includegraphics[width=2cm]{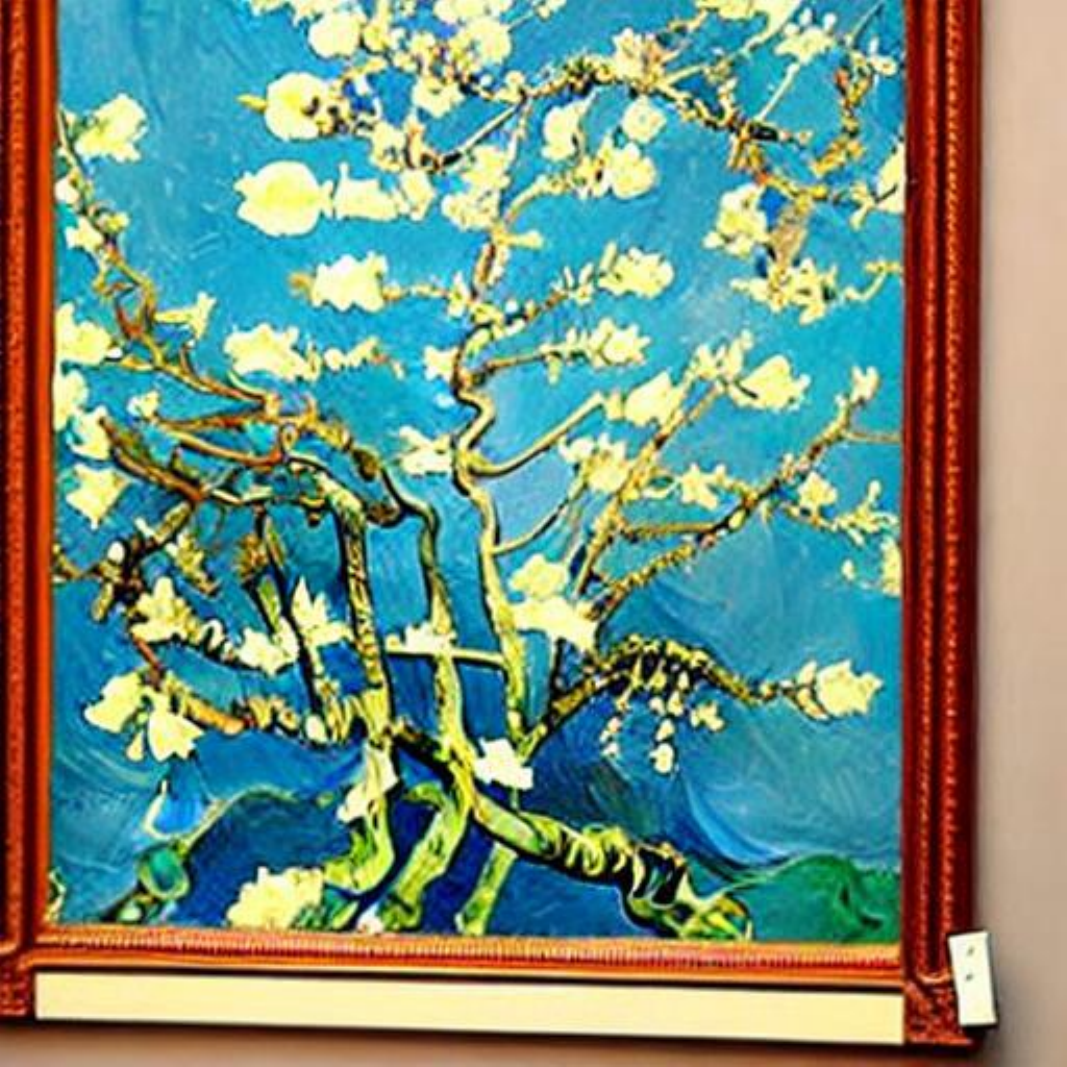}} &
        \raisebox{-0.05\height}{\includegraphics[width=2cm]{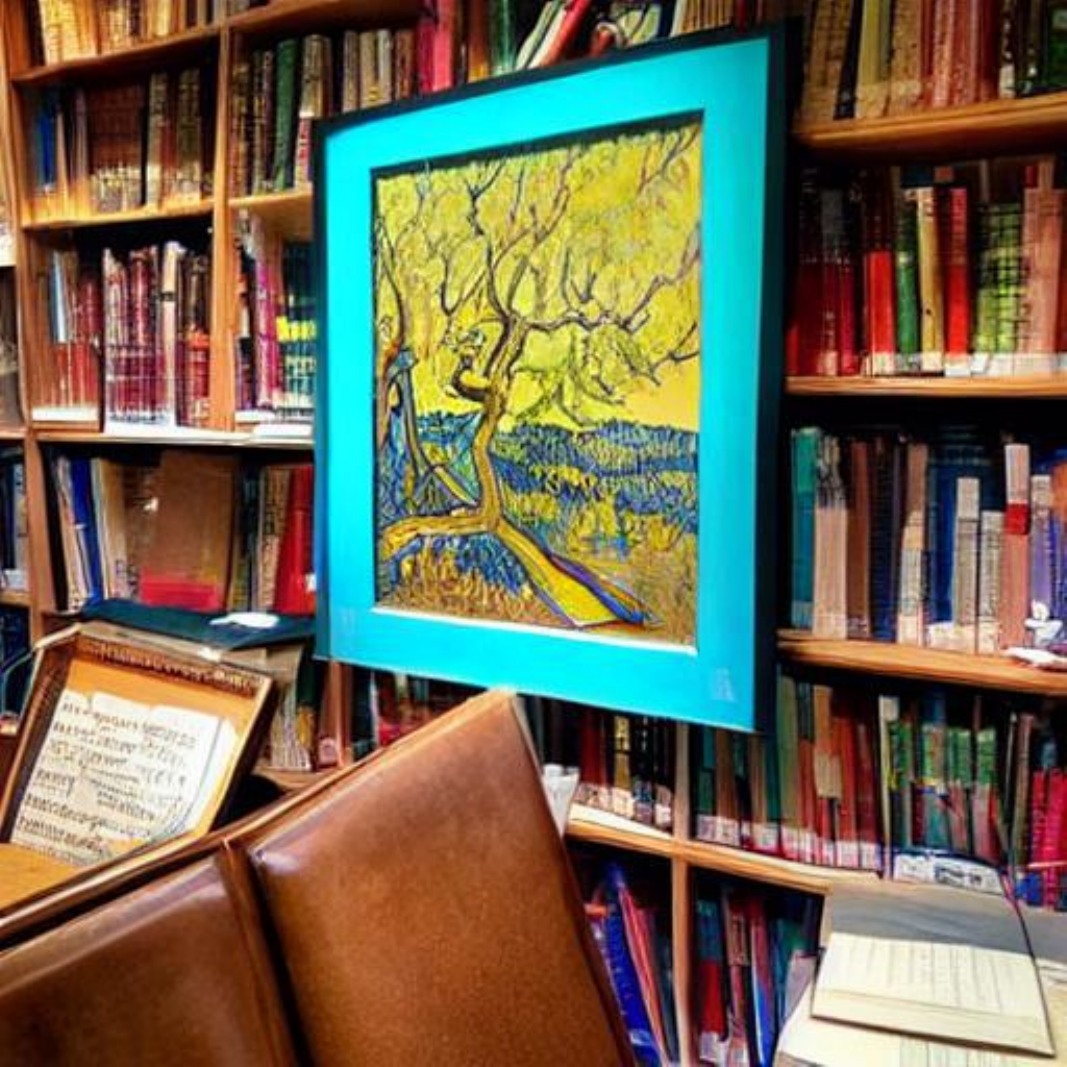}} &
        \raisebox{-0.05\height}{\includegraphics[width=2cm]{sec/almond_blossoming/almond_blossoming.pdf}}\\
        \hline
        \raggedright \small A digital art piece animating the life cycle of \textcolor{red}{almond blossoming}, each frame a celebration of growth and artistry. & 0.4621 & 5.4\% &
        \raisebox{-0.05\height}{\includegraphics[width=2cm]{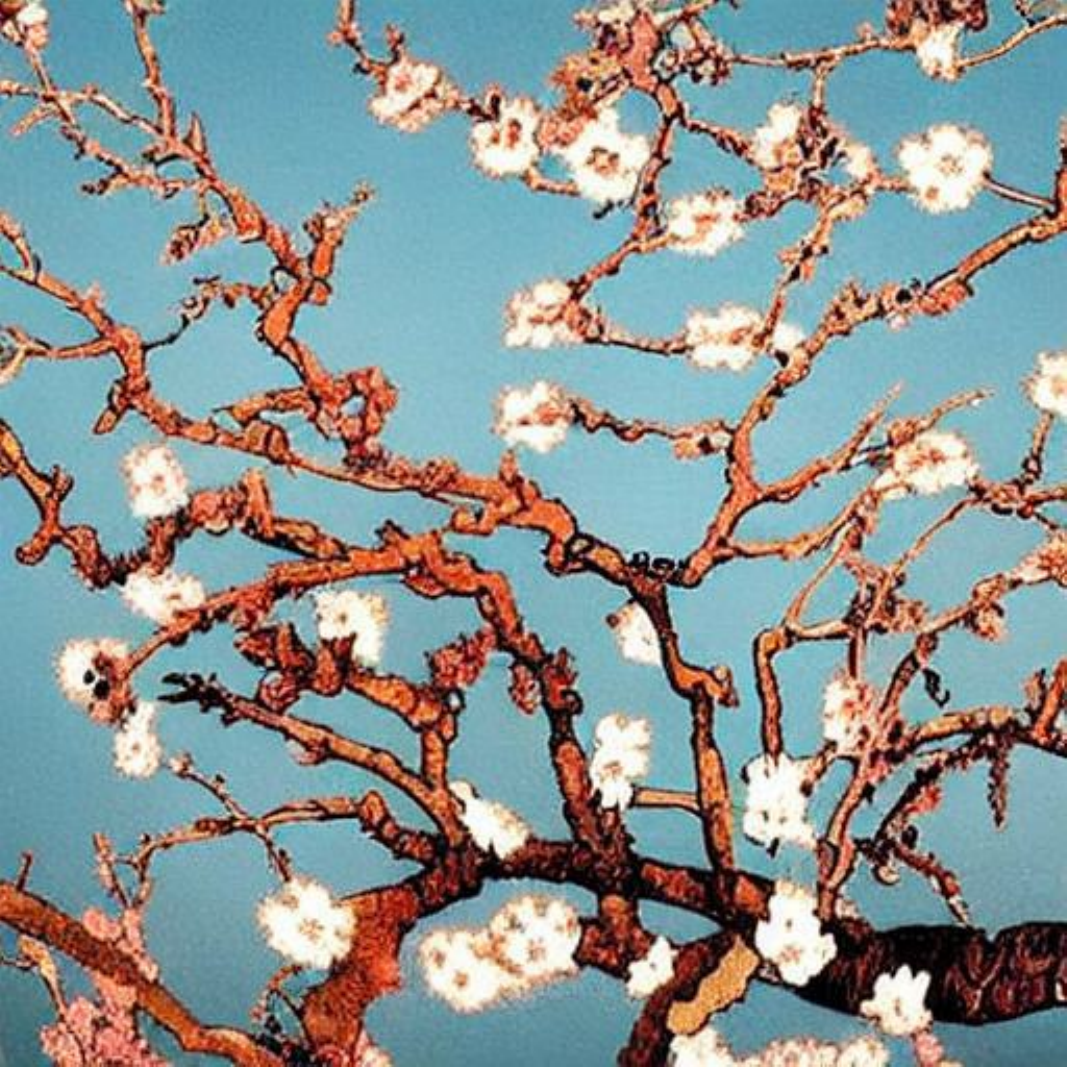}} &
        \raisebox{-0.05\height}{\includegraphics[width=2cm]{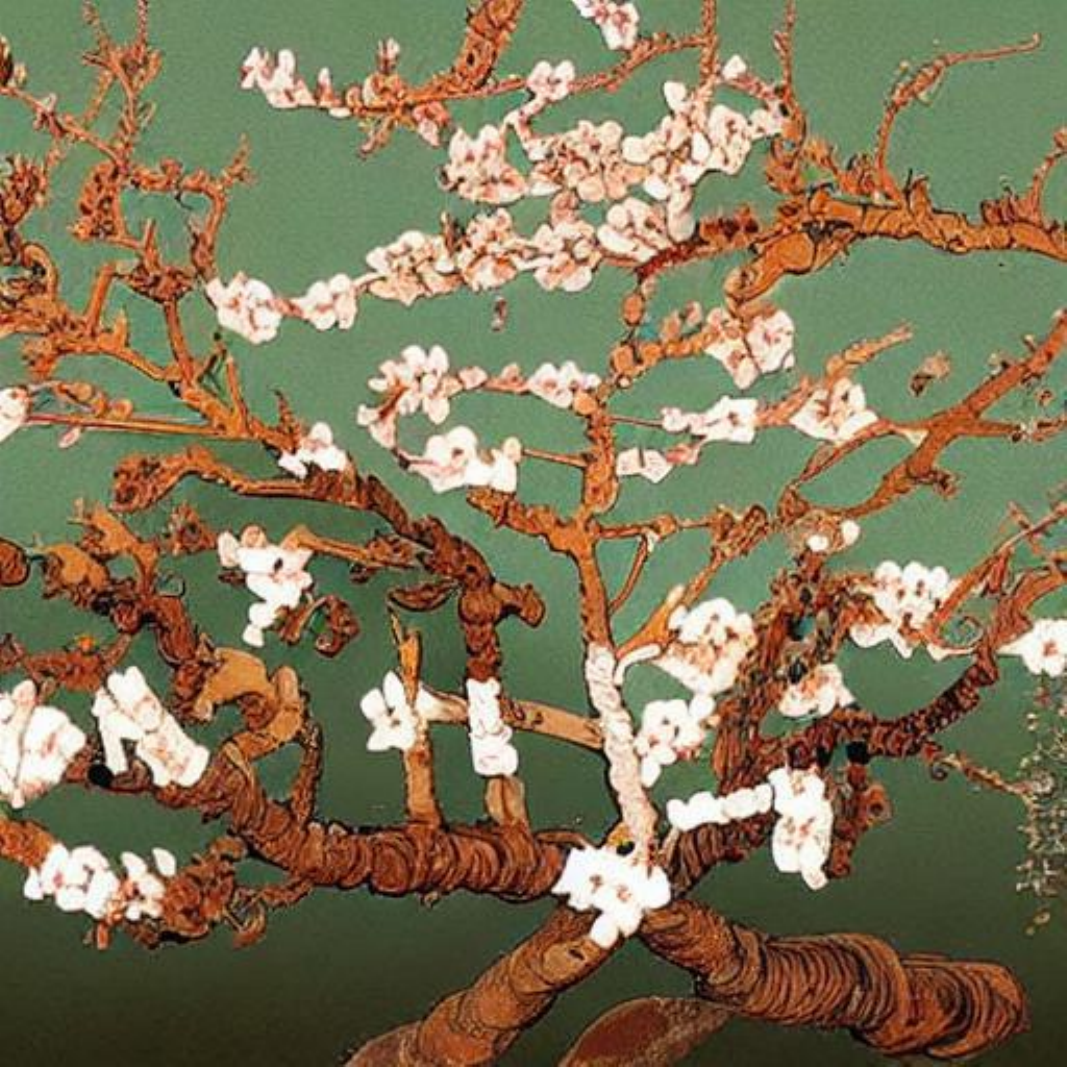}} &
        \raisebox{-0.05\height}{\includegraphics[width=2cm]{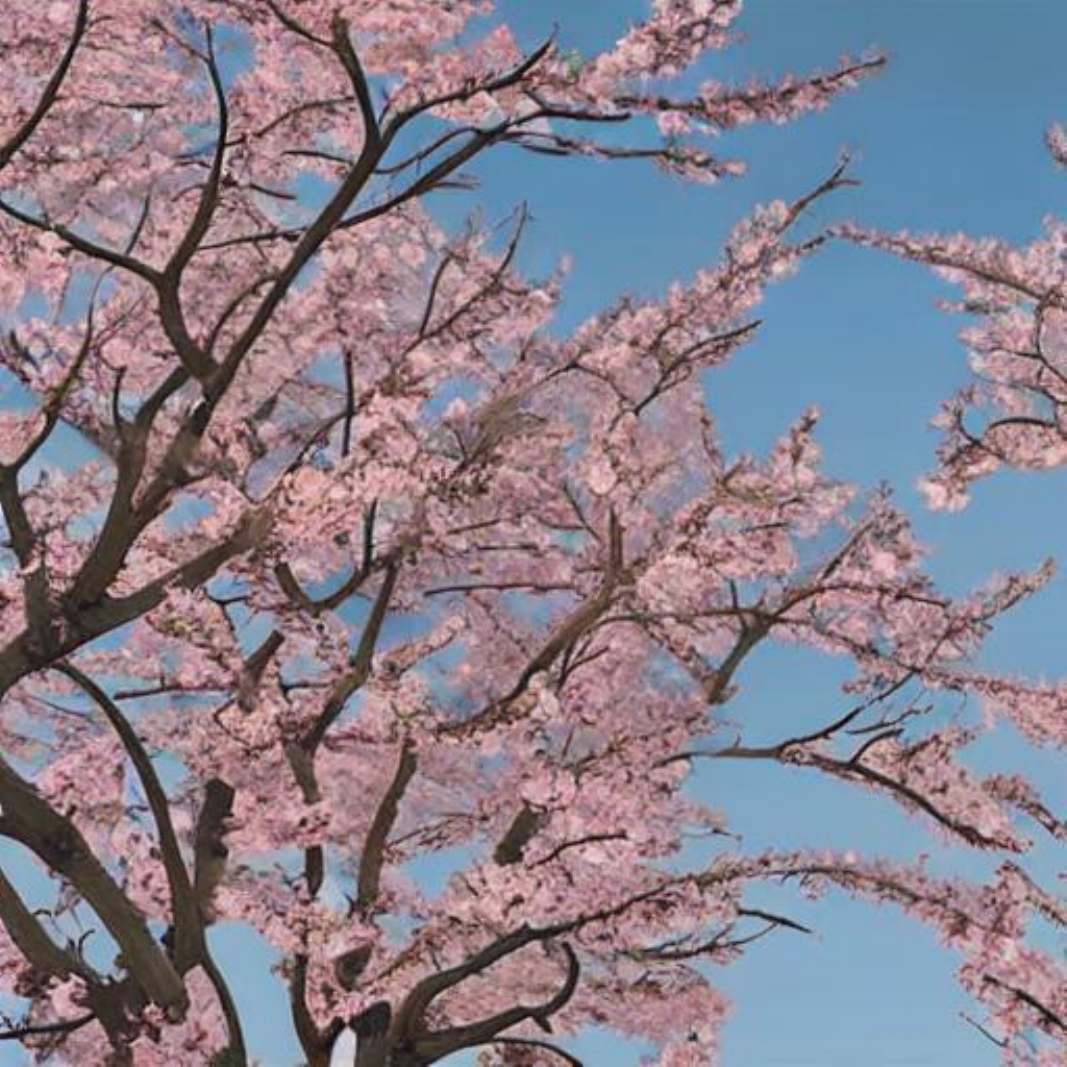}} &
        \raisebox{-0.05\height}{\includegraphics[width=2cm]{sec/almond_blossoming/almond_blossoming.pdf}}\\
        \hline
    \end{tabular}
    \end{adjustbox}
    \caption{Captions containing the terms “Van Gogh”, “blossoming”, and “almond”, alongside their respective generated images for various seed values.}
    \label{fig:blossoming}
\end{figure*}

\begin{figure*}[h]
    \centering
    \begin{adjustbox}{width=16.5cm,center}
    \begin{tabular}{>{\centering\arraybackslash}m{4cm}>{\centering\arraybackslash}m{1.5cm}>{\centering\arraybackslash}m{2cm}*{3}{m{2cm}}m{2cm}}
        \hline
        \multirow{2}{*}{\centering \textbf{Prompt}} & \multirow{2}{*}{\parbox{1.5cm}{\centering \textbf{Text\\Similarity}}} & \multirow{2}{*}{\parbox{2cm}{\centering \textbf{Image Sim $>$ 0.85 (\%)}}} & \multicolumn{3}{c}{\textbf{Example Generated Images}} & \multirow{2}{*}{\parbox{2cm}{\centering \textbf{Original Image}}} \\
        \cline{4-6}
        & & & \centering \textbf{$>$ 0.85} & \centering \textbf{0.75-0.85}  & \centering \textbf{$<$0.75} & \\
        \hline
        \raggedright \small Van Gogh sunflowers & 1.0000 & 83.4\% & 
        \raisebox{-0.05\height}{\includegraphics[width=2cm]{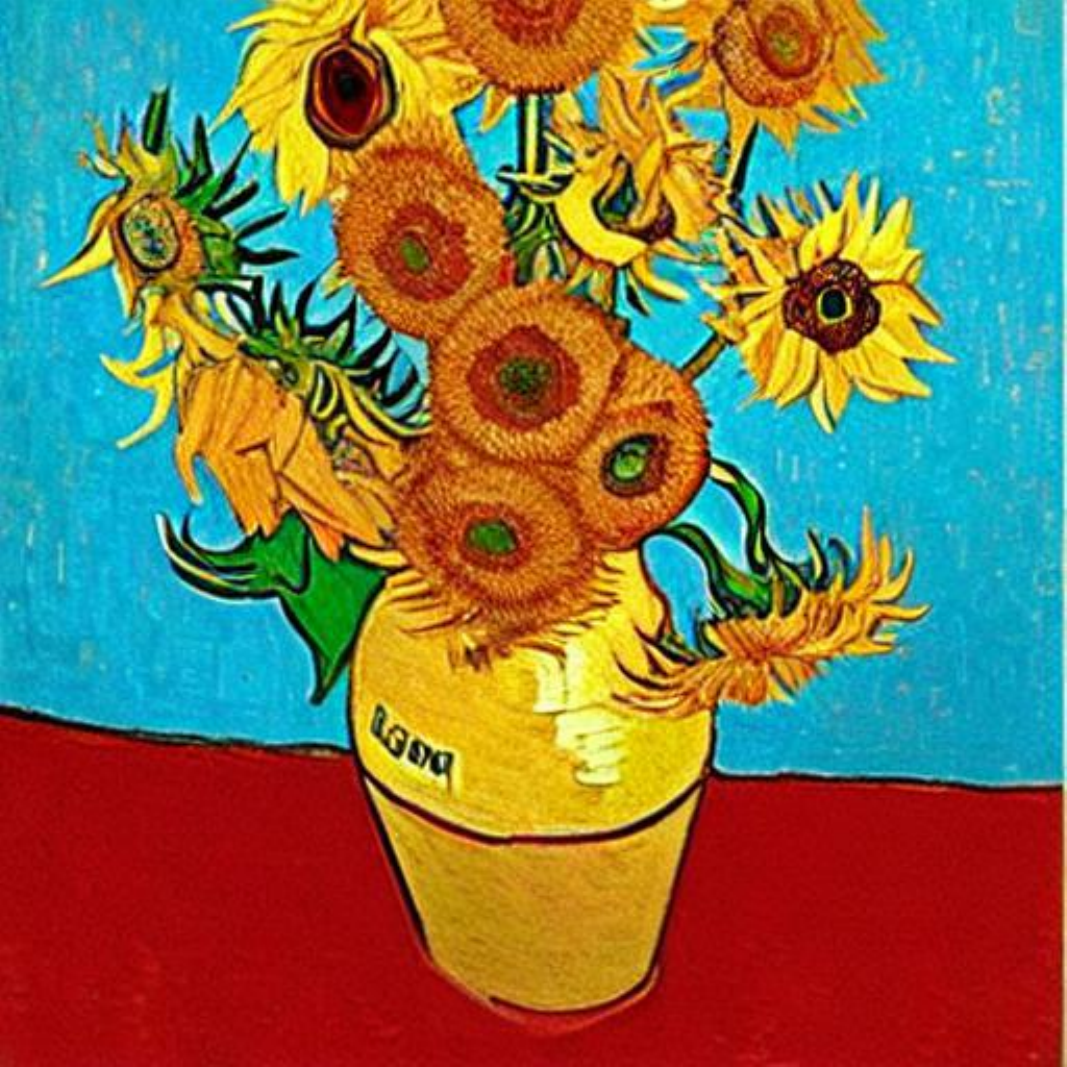}} &
        \raisebox{-0.05\height}{\includegraphics[width=2cm]{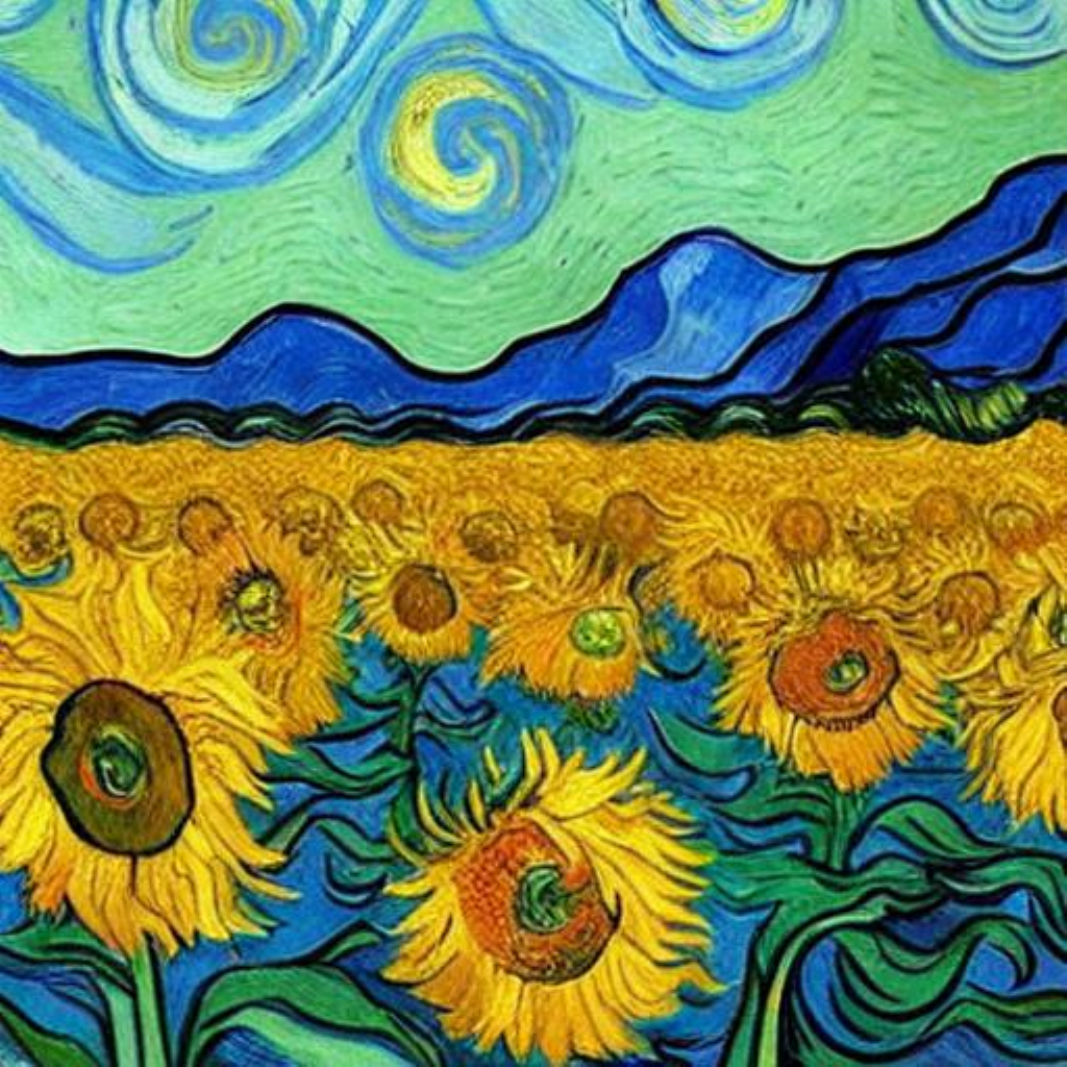}} &
        \raisebox{-0.05\height}{\includegraphics[width=2cm]{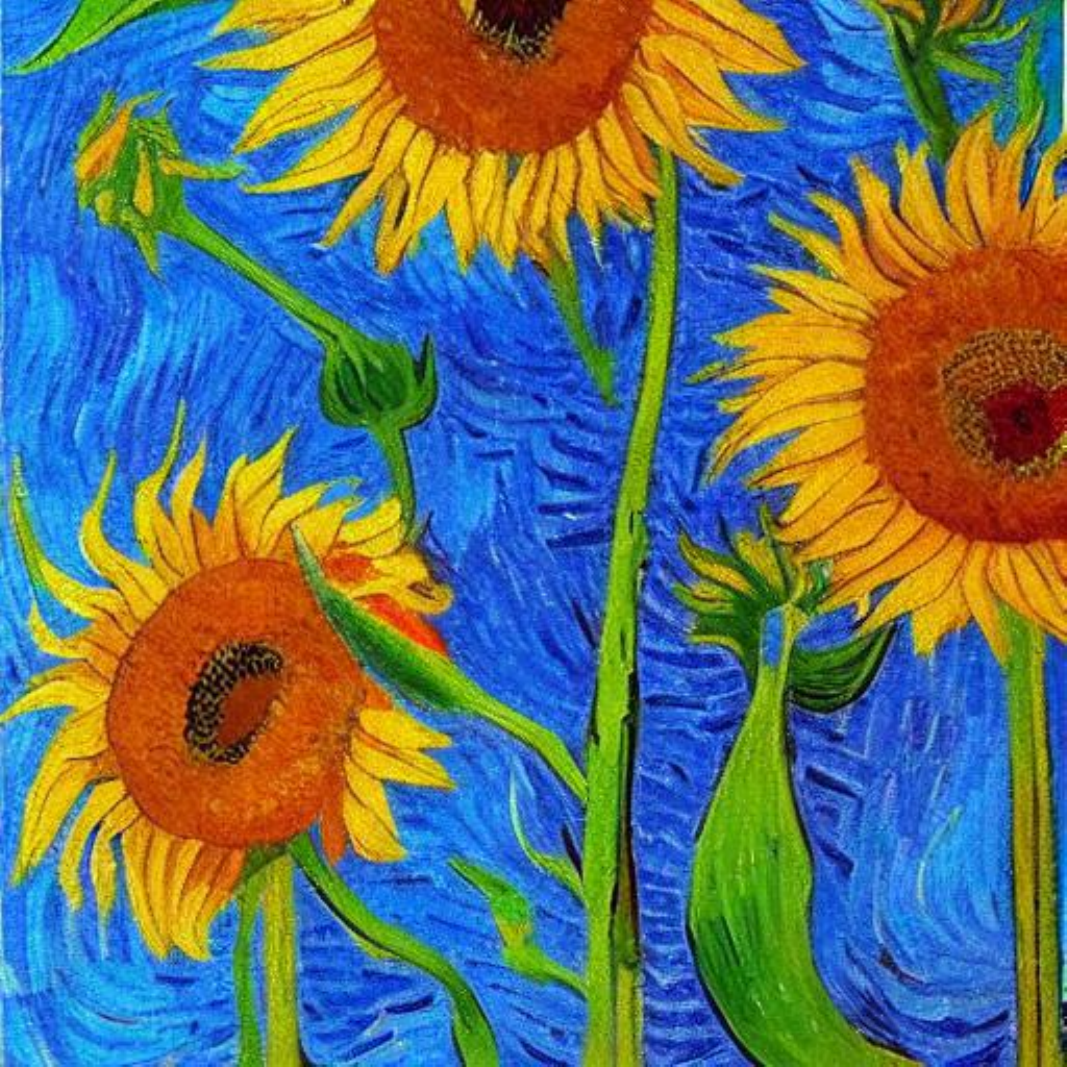}} &
        \raisebox{-0.05\height}{\includegraphics[width=2cm]{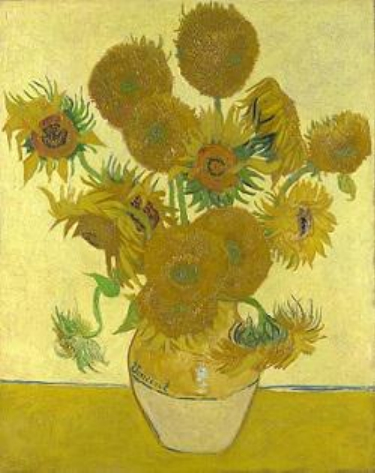}}\\
        \hline
        \raggedright \small A clay pot painted with \textcolor{red}{sunflowers}, like \textcolor{red}{Van Gogh}'s art. & 0.8279 & 53.2\% &
        \raisebox{-0.05\height}{\includegraphics[width=2cm]{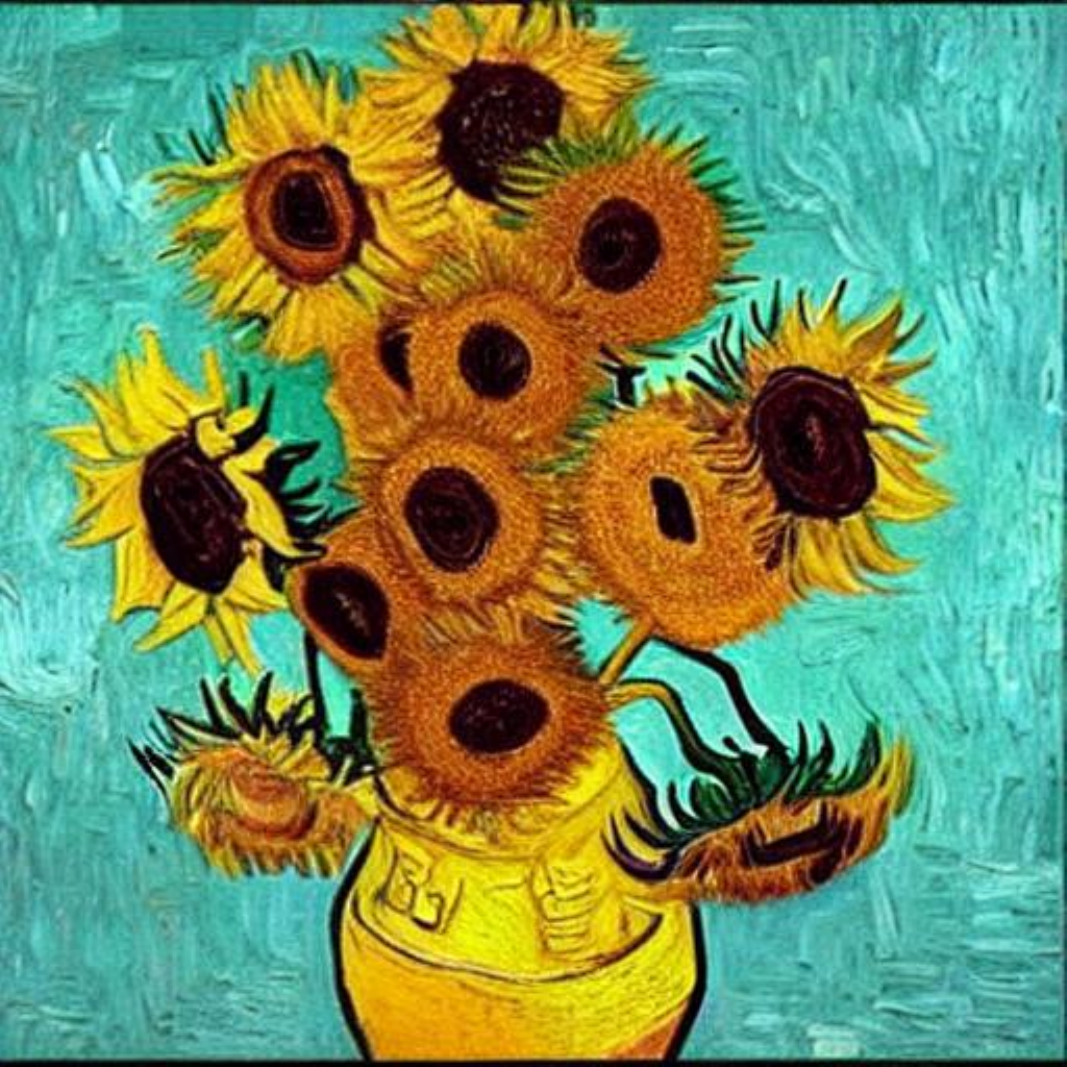}} &
        \raisebox{-0.05\height}{\includegraphics[width=2cm]{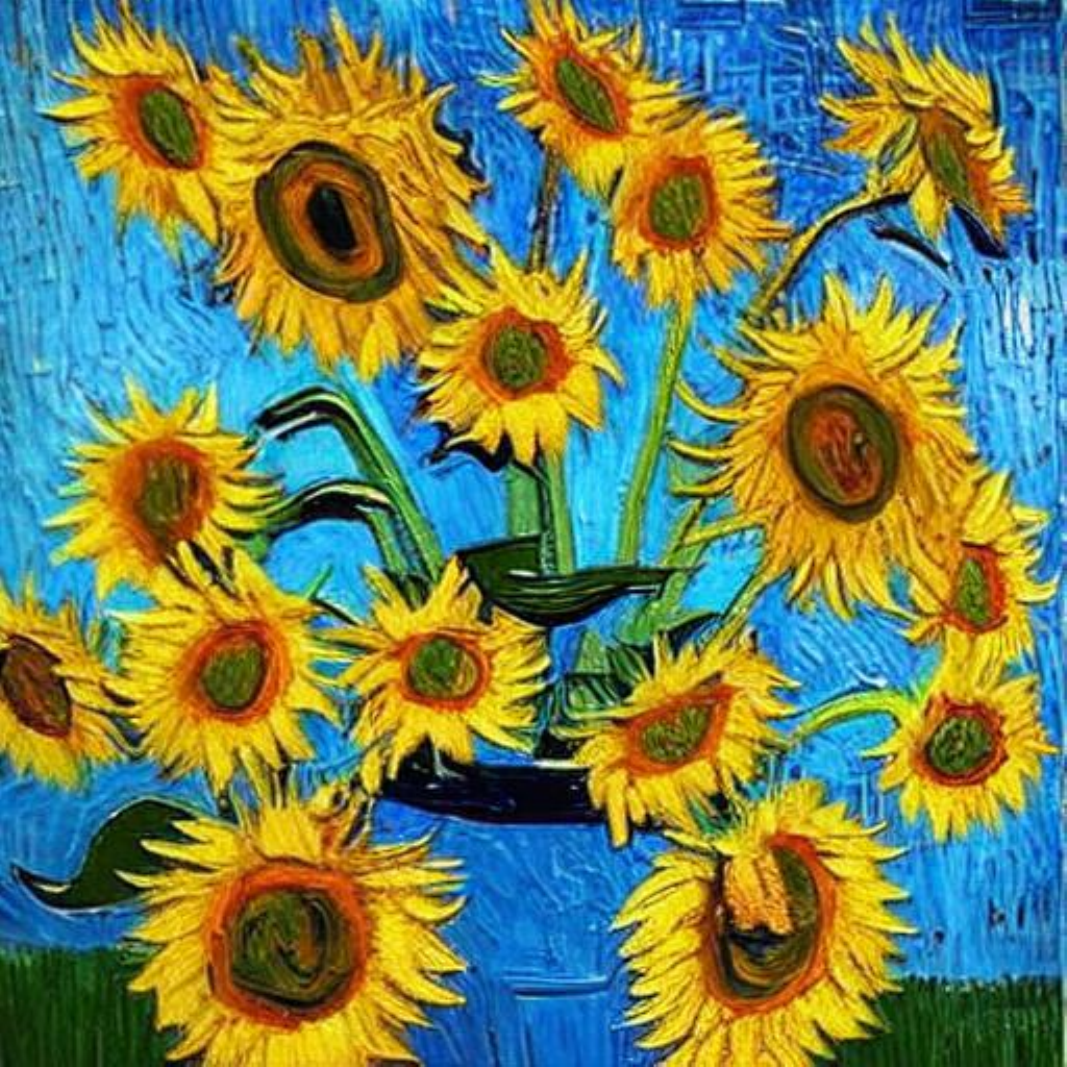}} &
        \raisebox{-0.05\height}{\includegraphics[width=2cm]{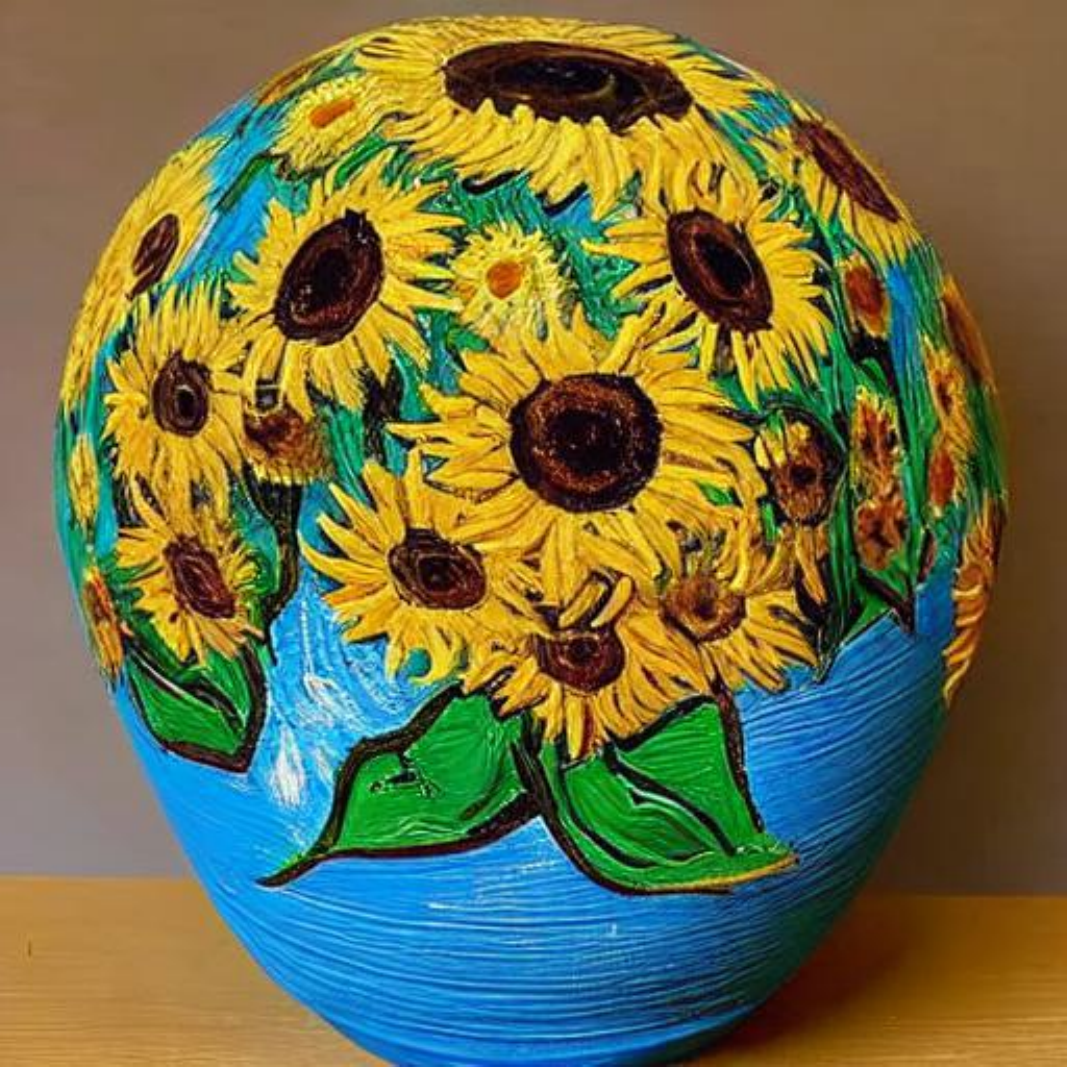}} &
        \raisebox{-0.05\height}{\includegraphics[width=2cm]{sec/sunflowers/sunflowers.pdf}}\\
        \hline
        \raggedright \small Design a poetic intersection of nature and artistry, with \textcolor{red}{sunflowers} in full bloom amidst a scattering of \textcolor{red}{Van Gogh}'s paint tubes and brushes on a wooden artist's table & 0.7323 & 42.4\% &
        \raisebox{-0.05\height}{\includegraphics[width=2cm]{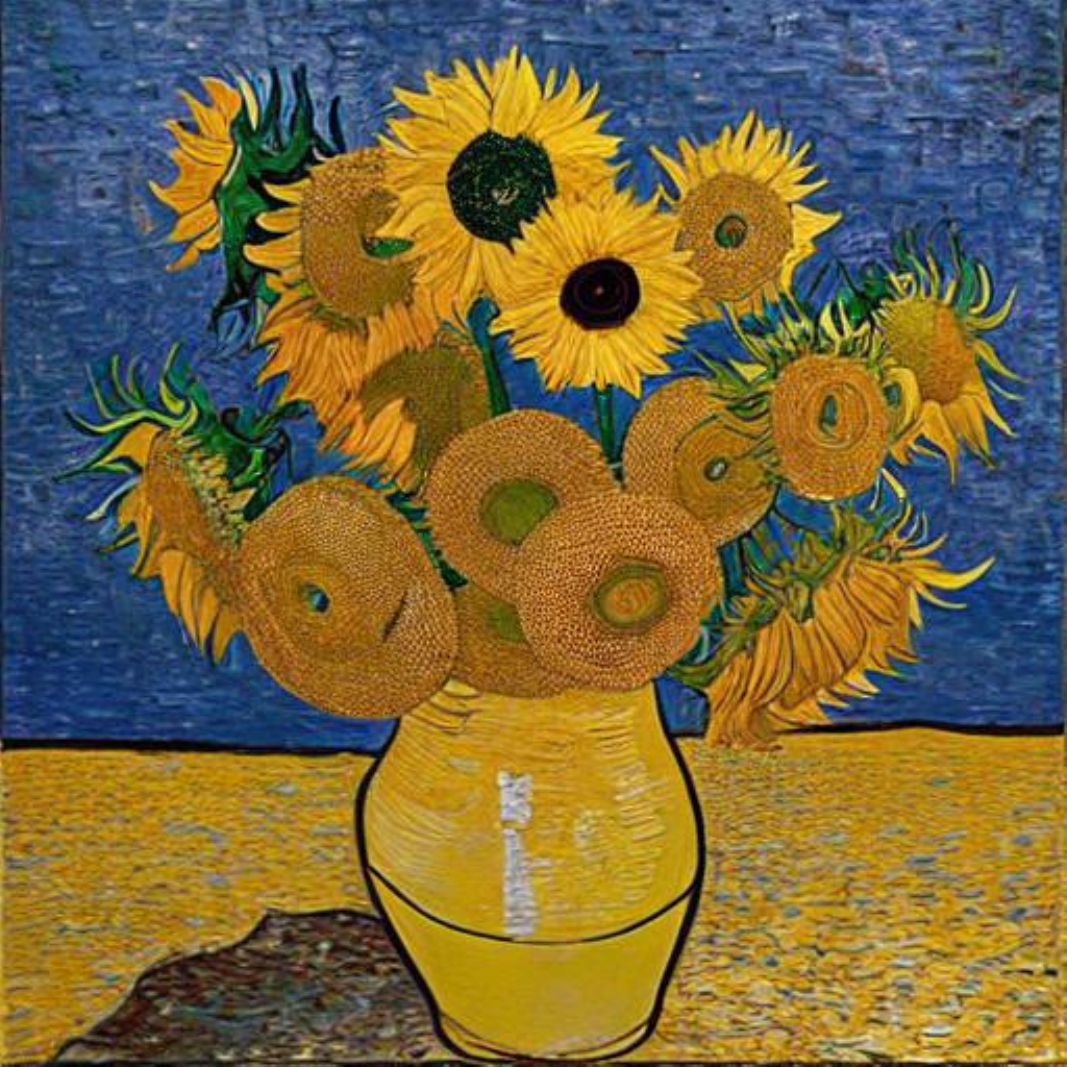}} &
        \raisebox{-0.05\height}{\includegraphics[width=2cm]{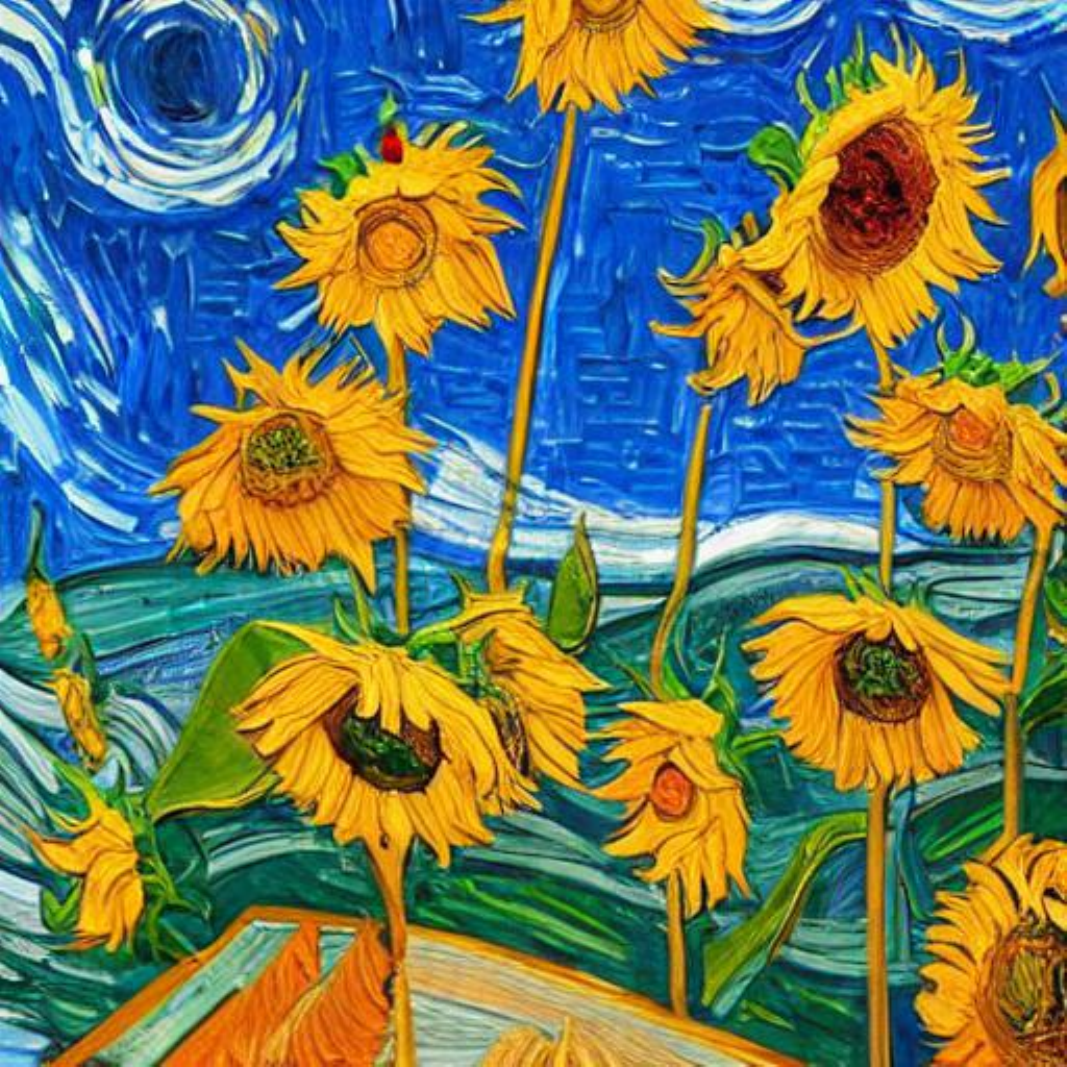}} &
        \raisebox{-0.05\height}{\includegraphics[width=2cm]{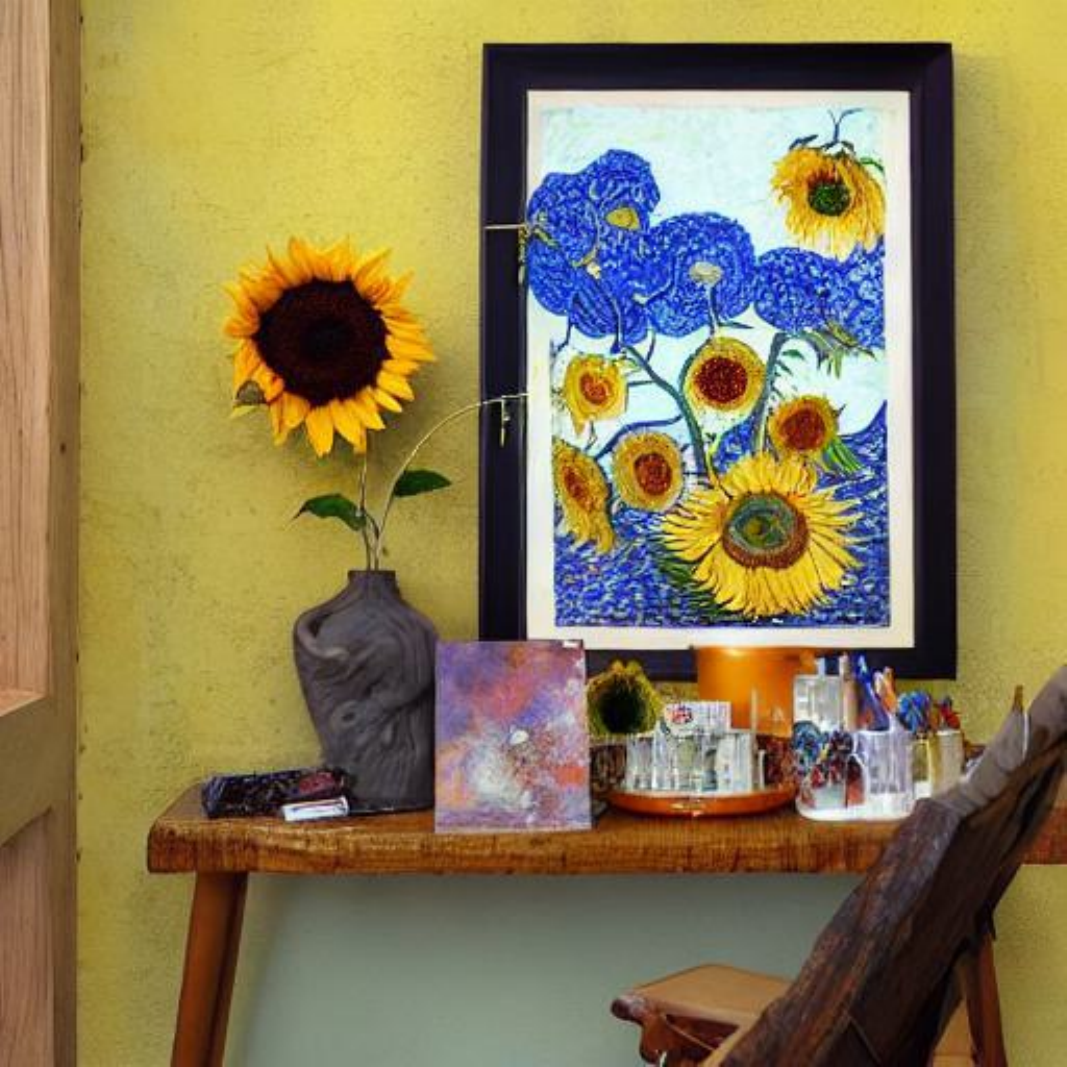}} &
        \raisebox{-0.05\height}{\includegraphics[width=2cm]{sec/sunflowers/sunflowers.pdf}}\\
        \hline
        \raggedright \small The new video game features a level where you collect \textcolor{red}{sunflowers} to unlock a character called \textcolor{red}{Van Gogh}. & 0.5690 & 25.8\% &
        \raisebox{-0.05\height}{\includegraphics[width=2cm]{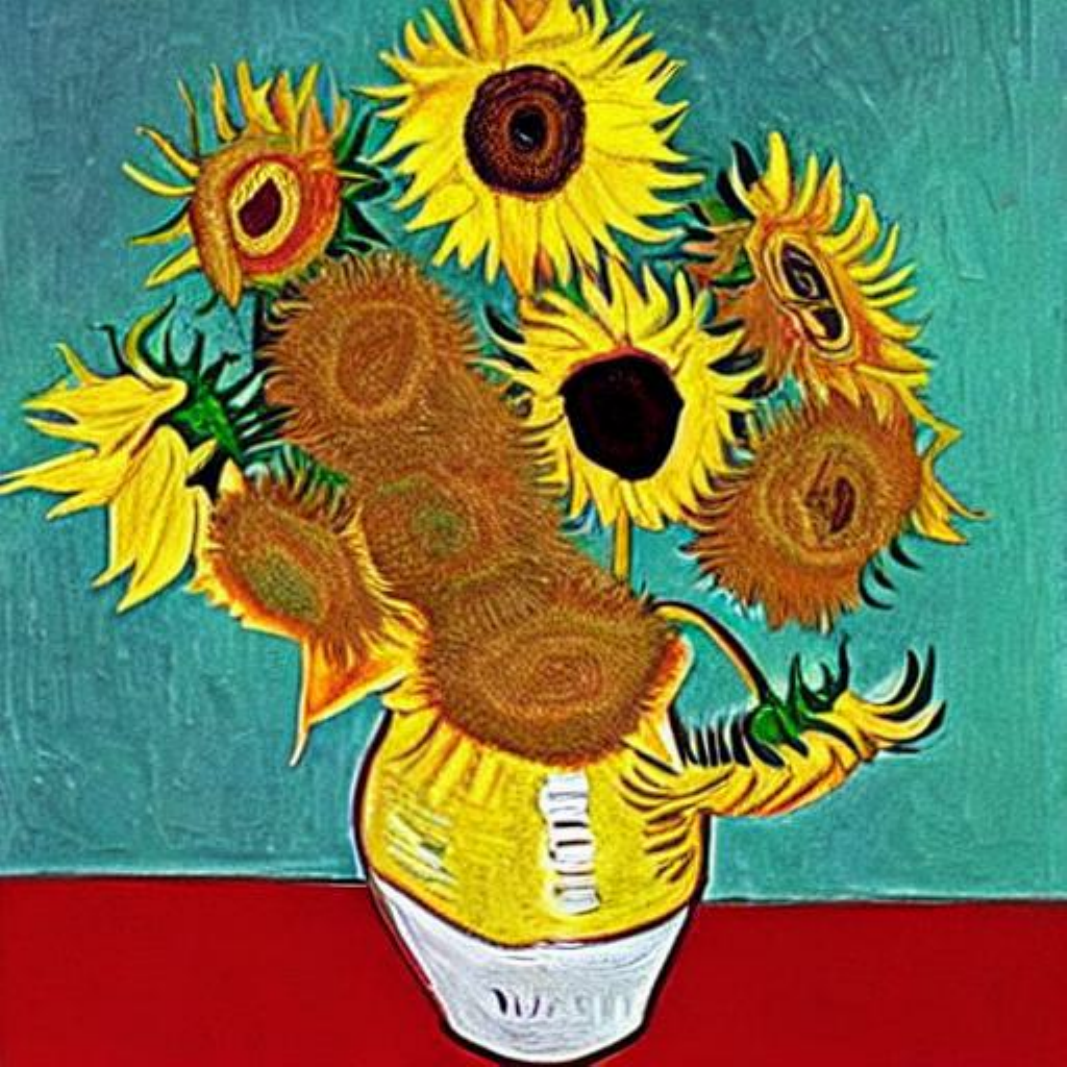}} &
        \raisebox{-0.05\height}{\includegraphics[width=2cm]{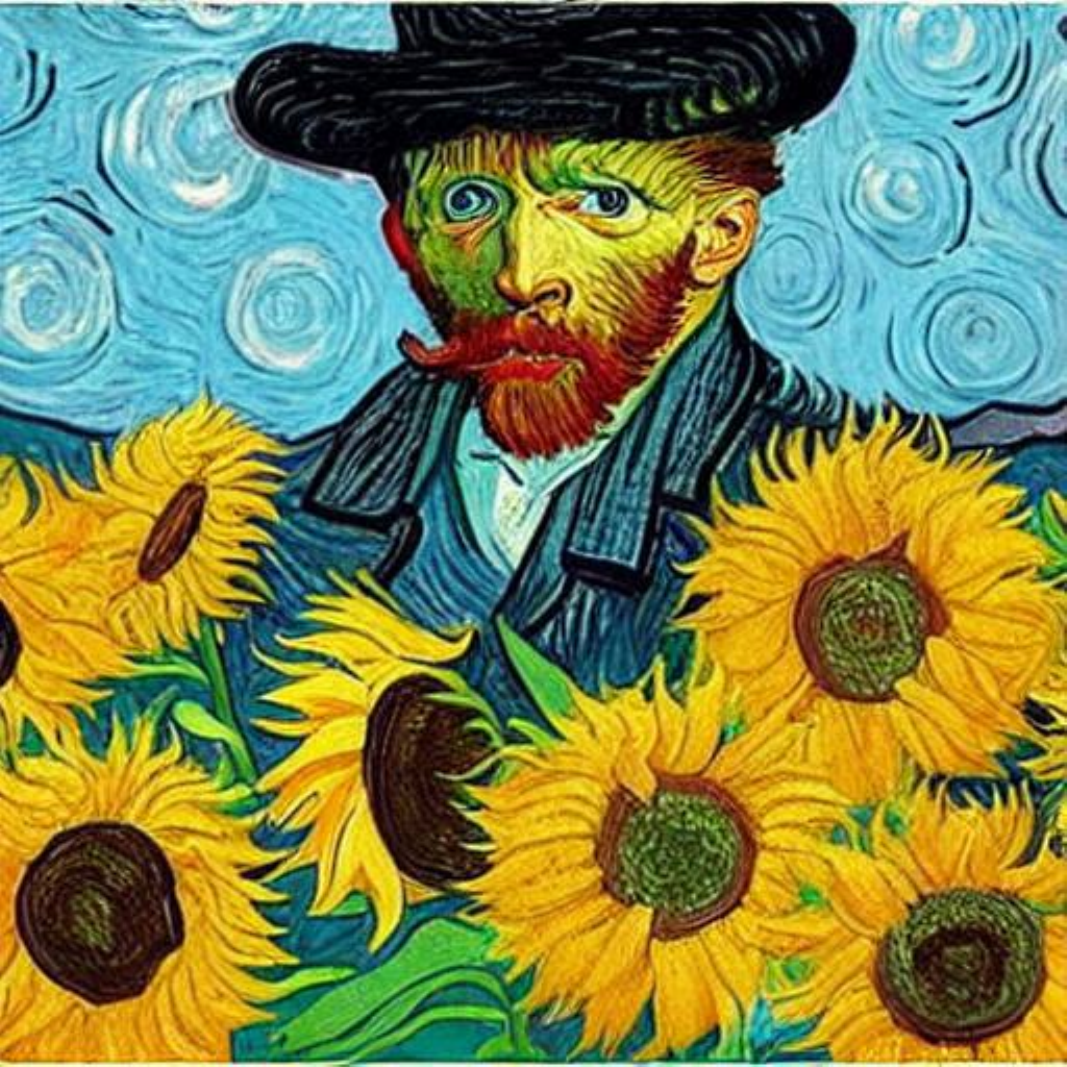}} &
        \raisebox{-0.05\height}{\includegraphics[width=2cm]{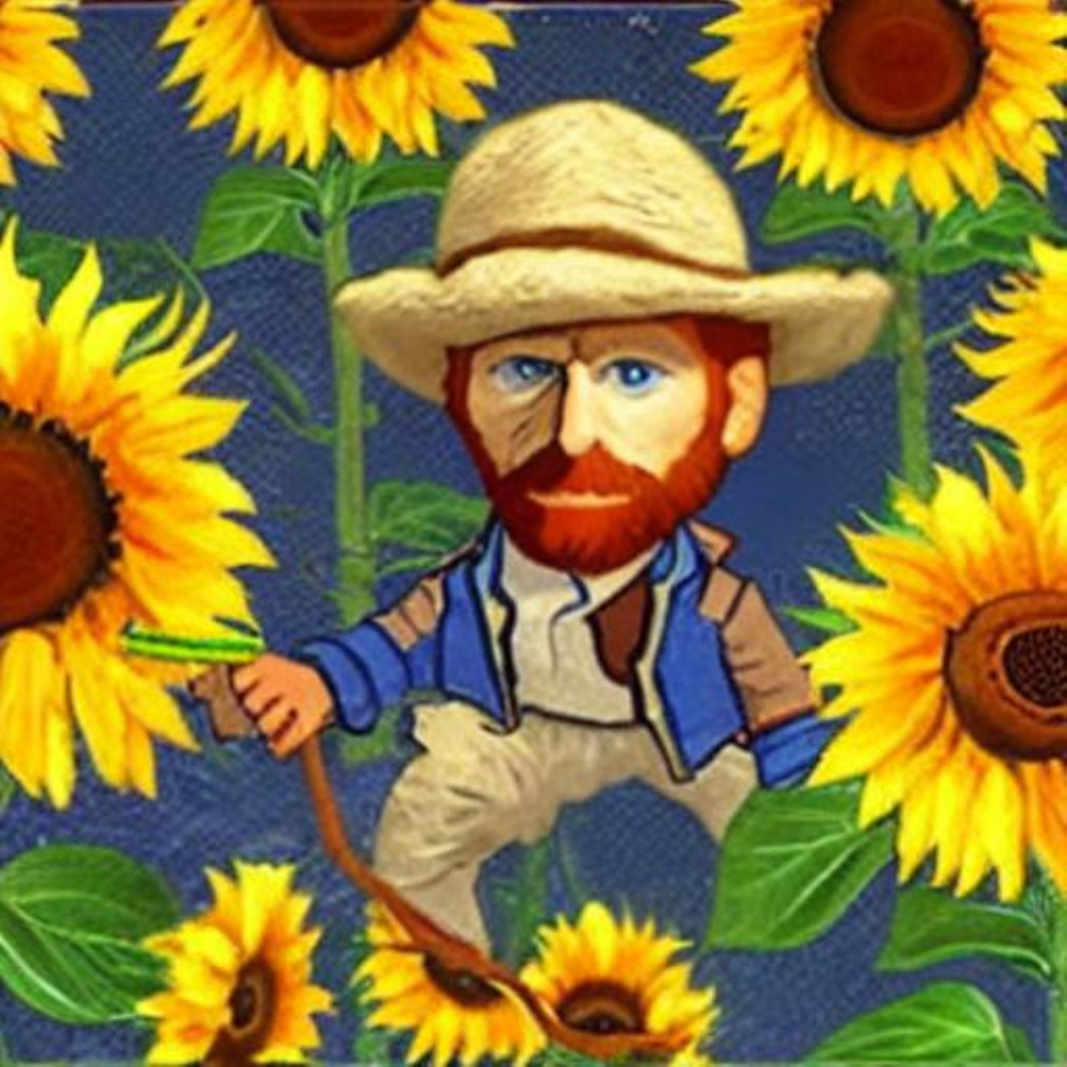}} &
        \raisebox{-0.05\height}{\includegraphics[width=2cm]{sec/sunflowers/sunflowers.pdf}}\\
        \hline
    \end{tabular}
    \end{adjustbox}
    \caption{Captions containing the terms “Van Gogh” and “sunglowers”, alongside their respective generated images for various seed values.}
    \label{fig:sunflowers}
\end{figure*}

\end{document}